\documentclass[twocolumn,tighten,numberedappendix]{aastex63}
\usepackage{amsmath}

\usepackage{lineno}

\def\simlt{\lower.5ex\hbox{$\; \buildrel < \over \sim \;$}}
\def\simgt{\lower.5ex\hbox{$\; \buildrel > \over \sim \;$}}

\def\beq{\begin{equation}}
\def\eeq{\end{equation}}
\def\ba{\begin{eqnarray}}
\def\ea{\end{eqnarray}}
\def\bB{\boldsymbol{B}}
\def\bE{\boldsymbol{E}}

\def\bj{\boldsymbol{j}}
\def\bv{\boldsymbol{v}}

\def\bk{\boldsymbol{k}}

\def\Sect{{\rm Section}} 
\def\Sects{{\rm Sections}} 

\def\Eq{Equation}
\def\Eqs{Equations}

\def\sT{\sigma_{\rm T}}

\def\Lw{L_{\nu}}
\def\Bw{B_{\rm w}}

\def\E{{\cal E}}
\def\N{{\cal N}}

\def\Grel{\Gamma_{\rm rel}}
\def\RLC{R_{\rm LC}}

\def\EFRB{{\cal E}_{\rm FRB}}

\def\omL{\omega_{\rm L}}
\def\bBw{\boldsymbol{B}_{\rm w}}

\def\tn{\tilde{n}}
\def\be{\boldsymbol{e}}

\def\bsh{\beta_{\rm sh}}

\def\tn{\tilde{n}}
\def\vgr{v_{\rm gr}}

\def\omB{\omega_B}

\def\Bbg{B_{\rm bg}}
\def\bBbg{\boldsymbol{B}_{\rm bg}}
\def\Ebg{E_{\rm bg}}

\def\tB{\tilde{B}}

\def\bb{\boldsymbol{\beta}}

\def\gD{\gamma_{\rm D}}
\def\bD{\beta_{\rm D}}

\def\bs{\beta_{\rm s}}
\def\gs{\gamma_{\rm s}}

\def\trho{\tilde{\rho}}

\def\xiin{\xi_{\rm i}}

\def\xiinc{\xiin^{\rm c}}

\def\trhou{\trho_{\rm u}}
\def\trhod{\trho_{\rm d}}

\def\c{\kappa}
\def\cu{\c_{\rm u}}
\def\cd{\c_{\rm d}}

\def\sigbg{\sigma_{\rm bg}}
\def\rhobg{\rho_{\rm bg}}
\def\nbg{n_{\rm bg}}

\def\Ttot{T}
\def\Ptot{P}
\def\Htot{H}

\def\Tp{T_{\rm p}}
\def\Hp{H_{\rm p}}
\def\Pp{P_{\rm p}}
\def\Up{U_{\rm p}}
\def\TEM{T_{\rm f}}
\def\HEM{H_{\rm f}}
\def\PEM{P_{\rm f}}
\def\Abg{A_{\rm bg}}

\def\Em{E_0}
\def\Km{K_0}
\def\Rm{R_{\times}}
\def\Bm{B_{\times}}
\def\sigm{\sigma_{\times}}

\def\gu{\gamma_{\rm u}}
\def\gd{\gamma_{\rm d}}
\def\betau{\beta_{\rm u}}
\def\betad{\beta_{\rm d}}
\def\Cu{C_{\rm u}}
\def\Cd{C_{\rm d}}

\def\tcross{t_{\rm cross}}
\def\cc{\c_{\rm c}}
\def\xic{\xi_{\rm c}}
\def\rc{r_{\rm c}}

\def\tc{t_{\rm c}}
\def\gc{\gamma_{\rm c}}

\def\e{\varepsilon}
\def\ed{\e_{\rm d}}
\def\eu{\e_{\rm u}}

\def\rK{r_\star}

\def\D{D}
\def\K{K}

\def\tv{t_{\rm v}}

\def\uus{u'_{\rm u}}
\def\uds{u'_{\rm d}}
\def\gus{\gamma'_{\rm u}}
\def\gds{\gamma'_{\rm d}}
\def\betaus{\beta'_{\rm u}}
\def\sigu{\sigma_{\rm u}}
\def\sigd{\sigma_{\rm d}}

\def\kk{k}
\def\hu{h_{\rm u}}
\def\hd{h_{\rm d}}
\def\pu{p_{\rm u}}
\def\pd{p_{\rm d}}

\def\ediss{\e_{\rm diss}}

\def\LMHD{L_{\rm MHD}}

\def\RF{R_{\rm F}}
\def\rsh{r_\times}

\def\bbD{\bb_{\rm D}}

\def\me{m}
\def\Aw{A_{\rm w}}
\def\bu{\boldsymbol{u}}

\def\gD{\gamma_{\rm D}}
\def\uD{u_{\rm D}}
\def\buD{\boldsymbol{u}_{\rm D}}

\def\tb{\tilde{\beta}}
\def\tg{\tilde{\gamma}}
\def\tu{\tilde{u}}
\def\tbu{\tilde{\boldsymbol{u}}}

\def\BMHD{B_{\rm MHD}}

\def\gRRL{\gamma_{\rm RRL}}

\def\etr{\e_{\rm tr}}

\def\KF{{\cal K}}
\def\omp{\omega_{\rm p}}
\def\nup{\nu_{\rm p}}
\def\dEe{\dot{\cal E}_e}
\def\Rh{R_{\rm h}}
\def\Lh{L_{\rm h}}
\def\Wp{W_{\rm p}}
\def\Ep{E_{\rm p}}

\newbox\grsign \setbox\grsign=\hbox{$>$} \newdimen\grdimen \grdimen=\ht\grsign
\newbox\simlessbox \newbox\simgreatbox \newbox\simpropbox
\setbox\simgreatbox=\hbox{\raise.5ex\hbox{$>$}\llap
     {\lower.5ex\hbox{$\sim$}}}\ht1=\grdimen\dp1=0pt
\setbox\simlessbox=\hbox{\raise.5ex\hbox{$<$}\llap
     {\lower.5ex\hbox{$\sim$}}}\ht2=\grdimen\dp2=0pt
\setbox\simpropbox=\hbox{\raise.5ex\hbox{$\propto$}\llap
     {\lower.5ex\hbox{$\sim$}}}\ht2=\grdimen\dp2=0pt
\def\simgt{\mathrel{\copy\simgreatbox}}
\def\simlt{\mathrel{\copy\simlessbox}}

\begin{document}

\title{Damping of strong GHz waves near magnetars and the origin of fast radio bursts}

\email{amb@phys.columbia.edu}

\author{Andrei M. Beloborodov}
\affiliation{Physics Department and Columbia Astrophysics Laboratory, Columbia University, 538  West 120th Street New York, NY 10027,USA}
\affil{Max Planck Institute for Astrophysics, Karl-Schwarzschild-Str. 1, D-85741, Garching, Germany}

\begin{abstract}
We investigate how a fast radio burst (FRB) emitted near a magnetar would propagate through its surrounding dipole magnetosphere at radii $r=10^7$-$10^9$\,cm. First, we show that a GHz burst emitted in the O-mode with luminosity $L\gg 10^{40}\,$\,erg/s is immediately damped for all propagation directions except a narrow cone along the magnetic axis. Then we examine  bursts in the X-mode. GHz waves propagating near the magnetic equator behave as magnetohydrodynamic (MHD) waves if they have $L\gg 10^{40}\,$erg/s. The waves develop plasma shocks in each oscillation and dissipate at $r\sim 3 \times 10^8\,L_{42}^{-1/4}$\,cm. Waves with lower $L$ or propagation directions closer to the magnetic axis do not obey MHD. Instead, they interact with individual particles and require a kinetic description. The kinetic interaction quickly accelerates particles to Lorentz factors $10^4$-$10^5$ at the expense of the wave energy, which again results in strong damping of the wave. In either propagation regime, MHD or kinetic, the dipole magnetosphere surrounding the FRB source acts as a pillow absorbing the radio burst and re-radiating the absorbed energy in X-rays. These results constrain the origin of observed FRBs. We argue that the observed FRBs avoid damping because they are emitted by relativistic outflows from magnetospheric explosions, so that the GHz waves do not need to propagate through the outer equilibrium magnetosphere surrounding the magnetar.
\end{abstract}





\section{Introduction}

Fast radio bursts (FRBs) are among the most mysterious astrophysical phenomena. They are detected at GHz frequencies from large cosmological distances. The bursts have huge luminosities up to $\sim 10^{44}$\,erg/s and millisecond durations \citep{Petroff19}. 

The short durations suggest that FRBs are  generated by compact objects. In particular, magnetars are natural candidates, as they are well known as prolific X-ray bursters \citep{Kaspi17}. Evidence for the magnetar-FRB association has been provided by the detection of millisecond GHz bursts from SGR~1935+2154, a known magnetar in our Galaxy \citep{Bochenek20,CHIME20}, although the bursts were weaker than the cosmological FRBs. The radio bursting mechanism is not established (see \cite{Lyubarsky21,Zhang22} for a review).

Useful  constraints on the FRB origin can be found by examining propagation of radio waves through the plasma magnetosphere surrounding magnetars. In particular, if the GHz source sits in the ultrastrong inner magnetosphere (which confines and powers the source) then the observed emission must be able to escape through the surrounding outer magnetosphere. Can the radio wave actually escape?

Two linear polarization modes are possible for electromagnetic waves in the magnetosphere: the O-mode and the X-mode \citep{Arons86}. As shown in \Sect~\ref{O-mode}, FRBs emitted in the O-mode experience immediate damping. For the X-mode, the propagation problem is more subtle, and it is solved in the remaining \Sects~\ref{MHD}-\ref{kinetic}. Hereafter by ``radio waves'' we mean the X-mode electromagnetic waves.  

A dangerous region for the X-mode wave is where the background dipole magnetic field $\Bbg$ decreases to $\sim \Em$ (the wave amplitude). In particular, calculation of the plasma response to a sine radio wave with $\Em>\Bbg$ shows its strong damping \citep{Beloborodov21b,Beloborodov22}. The wave quickly accelerates plasma particles up to the radiation reaction limit, and the particles radiate the received energy in the gamma-ray band. Effectively, the plasma scatters the radio wave to gamma rays, and then its energy converts to an avalanche of $e^\pm$ pairs. This calculation did not address how the oscillating wave reached the outer magnetosphere where $\Bbg<E_0$, but demonstrated that if it did then it would not survive. 

The present paper investigates the full evolution of radio waves emitted at small radii (where $\Bbg\gg\Em$) and propagating to the outer region where $\Bbg<\Em$. At small radii the X-mode wave has no problem with propagation  --- it is well described as a vacuum electromagnetic wave superimposed on the dipole background. This description fails where $\Bbg/\Em$ decreases to $\sim 1$. Here, the electromagnetic invariant $B^2-E^2$ approaches zero, and a dramatic transition occurs in the wave evolution. 

Kinetic plasma simulations of this transition show that the wave launches shocks in the background plasma \citep{Chen22b}. One can demonstrate the formation of shocks and track their evolution using the MHD framework (\citealt{Beloborodov23},  hereafter Paper~I). The MHD description holds for waves of sufficiently low frequencies, and then the X-mode radio wave behaves as a compressive MHD mode, called ``fast magnetosonic.'' Paper~I focused on kHz magnetosonic waves and showed that they evolve into monster radiative shocks, with Lorentz factors exceeding $10^5$. 

Remarkably, MHD description also holds for GHz radio waves of sufficiently high power $L>\LMHD$. As shown in the present paper, $\LMHD$ happens to be in the range relevant for FRB luminosities and its value $\LMHD(\theta)$ depends on the wave propagation angle $\theta$ relative to the magnetic dipole axis. In particular, near the magnetic equator the condition $L>\LMHD$ is satisfied by typical extragalactic FRBs. This fact allows us to solve for the wave propagation in the equatorial region using the MHD framework. Then, we examine waves around the magnetic axis, where $\LMHD>L$ and a kinetic description is required. We find that in both regimes, MHD and kinetic, the GHz waves are damped.


\section{Damping of O-modes}
\label{O-mode}

We will investigate propagation of waves from a putative GHz source through the surrounding $e^\pm$ dipole magnetosphere. The source sits sufficiently close to the magnetar, where $\Bbg$ far exceeds the amplitude $E_0$ of emitted waves,
\beq
\label{eq:E0_}
    E_0\ll\Bbg.
\eeq
This requires a source radius $r\ll 3\times 10^8\,\mu_{33}^{1/2}L_{42}^{-1/4}$\,cm, where $\mu$ is the dipole moment of the magnetosphere and $L=cr^2\Em^2/2$ is the emitted wave power. Condition~(\ref{eq:E0_}) corresponds to the local magnetospheric energy density $\Bbg^2/8\pi$ exceeding the wave energy density $E_0^2/8\pi$, as expected in any scenario picturing a FRB source confined in the magnetosphere and powered by the magnetosphere. The dipole magnetosphere forms the background for the emitted waves. Its most important property is the huge magnetization parameter $\sigbg\equiv \Bbg^2/4\pi\rhobg c^2\gg 1$, where $\rhobg$ is the plasma mass density.

\subsection{Review of wave modes}

Consider a harmonic wave described by its wavevector $\bk$, frequency $\omega=2\pi\nu$, amplitude $E_0$, and polarization. The frequency and amplitude determine the dimensionless strength parameter, 
\beq
\label{eq:a0}
  a_0=\frac{eE_0}{\me c \omega}\approx 2.3\times 10^5\, r_{8}^{-1} L_{42}^{1/2}\nu_9^{-1}.
\eeq
The oscillating wave fields $\bE_{\rm w}=\bE$ and $\bBw=\bB-\bBbg$ satisfy the induction equation $\partial_t\bB_{\rm w}=-c\nabla\times\bE$:
\beq
   \omega\bB_{\rm w}=c\bk\times\bE.
\eeq
The electric charge density $\rho_e$ and current density $\bj$ in the wave are determined by the Maxwell equations,
\beq
  4\pi\rho_e=i\bk\cdot\bE, \qquad 4\pi\bj=ic\bk\times\bBw+i\omega\bE.
\eeq
As long as the Larmor frequency of plasma particles $\omL$ far exceeds $\omega$, they respond to the wave through the guiding center motion, including the $\bE\times\bB$ drift with velocity $\bv_{\rm D}=c\bE\times\bB/B^2$ (whose contribution to the electric current is $\bj_{\rm D}=\bv_{\rm D}\rho_e$).

Linear wave modes in the limit of $\sigbg\rightarrow\infty$ and $\omL/\omega\rightarrow \infty$ have been examined by \cite{Arons86}. The modes have two possible polarizations: $\bE$ oscillates either along vector $\boldsymbol{n}=\bk\times\bBbg$ (X-mode) or perpendicular to $\boldsymbol{n}$ (O-mode). Both modes have a group speed nearly equal to the speed of light, $\vgr/c\approx 1-\sigbg^{-1}$. For the X-mode, $\bk\cdot\bE=0$ implies $\rho_e=0$ and $\bj_{\rm D}=0$. The X-mode excites a negligible electric current and propagates as in vacuum despite the presence of plasma, with the dispersion relation $\omega\approx ck$. By contrast, the O-mode may have a component of $\bE$ parallel to $\bBbg$ (which will be denoted $E_\parallel$) and drive strong plasma response. This section focuses on the O-modes.

In the limit of a low plasma density (vacuum), the O-mode would have
\beq
  E_\parallel=E\sin\alpha \qquad \mbox{(O-mode~in~vacuum)}, 
\eeq
where $\alpha$ is the angle between $\bk$ and $\bBbg$. We will assume $\alpha\neq 0$.\footnote{The case of $\alpha=0$ is degenerate: the O-mode behaves as an X-mode. However, sustaining $\alpha=0$ along the ray is possible only for waves propagating exactly along the magnetic dipole axis.}
In the presence of $e^\pm$ plasma, charged particles can freely slide along the magnetic field lines and tend to screen $E_\parallel$. Therefore, the O-mode dispersion relation $\omega(\bk)$ \citep{Arons86} depends on the plasma density $n$ or the corresponding plasma frequency, 
\beq
\label{eq:omp}
   \nup=\frac{\omp}{2\pi} = \left(\frac{e^2 n}{\pi \me}\right)^{1/2} \approx 0.9\,n_{10}^{1/2}\,{\rm GHz},
\eeq
where $\me$ is the electron/positron mass. A typical $e^\pm$ density distribution around magnetars is $n(r)\sim 10^{10}\,r_9^{-3}$\,cm$^{-3}$ \citep{Beloborodov20}. It implies $\omp>\omega$ at radii $r<10^{9}$\,cm. 

If plasma succeeds in screening, $E_\parallel=0$, the O-modes become relativistic Alfv\'en waves of  ideal MHD, which satisfy $\bE\cdot\bB=0$ and $\bE\cdot\bj=0$. At $\sigbg\gg 1$, the Alfv\'en waves have $\Bw\approx E$ and $\omega(\bk)\approx ck_\parallel$. Their group velocity $\bv_{\rm gr}=\nabla_{\bk}\omega$ is parallel to $\bBbg$, so Alfv\'en waves are ducted along the magnetic field lines and cannot escape the closed magnetosphere. One can also show that the oscillating electric current supporting the Alfv\'en wave is parallel to $\bBbg$ and given by
\begin{equation}
\label{eq:jw}
   j_{\rm A} \approx\frac{i\omega}{4\pi}\,E\sin\alpha.
\end{equation}

The condition for small-amplitude O-modes to behave as Alfv\'en waves with $E_\parallel=0$ is $\omp>\omega$. However, this screening condition holds only for waves with $a_0\ll 1$.\footnote{The screening condition for  O-modes with $a_0\ll 1$ is derived from the non-relativistic plasma response to the unscreened $E_\parallel$: a harmonic wave of $E_\parallel$ generates $e^\pm$ velocities $v_\parallel^\pm=\pm ieE_\parallel/\me\omega = \pm i c a_0 \sin\alpha$ and creates current $j=ie^2nE_\parallel/\me\omega=i\omp^2E_\parallel/4\pi\omega$.
If $j>j_{\rm A}$, the assumption of unscreened $E_\parallel$ becomes inconsistent, i.e. $E_\parallel$ is screened. This occurs if  $\omp>\omega$.}
For waves with $a_0> 1$, the screening $j=j_{\rm A}$ depends not only on $\omp/\omega$ but also on the wave amplitude $E_0$. This occurs because $j_{\rm A}\propto E_0$ while the maximum speed $c$ limits the electric current to $j<cen$, so $j$ is no longer proportional to $E_0$. O-modes with $\omega<\omp$ behave as Alfv\'en waves at small amplitudes $E_0$, and become ``charge starved'' at sufficiently large $E_0$ when $j_{\rm A}>enc$. Note that this can happen even when $\Em\ll\Bbg$. Charge starvation was previously discussed for Alfv\'en waves excited by neutron star quakes \citep{Thompson98,Bransgrove20,Kumar20b,Chen22a,Kumar22}. 

GHz O-modes enter charge-starvation much easier than kHz waves, as one can see from 
\beq
   \kappa\equiv\frac{|j_{\rm A}|_{\max}}{ecn}\approx \frac{\omega E_0\sin\alpha}{4\pi e cn}=\frac{\omega^2}{\omp^2}\,a_0\sin\alpha.
\eeq
Using $a_0$ from \Eq~(\ref{eq:a0}) and the typical $\omp$ from \Eq~(\ref{eq:omp}), one finds that the strong GHz O-modes have $\kappa>1$ unless $\alpha$ is nearly zero (i.e. $\bk\parallel\bBbg$, which is not sustainable in a curved $\bBbg$). Note that the regime of $\kappa>1$ does not always imply that the wave develops a strong $E_\parallel$: numerical experiments with Alfv\'en waves launched into a uniform background with $\kappa>1$ show that they remain Alfv\'en waves --- they sustain $j_{\rm A}$ by advecting and compressing a charge-separated plasma \citep{Chen22a}. However, in a non-uniform background (the dipole magnetosphere) Alfv\'en waves are sheared and their wavevector becomes increasingly oblique to $\bBbg$, i.e. angle $\alpha$ grows with time \citep{Bransgrove20,Chen22a}. Then, the plasma advected by the wave can  become insufficient to sustain $j_{\rm A}$ and a strong $E_\parallel$ develops.

Entering the unscreened regime with $E_\parallel\approx E\sin\alpha$ is a necessary condition for the O-mode escape, as otherwise the wave is ducted along the closed magnetic field lines. Therefore, in the remainder of this section we focus on O-modes that have developed the unscreened $E_\parallel$. Their dispersion relation is similar to vacuum electromagnetic waves and, like the X-modes, they can have $\bv_{\rm gr}\nparallel\bBbg$, so they are no longer ducted along $\bBbg$. This condition is necessary, but not sufficient for the O-mode escape, as escape also requires the waves to avoid damping.

\subsection{Energy losses of unscreened O-modes}

The escape of unscreened O-modes is hindered by their damping: the wave experiences energy losses because of particle acceleration by the $E_\parallel$. The $E_\parallel$ oscillates with amplitude $E_0\sin\alpha$ and accelerates the plasma particles in each oscillation up to the Lorentz factor
\beq
  \gamma\approx a_0\sin\alpha.
\eeq 
The particles with high $\gamma$ emit radiation at the expense of the electromagnetic energy of the O-mode. The power radiated by each particle $\dEe$ is determined by its motion in the wave and may be evaluated as follows.

The accelerated particles remain strongly magnetized in the inner magnetosphere, since their Larmor frequency $\omL=\omB/\gamma$ far exceeds the wave frequency $\omega$. The motion of magnetized particles is a  combination of sliding along $\bB$ with velocity $\bb_\parallel$ and drifting perpendicular to $\bB$ with $\bb_\perp=\bE\times \bB/B^2$ (other, slower drifts of the guiding center will be neglected below). Since the O-mode magnetic field $\bBw=\bB-\bBbg$ oscillates perpendicular to $\bBbg$, the direction of $\bB$ is tilted by the oscillating angle $\psi$ given by $\tan\psi=\Bw/\Bbg$. So, the direction of $\bb_\parallel\parallel\bB$ oscillates in the $\bBbg$-$\bBw$ plane with frequency $\omega$ and amplitude $\psi_0=\arctan(E_0/\Bbg)\approx E_0/\Bbg$. The oscillation amplitude of $\bb_\perp$ is of the same order or smaller, depending on $\alpha$. As a result, the particle velocity vector $\bb$ periodically deviates from $\bBbg$ by an angle $\sim \psi_0$, and the particle executes a curved orbit in each wave oscillation, with a characteristic curvature radius
\beq
  r_c\sim \frac{c}{\omega\psi_0} \sim \frac{c\Bbg}{\omega E_0}.
\eeq
The power of curvature radiation emitted by the relativistic particle \citep{Landau75} is 
\beq
  \dEe=\frac{2c\, e^2\gamma^4}{3r_c^2}\sim \frac{2e^2}{3c} \omega^2 a_0^4\,\frac{\Em^2}{\Bbg^2} \sin^4\!\alpha.
\eeq
This power is enormous, and the wave may avoid losses only if $\alpha$ is small, which is sustainable only near the magnetic dipole axis. The corresponding constraint on the propagation direction defines a narrow escape cone as shown below. 

Plasma with density $n$ emits energy with rate $n\dEe$, and so the wave experiences irreversible loss of energy on the timescale 
\beq
   t_{\rm damp} = \frac{E_0^2}{8\pi n\dEe}  \sim \frac{3c\, \sigbg}{4r_e  \omega^2 a_0^4 \sin^4\!\alpha},
\eeq 
where $r_e=e^2/\me c^2$. The efficiency of damping can be seen by comparing the wave travel time $r/c$ with $t_{\rm damp}$. 

Consider a wave with a radial wavevector $\bk$ at angle $\theta$ relative to the magnetic dipole axis. Angle $\alpha$ between $\bk$ and $\bBbg$ satisfies $2\tan\alpha=\tan\theta$ for a dipole $\bBbg$. Using \Eq~(\ref{eq:a0}) we find at $\theta<1$
\beq
   \frac{ct_{\rm damp}}{r}\sim \frac{3 \me^2 c^4\omega^2\sigbg r^3}{r_e^3  L^2 \theta^4}.
\eeq 
The escape condition $ct_{\rm damp}/r\gtrsim 1$ gives an upper limit for the luminosity of the escaping O-mode,
\beq
\label{eq:L_Omode}
   L(\theta)\simlt \frac{\sqrt{3}\, \me c^2\omega(\sigbg r^3)^{1/2}}{r_e^{3/2} \theta^2}.
\eeq
For magnetars in a persistent (non-bursting) state, one expects $\sigbg r^3\sim 10^{34}$\,cm$^3$ as a typical value (Paper~I). Such values of $\sigbg$ would allow the escape of O-mode with luminosity $L$ within a cone of $\theta\simlt 0.08\,\nu_9^{1/2}\,L_{42}^{-1/2}$. The corresponding maximum solid angle for escape is
\beq 
\label{eq:Omode_escape1}
  \frac{\delta\Omega_{\rm max}}{2\pi}=\frac{\theta_{\rm max}^2}{2}\sim 3\times 10^{-3}\, \frac{\nu_9}{L_{42}}\left(\frac{\sigbg r^3}{10^{34}\,{\rm cm}^3}\right)^{1/2}.
\eeq
This constraint is strongest for FRBs with the highest $L\simgt 10^{43}\,$erg\,s$^{-1}$ and for lowest frequencies ($\nu\approx  0.1\,$GHz is the lowest observed in FRBs so far).

The value of $\sigbg$ entering the limit~(\ref{eq:Omode_escape1}) has been normalized to an optimistically large value. Its actual value can be dramatically reduced by $e^\pm$ creation that accompanies curvature emission by the accelerated particles. Note that the spectrum of curvature emission declines exponentially at frequencies above $\omega_c= (3/2)\gamma^3 c/r_c$, so the characteristic energy of emitted photons is
\beq
  \hbar\omega_c\sim\frac{3}{2} \hbar\omega\frac{E_0}{\Bbg} a_0^3\sin^3\!\alpha.
\eeq
If $\hbar\omega_c\simgt\me c^2$, a huge number of curvature photons convert to $e^\pm$ pairs via photon-photon collisions. It is easy to verify that this process gives plasma density sufficient to put the O-mode back into the Alfv\'en wave regime (the wave is no longer charge starved), preventing its escape. 

Thus, a necessary condition for O-mode escape is $\hbar\omega_c\ll\me c^2$. The wave needs to avoid the avalanche of copious $e^\pm$ creation by the photons in the far exponential tail of curvature emission, which requires $\hbar\omega_c\simlt 0.1\me c^2$. This gives the constraint
\beq
   \theta\simlt \frac{2}{a_0}\left(\frac{0.1\me c^2}{\hbar\omega}\frac{\Bbg}{E_0}\right)^{1/3}
   \approx 0.05\, \frac{r_8^{1/3}\mu_{33}^{1/3}}{\nu_9^{2/3}L_{42}^{2/3}}.
\eeq
One now finds the maximum solid angle for O-mode escape from a source sitting at a radius $r<10^8\,$cm,  
\beq 
\label{eq:Omode_escape2}
  \frac{\delta\Omega_{\rm max}}{2\pi}=\frac{\theta_{\rm max}^2}{2}\approx 10^{-3}\, \frac{\mu_{33}^{2/3}}{\nu_9^{4/3}L_{42}^{4/3}}.
\eeq
This constraint is independent of the background plasma density $\rhobg$ (or the corresponding $\sigbg$). It becomes particularly tight for the brightest FRBs with $L\simgt 10^{43}$\,erg\,s$^{-1}$ and the highest frequencies observed in FRBs so far, $\nu=5-8$\,GHz.


\section{Radio waves in MHD regime}
\label{MHD}

The remaining sections will focus on propagation of X-modes. They propagate freely with negligible coupling to the magnetospheric plasma when $E_0\ll\Bbg$. The X-modes experience no damping until nonlinear effects develop. These effects appear where linear propagation would violate the condition $E^2<B^2$.

\subsection{Formulation of the problem}

Suppose a GHz wave packet is emitted near the magnetar and then expands to larger radii $r$ where the magnetosphere is initially unperturbed. We are interested in the evolution of the packet at $r\sim 10^8-10^9\,$cm where $E^2$ approaches $B^2$ and the linear vacuum-like propagation ends. These radii are still well inside the light cylinder $\RLC=c/\Omega\sim 10^{10}\,$cm, so rotation of the magnetosphere is slow, $\Omega r\ll c$, and may be neglected. The unperturbed magnetosphere here may be described as a dipole $\bBbg$.  

The unperturbed outer magnetosphere in front of the wave is populated with mildly relativistic electrons and positrons (their speeds are reduced by drag exerted by the magnetar radiation, see \cite{Beloborodov13a}). Then, the energy density of the background plasma is comparable to its rest mass-density $\rhobg c^2$. This density enters the definition of the magnetization parameter,
\beq
\label{eq:sigbg}
   \sigbg\equiv \frac{\Bbg^2}{4\pi \rhobg c^2} \approx \frac{\D}{r^3}, 
   \qquad D=\frac{\mu^2}{4\pi \N \me c^2},
\eeq
where $\me$ is the electron mass, and $\mu$ is the magnetic dipole moment of the magnetar. The dimensionless parameter $\N\equiv r^3\rhobg/\me$ is approximately constant with radius $r$; its typical expected value is $\N\sim 10^{37}$ \citep{Beloborodov20}. 

We will consider a wave packet far from its source. It occupies a thin shell $\delta r/r\ll 1$ and has a nearly radial wavevector $\bk$, so the packet behaves locally as part of an axisymmetric wave ($\partial_\phi\approx 0$). Magnetosonic waves have a toroidal electric field $\bE\parallel \bk\times\bBbg$, and we define
\beq
   E\equiv -E_{\phi},
\eeq
using the normalized basis $\be_r,\be_{\theta},\be_{\phi}$ of the spherical coordinate system $r,\theta,\phi$ with the polar axis along the magnetospheric dipole moment $\boldsymbol{\mu}$. Our calculation will track the propagation of the spherically expanding wave packet. As a concrete example, we will consider a radio wave launched with an initial sine profile
\beq 
\label{eq:initial}
   E(\xi)=\Em\sin(\omega\xi), \qquad 0<\xi\equiv t-\frac{r}{c}<\tau.
\eeq
The packet has a short duration $\tau\simlt 1$\,ms, and we are interested in its propagation at radii $r\gg c\tau$.

As long as the magnetospheric particles exposed to the wave remain magnetized, i.e. their Larmor frequency far exceeds the wave frequency $\omega$, the radio wave obeys MHD and can be thought of as a fast magnetosonic wave (the validity of MHD description will be discussed in detail in \Sect~\ref{MHD_kinetic}). Particle motion in the MHD wave can be thought of as the drift of the Larmor orbit. The $e^\pm$ drift velocity $\bb_\pm$ has a charge-symmetric (MHD) component $\bb$ and a small antisymmetric component $\pm\bb_p$ (polarization drift), which sustains the electric current $\bj=en\bb_p$. There is no need for explicit calculations of these drifts in response to the electromagnetic wave. Instead, MHD describes the evolution of fields $\bE$ and $\bB$ by treating the plasma as a perfectly conducting fluid, which satisfies $\bE+\bv\times\bB/c=0$. The fluid is described by its velocity $\bv=\bb c$ and mass density $\rho=\me n$. The unperturbed static background corresponds to $\bE=0$, $\bB=\bBbg$, $\bv=0$, and $\rho=\rhobg$. 

At small radii, where $\Em/\Bbg\ll 1$, the wave  propagates without deformation. It has the speed $v_{\rm wave}/c=1-\sigbg^{-1}\approx 1$, and the MHD wave is equivalent to a vacuum electromagnetic wave superimposed on $\Bbg$. The linear propagation ends where the linear superposition hits the condition $E^2=B^2$, which corresponds to $v\rightarrow c$ (Paper~I). In the equatorial plane, this occurs at radius $\Rm$ where $\Em=\Bbg/2$,
\beq 
\label{eq:Rm}
  \Rm=\left(\frac{c\mu^2}{8L}\right)^{1/4}
  \approx 2.47\times 10^8\, \frac{\mu_{33}^{1/2}}{L_{42}^{1/4}}{\rm ~cm}.
\eeq
Here, $L=c r^2\Em^2/2$ is the wave power.

We wish to find the nonlinear evolution of the electromagnetic wave as it crosses radius $\Rm$. Note that the evolution occurs at a very high magnetization parameter $\sigbg$. In particular, at $\Rm$ one finds
\beq
\label{eq:sigm}
   \sigm\equiv\sigbg(\Rm) \approx 6.4\times 10^8\,\frac{\mu_{33}^{1/2}L_{42}^{3/4}}{\N_{37}}.
\eeq
Numerical examples shown below will assume a magnetar with a typical magnetic dipole moment $\mu=10^{33}$\,G\,cm$^3$ and plasma density parameter $\N=10^{37}$.

\subsection{Nonlinear wave equation}

Before describing the full problem of GHz waves in a hot plasma (heated by shocks), we start with waves in a cold plasma. This gives a quick introduction to the calculation method using characteristics.

The nonlinear evolution equation for magnetosonic waves with a spherical wave front (far from the source) is derived in Paper~I. At all polar angles $\theta$, the wave excites a pure toroidal current $\bj$ while sustaining zero charge density, $\rho_e=0$.  Plasma motion in the wave obeys the momentum and energy equations,
\beq
\label{eq:dynamics}
   \rho c^2\,\frac{d\bu}{dt} = \bj\times\bB, \qquad  
   \rho c^2 \frac{d\gamma}{dt} =  \bE\cdot\bj,
\eeq
where $\bu=\gamma\bb$, $\gamma=(1-\beta^2)^{-1/2}$, and the time derivative is taken along the fluid streamline: $d/dt=\partial_t+\bv\cdot\nabla$. Conservation of particle number (neglecting $e^\pm$ creation and annihilation) is stated by the continuity equation, 
\beq
\label{eq:momentum}
  \partial_\alpha F^\alpha = \partial_t n +\nabla\cdot(n\bv)=0,
\eeq
where $u^\alpha=(\gamma,\bu)$ is the fluid four-velocity, $F^\alpha=\tn u^\alpha$ is the four-flux of particle number, and $\tn=n/\gamma$ is the proper density.

We are interested here in wave packets with many oscillations and a short length $c\tau\ll r$. It is convenient to use coordinates $(t,\xi,\theta,\phi)$, so that the fast oscillation is isolated in the single coordinate $\xi=t-r/c$ (and variations with $t$ and $r$ at fixed $\xi$ are slow). The continuity equation in the short-wave limit gives
\beq
\label{eq:short1}
   F^\xi=(c-v_r)n=const=c \nbg,
\eeq
and energy conservation can be cast into the following form (Paper~I):
\beq
\label{eq:NWE}
    \partial_t(r^2E^2) =-4\pi r^2\rhobg c^2\left[\partial_\xi\gamma+\frac{r\partial_t\gamma+\beta_\theta\partial_\theta\gamma}{r(1-\beta_r)}\right].
\eeq 
Here, the derivative $\partial_t$ is taken at fixed $\xi,\theta$ (i.e. along the radial ray $r=ct+const$), and the derivative $\partial_\xi$ is taken at fixed $t,\theta$. The second term in the square brackets is small compared to $\partial_\xi\gamma$ unless $\gamma$ approaches $\sigbg^{1/3}$ (Paper~I). This typically does not occur in GHz waves, which develop less extreme $\gamma$ compared to kHz waves (as explained below), and then the energy equation simplifies to
\beq
\label{eq:NWE1}
    \partial_t(r^2E^2) =-4\pi r^2\rhobg c^2\, \partial_\xi\gamma.
\eeq
It describes the coupled evolution of $E(t,\xi,\theta)$ and $\gamma(t,\xi,\theta)$ for the waves propagating in the MHD regime.

We now focus on waves in the equatorial plane ($\theta=\pi/2$), assuming  equatorial symmetry. Waves at different polar angles will be investigated in \Sect~\ref{outside_equator}.

\subsection{Equatorial waves}

In the equatorial wave, the plasma oscillates with a radial drift speed $\bb=\bE\times\bB/B^2=\beta_r\be_r$, since $v_\theta=0$ by symmetry. Besides $E\equiv -E_\phi$ we will use the following notation:
\beq
  B\equiv B_\theta, \qquad \beta\equiv \frac{v_r}{c}=\frac{E}{B}.
\eeq
\Eq~(\ref{eq:short1}) gives the plasma compression factor in short waves, $n=(1-\beta)\nbg$.
The magnetic field is frozen in the fluid and compressed by the same factor,
\beq
\label{eq:continuity_short}
   \frac{n}{\nbg}=\frac{\rho}{\rhobg}=\frac{B}{\Bbg}=(1-\beta)^{-1}.
\eeq 
All MHD quantities in the equatorial wave can now be expressed in terms of $\beta$, including the electric field,
\beq 
\label{eq:E1}
  E=\beta B=\frac{\beta\Bbg}{1-\beta}.
\eeq
Substituting \Eq~(\ref{eq:E1}) into \Eq~(\ref{eq:NWE1}) and using $d\gamma=\gamma^3\beta d\beta$, one obtains 
\beq
\label{eq:evol_gamma}
   \frac{2\sigbg\,\partial_t\gamma}{\gamma^3(1-\beta)^3}
   +\partial_\xi\gamma = \frac{4c\,\sigbg \beta^2}{r\,(1-\beta)^2}. 
\eeq

A convenient MHD variable is the compression of proper density $\trho=\rho/\gamma$ relative to its background value $\rhobg$,
\beq
\label{eq:c}
    \c \equiv \frac{\trho}{\rhobg} = \frac{\tB}{\Bbg} =\frac{\sigma}{\sigbg} = 
     \sqrt{\frac{1+\beta}{1-\beta}},
\eeq
where $\trho$ and $\tB$ are measured in the fluid rest frame, and $\sigma\equiv \tB^2/4\pi\trho c^2$. \Eq~(\ref{eq:evol_gamma}) rewritten in terms of $\c$ becomes 
\beq
\label{eq:cold_wave}
   2\sigbg \c^3 \partial_t \c +  \partial_\xi \c  =  \frac{2c}{r}\,\sigbg\c^2(\c^2-1).
\eeq
Using the method of characteristics, we express this equation as
\beq
 \label{eq:evol}
  \left.\frac{d\c}{dt}\right|_{C^+}  = \frac{c}{r}\left(\c-\c^{-1}\right),
\eeq
where the derivative is taken along curves $C^+$ (characteristics) determined by the ratio of the coefficients of $\partial_t\xi$ and $\partial_t\c$ in \Eq~(\ref{eq:cold_wave}),
\beq
\label{eq:C+_cold}
  \frac{d\xi_+}{dt}= \frac{1}{2\sigbg\c^3}.
\eeq
 The characteristics $\xi_+(t)$ can also be described by their radial speed $\beta_+=c^{-1}dr_+/dt=1-d\xi_+/dt$.

Recall that \Eqs~(\ref{eq:evol}) and (\ref{eq:C+_cold}) are obtained assuming cold plasma. Appendix~A gives a more formal derivation using the stress-energy tensor of electromagnetic field + plasma, and shows that \Eq~(\ref{eq:evol}) also holds when the cold approximation is relaxed, i.e. the plasma is allowed to be relativistically hot. The shape of $C^+$ characteristics in this more general  case is described by
\beq
\label{eq:C+}
   \frac{d\xi_+}{dt} = \frac{1}{2\gs^2\c^2},
\eeq
where $\gs$ is the magnetosonic Lorentz factor defined in Appendix~A. Its value in a hot plasma is given by (see Appendix~\ref{gs})
\beq
\label{eq:gs}
  \frac{1}{\gs^2} = \frac{1}{\kk^2\sigma} \left[(\kk^2-1)\e+\frac{1}{\e^3}\right].
\eeq
Here, $\e=(\trho c^2+\Up)/\trho c^2$ is the dimensionless specific plasma energy (including rest-mass and thermal energy); $\kk=3$ if thermal motions of plasma particles are isotropic in the fluid frame, and $\kk=2$ if the thermal velocities are confined to the plane perpendicular to $\bB$. As shown below, the plasma is heated in shocks, which are mediated by Larmor rotation, and so heating occurs in the plane perpendicular to $\bB$. It is uncertain whether the plasma becomes isotropic far downstream of the shock; therefore, we allow both possibilities $\kk=2$ and $\kk=3$.

For waves in a cold plasma, $\e=1$ and $\gs^2=\sigma=\c\sigbg$. In this case, \Eq~(\ref{eq:C+}) is reduced to \Eq~(\ref{eq:C+_cold}).

\subsection{Bending of characteristics}

Shocks form because the $C^+$ characteristics in spacetime are bent from straight lines,  leading to collisions between them.  This bending is described by $d\xi_+/dt \neq 0$, and one can see from \Eq~(\ref{eq:C+}) that it is strongest when $\c$ is small. Note that $\c=\gamma(1+\beta)=[\gamma(1-\beta)]^{-1}$ is smallest where $\beta$ approaches $-1$ (i.e. the plasma drifts with a maximum Lorentz factor $\gamma_{\max}$ toward the star), which occurs where $E^2$ approaches $B^2$.

As explained in Paper~I (and in \Sect~\ref{analytical} below), $\gamma_{\max}$ and $\c_{\min}\approx(2\gamma_{\max})^{-1}$ are set by the ratio of the electromagnetic energy in one wave oscillation, $L/\nu$, to the plasma rest mass in the magnetosphere, $\sim 4\pi\N\me c^2$. This ratio scales with the wave frequency as $\nu^{-1}$, and here one can see the first big difference between GHz and kHz waves: $\gamma_{\max}$ is much lower in GHz waves. In particular, $\gamma$ stays far below $\gs$, and $\gs\c \gg 1$ holds across the wave. This implies that the $C^+$ characteristics propagate with $d\xi_+/dt\ll 1$ (see \Eq~\ref{eq:C+}), i.e. their speeds $dr_+/dt$ stay close to $c$. Thus, the bending of characteristics is a small parameter.

This feature of GHz waves allows one to easily find the evolution of $\c$ along $C^+$. Using $dt=(1-d\xi_+/dt)^{-1} dr/c$\, and\, $d\xi_+/dt \ll 1$, we find from \Eq~(\ref{eq:evol}),
\beq 
   \frac{d\c}{d\ln r} = \left(\c-\c^{-1}\right)\left[1+{\cal O}\left(\frac{d\xi_+}{dt}\right)\right].
\eeq
This equation implies that the radial dependence of $\c$ along each $C^+$ has the functional shape,
\beq
\label{eq:cC+}
     \c=\sqrt{1+2Kr^2}, 
\eeq
where $K=const$. The constant $K$ is different on different $C^+$ and set by the initial profile of the wave. Using $E=\beta B=\beta \Bbg/(1-\beta)$ and substituting $\beta=(\c^2-1)/(\c^2+1)$ we obtain the solution for $E$ along $C^+$, 
\beq
\label{eq:EK}
    E=\frac{\mu K}{r}.
\eeq
It is the same as in a vacuum wave, $E\propto r^{-1}$. We conclude that the presence of plasma influences the GHz wave propagation by slightly changing  the shape of $C^+$ characteristics while the evolution of $E(r)$ along each $C^+$ remains unchanged from the vacuum solution. 

Note that the small bending of characteristics, $d\xi_+/dt \ll 1$, can strongly deform the oscillations with wavelength $\lambda\ll r$. This occurs when the small deviation of $C^+$ from straight lines, $\delta r_+\sim c (d\xi_+/dt) t \ll r$, reaches a fraction of $\lambda$. Then,  characteristics collide, forming a discontinuity of the MHD quantities -- a shock.

\subsection{Coupling of wave evolution to thermal balance}

Next, we note another essential difference between kHz and of GHz waves. In kHz waves, the monster shocks have ultra-fast radiative losses. As a result, it turns out sufficient to use the cold approximation $\e\approx 1$, which gives $\gs^2\approx\sigma=\c\sigbg$. By contrast, for GHz waves, the plasma cooling time exceeds the wave oscillation period. This leads to accumulation of a large $\e$ along the wave train. It affects $\gs$, and the wave evolution becomes coupled to the plasma thermal balance. Thus, the wave problem requires a self-consistent solution for $\c(t,\xi)$ and $\e(t,\xi)$. The evolution of $\e(t,\xi)$ is governed by heating in shocks and synchrotron cooling, as described below.

\subsection{Shock heating}
\label{heat}

The plasma speed $\beta$ is discontinuous at the shock, as the upstream and downstream characteristics bring to the shock different values of $\beta$: $\betau\neq \betad$ (hereafter subscripts ``u'' and ``d'' refer to the immediate upstream and immediate downstream of the shock). The Lorentz factor of the upstream plasma relative to the downstream plasma, $\Grel$, is related to the shock compression factor $q=\trhod/\trhou=\cd/\cu$ (Appendix~\ref{jump}):
\beq
  \Grel = \gu\gd(1-\betau\betad)=\frac{1}{2}\left(q+q^{-1}\right).
\eeq
In Appendix~\ref{jump} we describe the shock jump conditions and derive the plasma energy per unit mass immediately downstream of the shock:
\beq
\label{eq:ed}
   \ed=\frac{b+\sqrt{b^2-(3\kk-1)q^4+2(\kk-1)q^2+\kk+1}}{(3\kk-1)q^2+\kk+1},
\eeq
\beq
\nonumber
  b \equiv \frac{q}{2} \left[(\kk+1)q^2+3\kk-1\right]\eu-\frac{q(q^2-1)}{2\eu},
\eeq
where $\kk=3$ for isotropic plasma and $\kk=2$ when particles are heated only in the plane perpendicular to $\bB$. Note that radiative losses do not affect the shock jump conditions, as the plasma cools on a timescale much longer than the Larmor time that sets the shock width (the opposite regime occurs in kHz waves, see Paper~I). 

The jump conditions also determine the shock speed,
\beq
\label{eq:vsh}
   1-\bsh = \frac{(\kk+1)\eu - \eu^{-1} - [(\kk+1)\ed-\ed^{-1}]/q}{\kk\,\sigbg\cu^3(q^2-1)}.
\eeq
For relativistic shocks with $q\gg 1$, this simplifies to 
\beq
\label{eq:vsh1}
   1-\bsh\approx \frac{2(\kk-1)\ed}{\kk\,\sigbg\cu^3 q^3},
     \quad    \ed\approx \frac{q[(\kk+1)\eu-\eu^{-1}]}{3\kk-1}.
\eeq

\subsection{Radiative losses}

The shock-heated plasma gradually loses energy to synchrotron emission. Thermal evolution of the plasma behind a shock obeys the first law of thermodynamics along the fluid streamline:
\beq
   d\e=-\Pp \, d\frac{1}{\trho c^2}-d\e_s=p\,d\ln\c-d\e_s,
\eeq
where $\Pp$ is the plasma pressure,
\beq
  p\equiv \frac{\Pp}{\trho c^2} = \frac{1}{\kk}\left(\e-\e^{-1}\right),
\eeq 
$\me c^2 d\e_s$ is the energy loss due to synchrotron emission,
\beq
\label{eq:des}
   d\e_s =\frac{\sT}{\me c}  \frac{\tB^2}{2\pi\kk}(\e^2-1)d\tilde{t},
\eeq
and $\tilde{dt}=dt/\gamma$ is the proper time of the fluid element. We here approximated the particle distribution function in the fluid frame as mono-energetic (each particle has the Lorentz factor $\tilde{\gamma}_e=\e$). Then, using the relations $\tB=\c\Bbg$ and $d\tilde{t}=\c\, d\xi$ along the fluid streamline, we obtain the equation for $\e(\xi)$,
\beq
\label{eq:e}
   d\e=\frac{\e^2-1}{\kk\e}\,d\ln\c - \frac{\sT\Bbg^2\c^3(\e^2-1)}{2\pi\kk\, \me c}\,d\xi,
\eeq
which can be integrated numerically along a given wave profile $\c(\xi)$. 

Energy emitted in the lab frame is $dE_s=\gamma\, d\e_s \me c^2$ per particle. The number of particles passing through the wave per unit time is $4\pi r^2 (\rhobg/\me) c$ (we have multiplied by $4\pi r^2$ to define the isotropic equivalent). Energy $E_s$ radiated per particle is distributed over $\xi$ in the wave as $dE_s/d\xi=\gamma\, \me c^2d\e_s/d\xi$. This gives the following distribution of the synchrotron power over $\xi$,
 \beq
   \frac{dL_s}{d\xi} =\frac{\sT\mu^4\gamma\c^3(\e^2-1)}{2\pi\kk\, \me D\, r^7}.
 \eeq

Solutions for MHD quantities along characteristics described by \Eqs~(\ref{eq:evol}) and (\ref{eq:C+}), with local $\gs$ calculated under the adiabatic assumption (\Eq~\ref{eq:gs}), become inaccurate if the plasma radiates a significant fraction of its energy $\e$ during one wave oscillation. We will monitor for this condition, which limits the applicability of our simulation method. Note that the solution may hold even when radiative losses have a strong net effect on the wave. For instance, the GHz wave packet with 30,000 oscillations (simulated below) eventually loses most of its energy to synchrotron emission, however it is approximately adiabatic in each oscillation. Radiative losses impact $\e$ that enters \Eq~(\ref{eq:C+}) through $\gs$, however the local fast-magnetosonic speed remains approximately adiabatic and determined by the local $\e$ according to \Eq~(\ref{eq:gs}).

\subsection{Numerical implementation}

One advantage of using characteristics, compared with grid-based MHD solvers, is the ability to track waves in a plasma with any large magnetization parameter $\sigma$. In addition, low computational costs of tracking characteristics allows one to follow radio bursts with a large number of oscillations $N$. 

The calculation starts at small radii where $\Bbg$ far exceeds  the wave electric field $E$, and the plasma oscillates with small $|\beta|\ll 1$, which implies a negligible modulation of plasma density, $|\c-1| \ll 1$. In this inner zone, the $C^+$ characteristics propagate with speed $\beta_+=1-{\cal O}(\sigbg^{-1})$, and so each characteristic keeps a constant coordinate $\xi=t-r/c=\xiin$. The initial wave profile $E(\xiin)$ is well defined in this inner zone of nearly vacuum propagation with $E\propto r^{-1}$. It is conveniently described by the function,  
\beq 
   K(\xiin)\equiv \frac{rE}{\mu}=\Km\sin(\omega\xiin), \qquad 
   \Km\equiv \frac{r\Em}{\mu}.
\eeq
When launching the wave, we set up an initially uniform grid in $\xiin$ of size $N_+$, and then use the $N_+$ characteristics to track the wave evolution. Typically, 500 characteristics per wave oscillation are sufficient (convergence has been verified by varying $N_+$). 

At each timestep $dt$, the displacement $d\xi_+$ of each characteristic $\xiin$ is determined by $d\xi_+/dt$ (\Eq~\ref{eq:C+}), which is controlled by the evolving values of $\c(t,\xiin)$ and $\e(t,\xiin)$ on $C^+$. The compression $\c(t,\xiin)$ evolves according to \Eq~(\ref{eq:evol}), and the plasma specific energy $\e(t,\xiin)$ is found by integrating the ordinary differential equation~(\ref{eq:e}) in $\xi$ when scanning though the array of $N_+$ characteristics.\footnote{Recall that we consider short wave packets, so that the plasma crosses the wave faster than the wave evolves. Then, the profile of $\e(\xi)$ can be calculated at fixed $t=const$. This approximation loses accuracy in low-power waves, which have extremely fast evolution at $\Rm$ (the model with $L=10^{40}\,$erg/s shown below), however this weakly affects the final result.}
The downstream energy $\ed$ of each shock is found from the jump conditions (\Sect~\ref{heat}). The propagation speed of each shock is determined by \Eq~(\ref{eq:vsh}).

After each timestep, the code examines the updated positions or the characteristics and any existing shocks, and determines which characteristics terminate at the shocks. The code also constantly watches for any new crossings of characteristics to detect formation of new shocks. We use an adaptive timestep to resolve any fast evolution in MHD quantities near $\Rm$. We have also implemented sub-timesteps in the leading oscillation of the wave, which is coldest and reaches the lowest $\c$, leading to more demanding timestep requirements. Note also that the density of characteristics $dN_+/d\xi$ drops ahead of shocks, where $\c$ is lowest and the high $d\xi_+/dt$ results in stretching the array of $C^+$ in $\xi$. To maintain sufficient spatial resolution everywhere in the wave, we use adaptive mesh refinement in $\xiin$ without changing the total number of 500 active (not terminated) characteristics in each oscillation. These technical tricks allow one to significantly speed up the simulation and trace the evolution of long wave trains. Sample wave trains presented below have $N=3\times 10^4$ oscillations, traced on a grid with $N_+=500 N=1.5\times 10^7$ characteristics. 

The wave evolution should conserve the total energy (electromagnetic + plasma + synchrotron losses), which provides a simple test. The simulations passed this test.


\section{MHD results for equatorial waves}
\label{numerical}

\begin{figure}[t]
\includegraphics[width=0.48\textwidth]{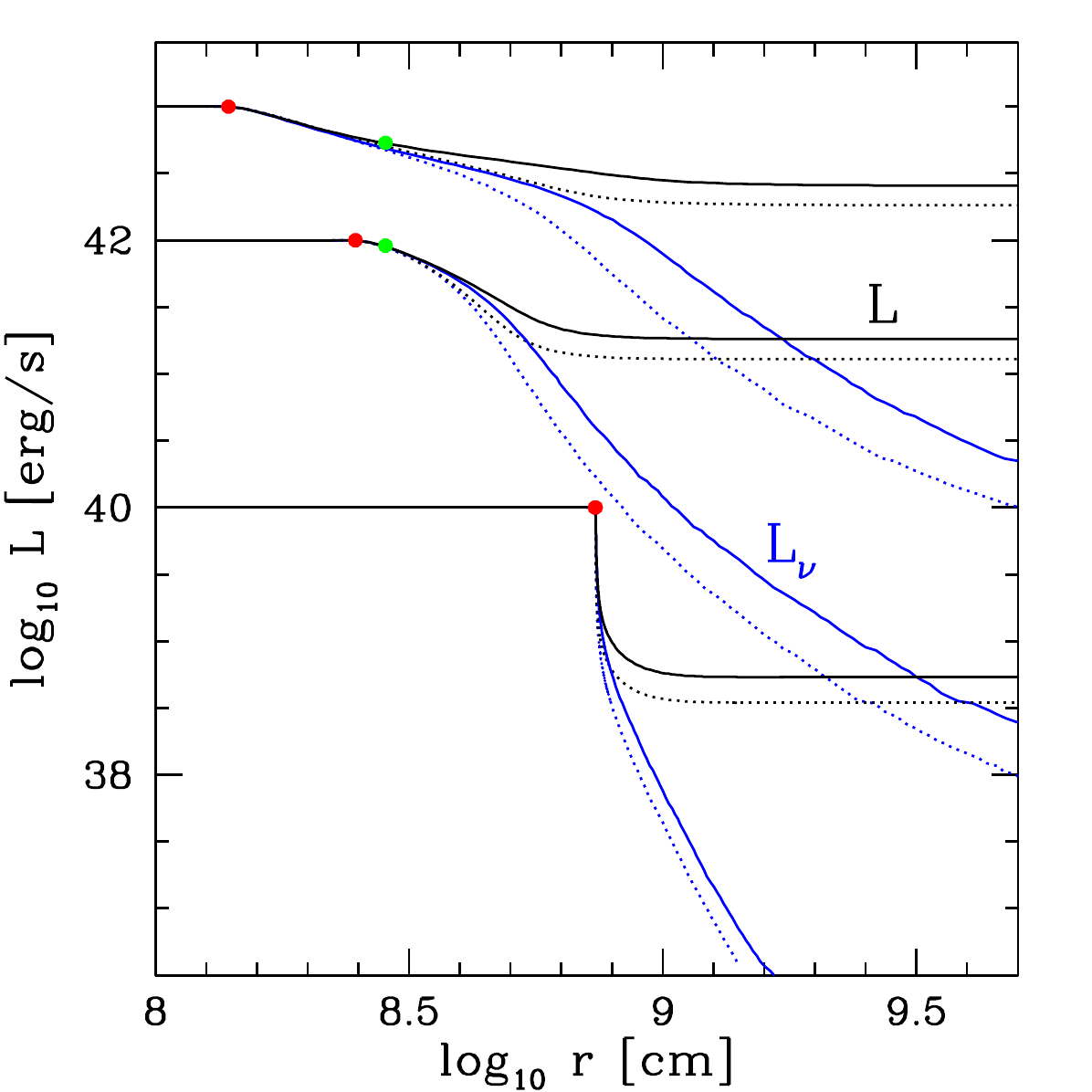} 
\caption{Evolution of the wave power $L$ with radius $r$, for wave packets with initial $L=10^{40}$, $10^{42}$, and $10^{43}$\,erg\,s$^{-1}$. In the sample numerical models the waves have frequency $\nu=0.3$\,GHz; similar results hold for waves in a broad range of GHz frequencies. Total Poynting flux $L$ (isotropic equivalent) is shown by black curves. Its oscillating component $L_\nu$ (blue curves) is of main interest for FRBs. The calculations have been performed assuming $\kk=3$ (isotropic plasma, dotted curves) and $\kk=2$ (thermal motions perpendicular to $\bB$; solid curves). Shocks form at radius $\rc$ indicated by the red dot; it is slightly smaller than $\Rm$ (see text). Green dot marks the heating transition radius $\Rh$ (\Eq~\ref{eq:Rh}).}
\label{fig:L}
 \end{figure}

Two models are discussed in detail below: waves with initial power $L=10^{42}\,$erg/s and $L=10^{40}\,$erg/s. Both simulated waves have frequency $\nu=\omega/2\pi=0.3\,$GHz and duration $\tau=0.1\,$ms. In addition, we will briefly discuss an example of waves with even higher initial power $L=10^{43}$\,erg/s. Figure~\ref{fig:L} shows the evolution of the wave power with radius found in the three simulations.
 
\subsection{Model~I: $L=10^{42}\,$erg/s}

As one can see in Figure~\ref{fig:L}, the wave experiences strong damping near radius $\Rm\approx 2.5\times 10^8$\,cm. The development of shocks in each oscillation results in plasma heating and synchrotron losses, reducing the wave energy by a factor of $\sim 10$ between $\Rm$ and $2\Rm$. Then, the electromagnetic packet evolves into a smooth, almost uniform, Poynting flux with unchanged duration $\tau=0.1\,$ms and strongly damped oscillations. The wave power $\Lw$ that is carried by the {\it alternating} component of the electromagnetic field with frequency $\nu$ is reduced below $10^{-3}$ of its original value. 

\begin{figure}[t]
\includegraphics[width=0.48\textwidth]{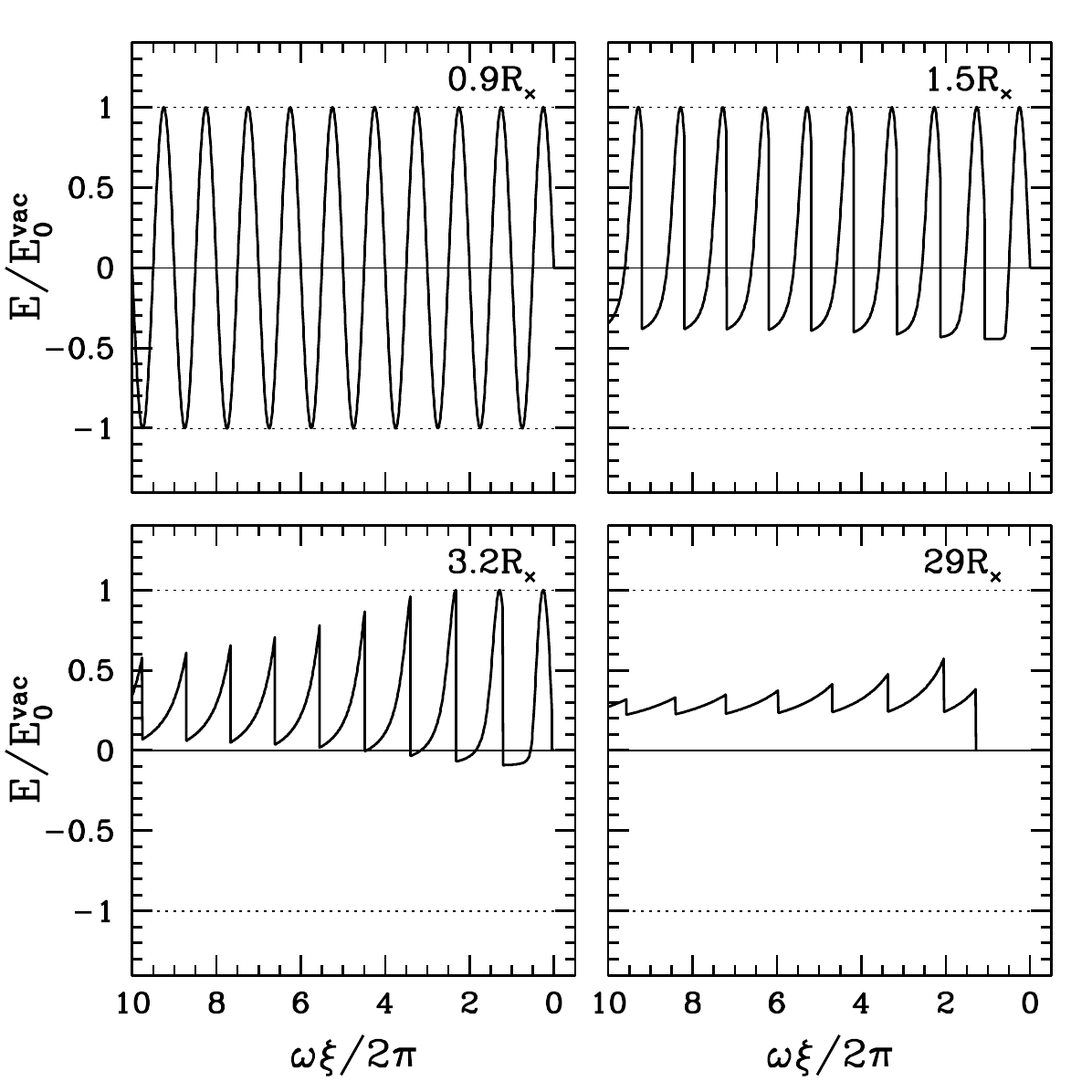} 
\caption{Evolution of the wave profile $E(\xi)$ in Model~I ($L=10^{42}\,$erg/s, $\nu=0.3\,$GHz, isotropic plasma). The snapshots were taken when the packet reached $r/\Rm=0.9$, 1.5, 2.6, and 29. For clarity only the leading 10 oscillations are shown.  Electric field $E$ is normalized to the amplitude $\Em^{\rm vac}$ that the wave would have if it propagated in vacuum.}
\label{fig:E_L42}
 \end{figure}

\begin{figure}[t]
\includegraphics[width=0.46\textwidth]{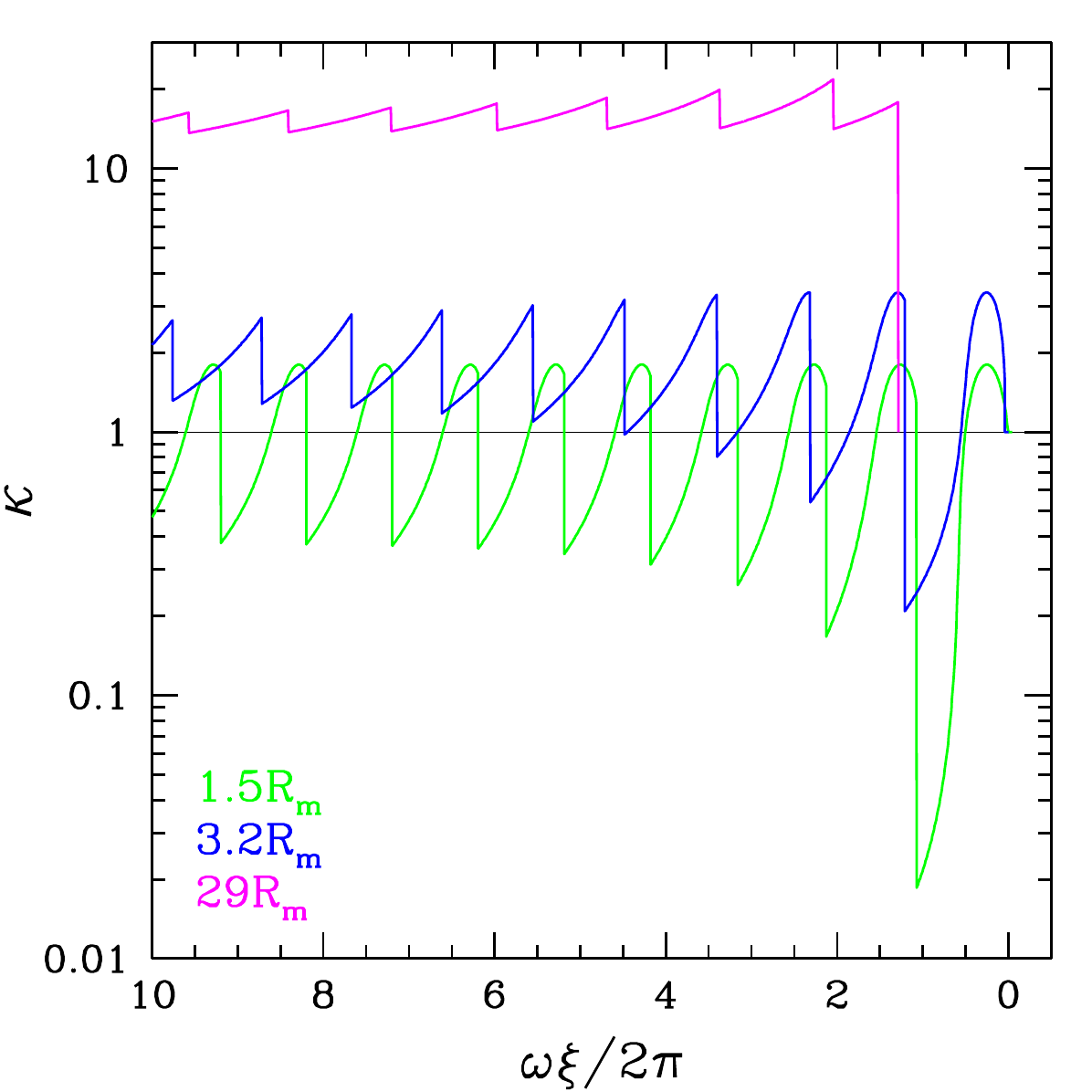} 
\caption{Evolution of the plasma compression profile $\c(\xi)$ in Model~I (same model and snapshot times as in Figure~\ref{fig:E_L42}). The vertical jump observed in each oscillation is a shock.}
\label{fig:kappa_L42}
 \end{figure}

\begin{figure}[t]
\includegraphics[width=0.46\textwidth]{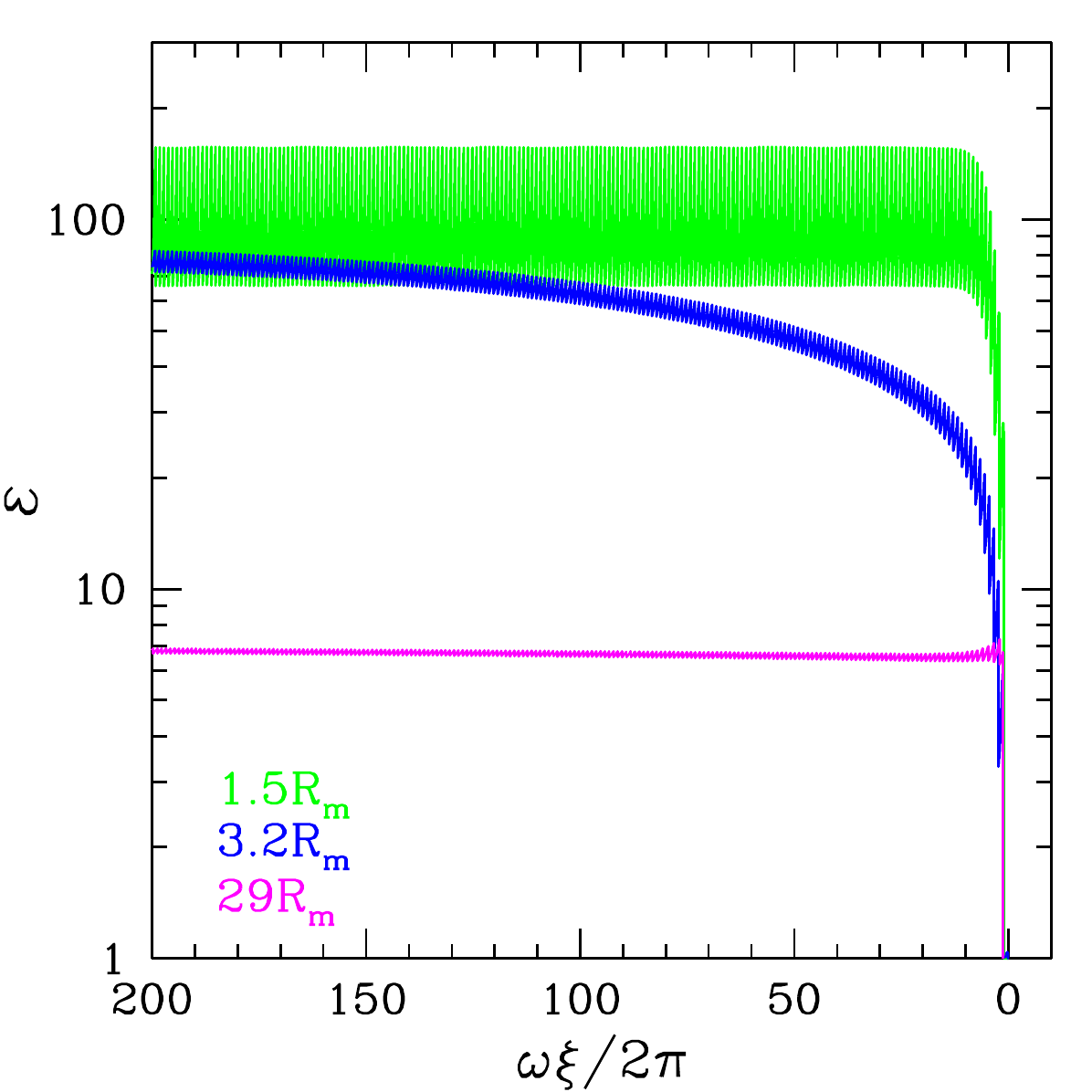} 
\caption{Evolution of the plasma internal energy $\e(\xi)$ in Model~I (same snapshot times as in Figure~\ref{fig:kappa_L42}). The larger number of oscillations (200) are shown to demonstrate the heating by the shock train and the saturation of $\e$ when synchrotron cooling offsets shock heating in each oscillation.}
\label{fig:heat_L42}
\end{figure}

\begin{figure}[t]
\includegraphics[width=0.46\textwidth]{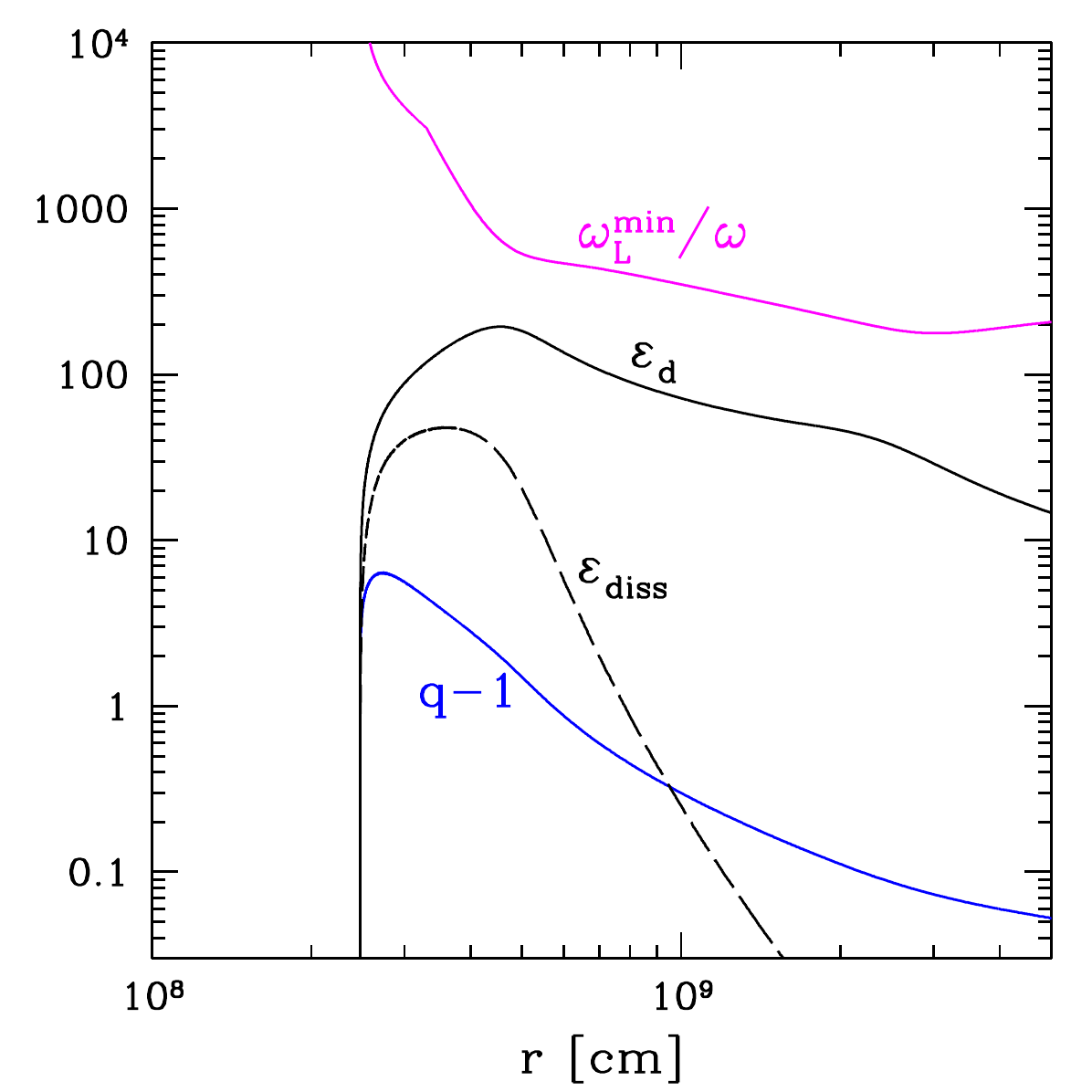} 
\caption{Evolution of a typical shock (measured 1000 oscillations from the leading edge of the wave train) in Model~I. All curves begin at the shock formation radius $\rc\approx 0.998\Rm$. The shown parameters are the shock compression factor $q$, downstream specific energy $\ed$, and specific dissipated energy $\ediss$. In addition, the figure shows Larmor frequency of the plasma particles $\omL$ normalized to the wave frequency $\omega$; $\omL$ oscillates in the wave and the figure shows the evolution of its minimum value.}
\label{fig:shock_L42}
 \end{figure}

The evolution of the wave profile is shown in Figure~\ref{fig:E_L42}. Showing the entire profile with $3\times 10^4$ oscillations would be impractical, so we limited the figure to the first 10 oscillations in the packet; this is sufficient to see the evolution. It has three phases:
\\
{\bf (1) Cold oscillations at $r<\Rm$.}  As the plasma flows through the wave, it performs $N=\tau\nu$ small-amplitude, harmonic oscillations with frequency $\nu$, and exits behind the packet. This simple behavior ends when the packet reaches $\rc\approx 0.998\Rm$. Then, caustics form, launching shocks in each oscillation. The shocks appear at the oscillation phases where $E$ is close to its minimum ($E\approx -\Em$) and $\c \ll 1$. 
\\
{\bf (2) Main dissipation phase at $r\simgt\Rm$.} The shocks reach a maximum strength at $r\approx 1.1 \Rm$. The shock in the first oscillation of the wave train is strongest (because it has a cold upstream), reaching the compression factor $q=\cd/\cu\sim 10^2$ (Figure~\ref{fig:kappa_L42}).  Subsequent shocks down the wave train occur in the plasma already heated in the leading shocks; therefore, they have smaller $q$. Figure~\ref{fig:shock_L42} shows the evolution of a typical shock located 1000 oscillations away from the leading edge of the wave. 

The oscillation of plasma compression $\c$ (Figure~\ref{fig:kappa_L42}) modulates the plasma temperature by adiabatic heating/cooling and, in addition, there is dissipative compression at each shock. In general, dissipation breaks periodicity: as the plasma moves through the wave train, it can accumulate heat $\ediss$ gained in each of the $N$ shocks. However, synchrotron losses offset the gradual growth of $\e$ and make the wave train approximately periodic, with $\e$ oscillating about a flat $\bar{\e}$ (Figure~\ref{fig:heat_L42}). The value of $\bar{\e}\approx 10^2$ is determined by the heating = cooling balance, as described in \Sect~\ref{balance} below.
\\
{\bf (3)  Evolution toward a uniform Poynting flux at $r\gg\Rm$.} By the time the packet reaches $r=2\Rm$ the shocks have erased the low-$\c$ regions, and the plasma oscillations now have a small Lorentz factor $\gamma\sim 1$ throughout the wave. The shock strength becomes sub-relativistic, $\Grel\sim 1$. The electric field $E$ oscillates with a decreasing amplitude about a positive average value $\bar{E}\sim 0.3\Em$. At $r\simgt 30\Rm$, $E(\xi)$ becomes nearly uniform across the entire wave.

We conclude that the oscillating GHz wave is absorbed in the magnetosphere. Part of its energy and momentum ($\sim 10$\%) is used to eject the outer magnetospheric layers, forming the Poynting flux that continues to expand freely. Most of the absorbed wave energy ($\sim 90$\%) converts to synchrotron emission from the heated plasma.

The simulation also demonstrates the gradual steepening of the wave profile at the leading edge of the packet. This leads to the formation of a strong forward shock when the packet reaches $\RF\approx 7\times 10^8\,$cm. It is consistent with the analytical expectation (Paper~I):
\beq
\label{eq:RF}
   \RF=\left(\frac{8c\sigm}{\omega\Rm}\right)^{1/6} \Rm 
   \approx 7 \times 10^8 {\rm ~cm}.
\eeq
The forward shock is the leading edge of the Poynting flux ejected from the magnetosphere, i.e. the wave packet effectively has become an ultrarelativistic blast wave that continues to expand into the external medium. The blast has thickness $c\tau$ and carries $\sim 10\%$ of the original wave energy.

\subsection{Model~II: $L=10^{40}\,$erg/s}

The main difference of the low-power wave is seen from Figure~\ref{fig:L}: it is damped in a narrow range of radii $\delta r$ when the wave packet approaches $\Rm\approx 7.8\times 10^8\,$cm. Shocks and damping develop somewhat before $\Rm$, at $\rc\approx 0.943\Rm$. The final outcome is that about 95\% of the wave energy is radiated away in synchrotron X-rays, and the remaining $5\%$ forms a smooth Poynting flux of duration $\tau$, with no GHz oscillations. 

\begin{figure}[t]
\includegraphics[width=0.46\textwidth]{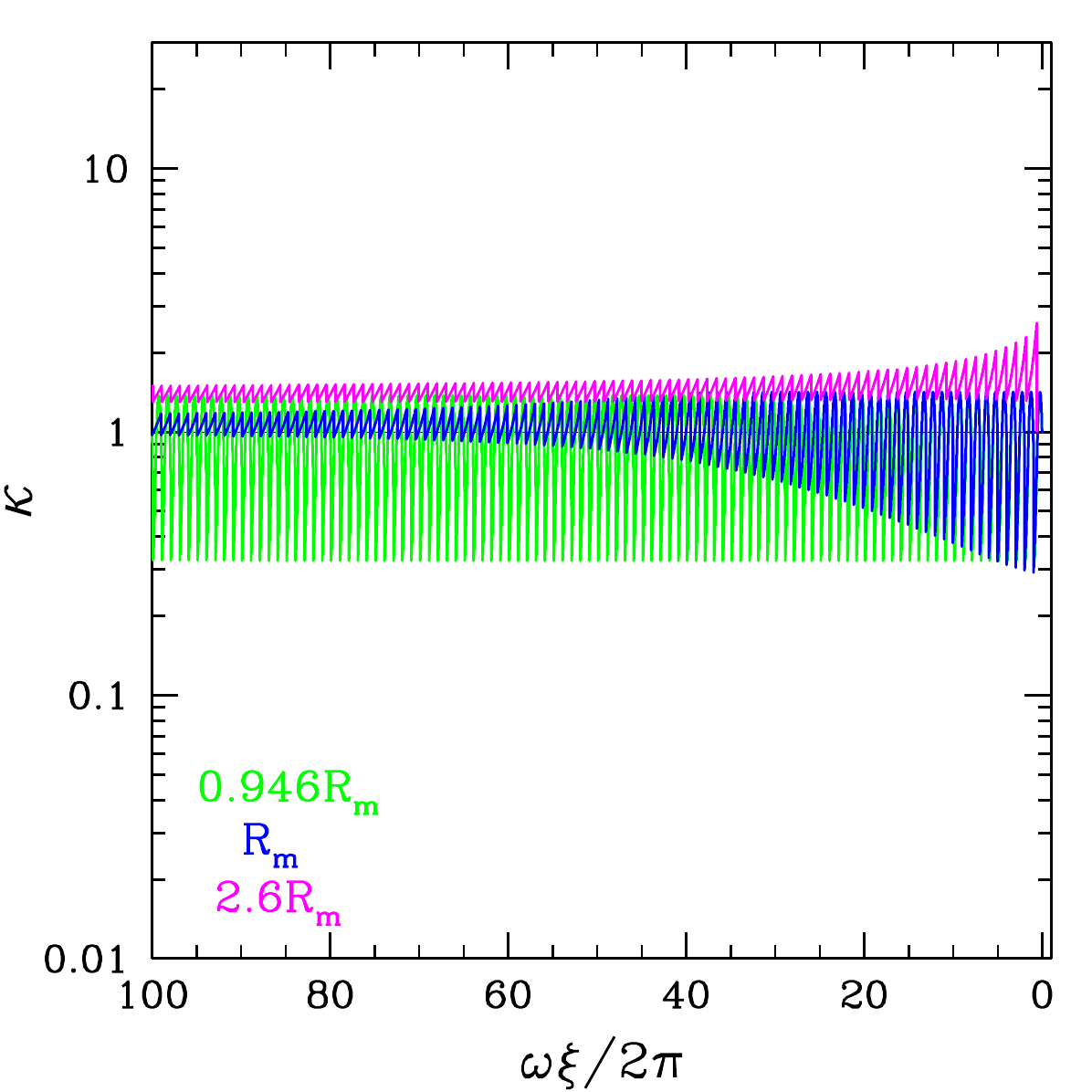} 
\caption{Oscillations of the plasma compression $\c(\xi)$ in Model~II ($L=10^{40}\,$erg/s, $\nu=0.3\,$GHz, isotropic plasma). Three snapshots are shown, when the wave packet reached $r/\Rm=0.946\Rm$, 1, 2.6. The figure shows the leading 100 oscillations (the simulated packet has $3\times 10^4$ oscillations).}
\label{fig:kappa_L40}
\end{figure}

\begin{figure}[t]
\includegraphics[width=0.46\textwidth]{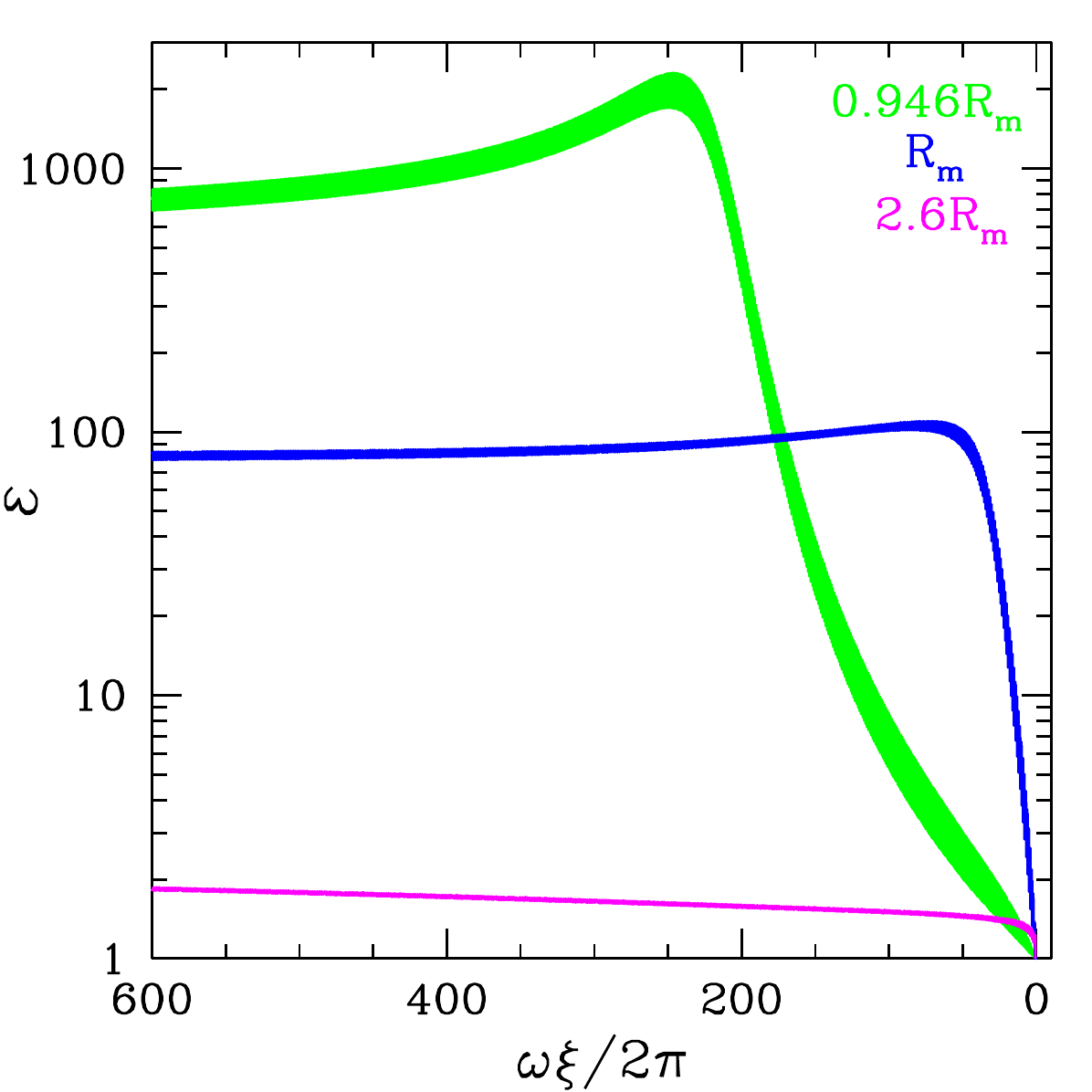} 
\caption{
Evolution of the plasma internal energy $\e(\xi)$ in Model~II (same snapshot times as in Figure~\ref{fig:kappa_L40}). The larger number (600) of oscillations are shown to demonstrate the heating by the shock train and the saturation of $\e$ when synchrotron cooling offsets shock heating in each oscillation.}
\label{fig:heat_L40}
\end{figure}

\begin{figure}[t]
\includegraphics[width=0.48\textwidth]{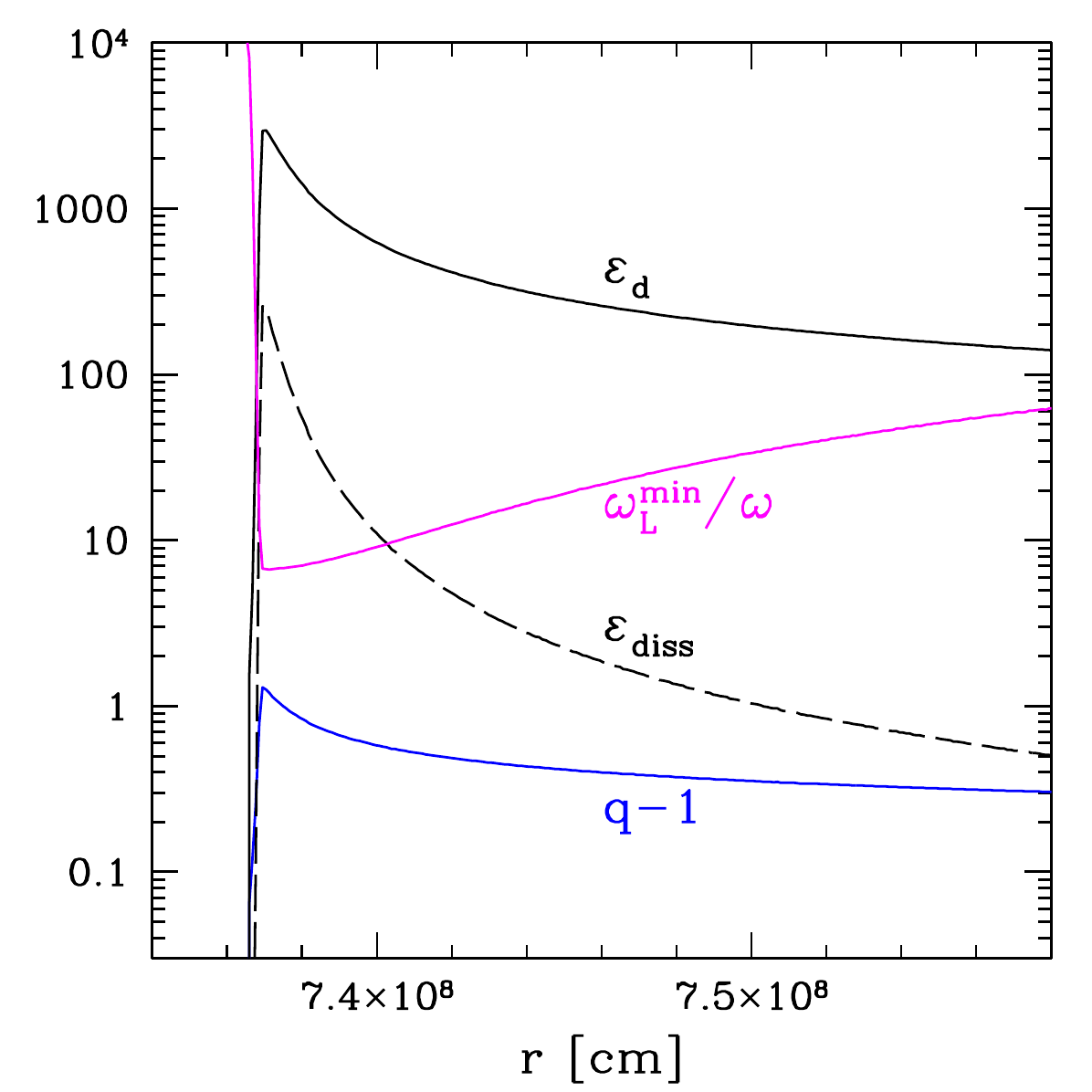} 
\caption{Evolution of a typical shock (measured 1000 oscillations from the leading edge of the wave train) in Model\,II. The shocks appear at $\rc\approx 0.943\Rm$, and the figure focuses on the narrow range $\delta r$ where the wave energy is dissipated. The shown parameters are the shock compression factor $q$, downstream specific energy $\ed$, and specific dissipated energy $\ediss$. In addition, the figure shows Larmor frequency of the plasma particles $\omL$ normalized to the wave frequency $\omega$; $\omL$ oscillates in the wave and the figure shows the evolution of its minimum value.}
\label{fig:shock_L40}
\end{figure}

Evolution of the wave profile is qualitatively similar to Model~I. There are two main quantitative differences:
(1) The shocks are weaker (Figure~\ref{fig:kappa_L40}), which implies a lower dissipated fraction $\ediss/\ed$ in each shock. 
(2) The plasma passing through the wave develops a much higher temperature ($\e$ reaches $\sim 2\times 10^3$, see Figure~\ref{fig:heat_L40}). This occurs because dissipation takes place at a larger $\Rm$ where the magnetic field is weaker and synchrotron cooling is much slower. This leads to $\ediss\sim 200$ during the main dissipation phase, much higher than in Model~I (compare Figures~\ref{fig:shock_L40} and \ref{fig:shock_L42}). Therefore, the wave damping occurs much faster. 

Our numerical method loses accuracy when radiative losses during one oscillation become comparable to the plasma energy (then the speed of characteristics $\bs$ significantly deviates from its adiabatic value). This does not occur in Model~II. Model~I is less accurate, in particular near the dissipation onset, when the plasma radiates in one oscillation $\sim 30\%$ of its thermal energy, in balance with heating $\ediss/\ed\sim 0.3$  (Figure~\ref{fig:shock_L42}).


\subsection{Analytical description}
\label{analytical}

The wave behavior shown by the numerical models can be understood analytically and described by approximate formulae. An important dimensionless parameter of the problem is 
\beq
\label{eq:zeta}
    \zeta\equiv\frac{\omega\Rm}{c \,\sigm}=\pi^2\,\frac{\me c^2\N \nu}{L}
    \approx 8\times 10^{-2}\,\N_{37}L_{42}^{-1}\nu_9.
\eeq
As shown in Paper~I, the plasma exposed to waves with $\zeta\ll 1$ develops Lorentz factor $\gamma\sim\zeta^{-1}$ at $r\sim\Rm$. The parameter $\zeta$ is tiny in powerful kHz waves investigated in Paper~I, and GHz waves have less extreme $\zeta$. Our first sample model ($\nu=0.3$~GHz and $L=10^{42}\,$erg/s) has $\zeta\approx 2.4\times 10^{-2}$. The second model ($\nu=0.3$\,GHz and $L=10^{40}\,$erg/s) has $\zeta>1$, which creates only mildly relativistic plasma motions in the wave. Note also that the plasma speed is related to compression $\c$ (\Eq~\ref{eq:c}), which implies $\gamma=(\c^2+1)/2\c$.

\subsubsection{Shock formation}

The caustic forms quickest on characteristics with $K<0$, which develop the smallest $\c$. The first shock forms in the first (leading) oscillation in the initially cold plasma. It occurs at coordinate $\xic$ with plasma compression factor $\cc$ at time $\tc$, all of which can be derived analytically. For waves with $\zeta\ll 1$ the result is (Paper~I) 
\beq
\label{eq:caustic}
    \cc\approx\frac{\sqrt{\zeta}}{24^{1/4}}, \qquad 
    \omega\xic \approx  \frac{3\pi}{2} + \frac{16\sqrt{\zeta}}{24^{3/4}}+\frac{c}{\Rm\omega},
\eeq
\beq
 \frac{c\tc}{\Rm} - 1\approx -\frac{\zeta}{3\sqrt{24}}+\frac{8\sqrt{\zeta}}{24^{3/4}}\frac{c}{\Rm\omega}+\frac{c^2}{4\Rm^2\omega^2}.
\eeq
Note that $\zeta\gg c/\omega\Rm$ in the GHz waves. In particular, for the 0.3-GHz wave with $L=10^{42}\,$erg/s, we find $c\tc/\Rm-1\approx -\zeta/3\sqrt{24}\approx -1.6\times 10^{-3}$, in agreement with the numerical simulation. Note also that Model~II is in the opposite regime, $\zeta>1$. In this case, the caustic forms without a strong drop in $\c$. 

Similar shocks develop in each oscillation of the wave train and eventually dissipate the wave energy into heat, most of which is radiated away.

 \subsubsection{Thermal balance in the shock-heated wave train} 
 \label{balance}

Each shock in the wave train heats the plasma passing through the wave. Without synchrotron cooling, the plasma specific entropy would monotonically grow with $\xi$ from the leading edge of the shock train toward its end. This growth occurs over many oscillations and gradually pushes the specific energy $\e$ to so high values that synchrotron cooling $Q\propto \e^2$ becomes important. Indeed, one can see in Figures~\ref{fig:heat_L42} and \ref{fig:heat_L40} that at large $\xi$ the growth of $\e(\xi)$ stops. Thus, the plasma passing through the wave enters a thermal balance: shock heating in each oscillation is offset by synchrotron cooling. 

The thermal balance may be stated by equating the specific energy dissipated at the shock $\ediss$ to the synchrotron losses (\Eq~\ref{eq:e}) integrated over one oscillation,
\beq
\label{eq:RRL}
     \oint\displaylimits_{2\pi/\omega} \frac{\sT\Bbg^2\c^3(\e^2-1)}{2\pi\kk\, \me c}\,d\xi = \ediss.
\eeq
The shock dissipates energy $\ediss=\ed-\e_{\rm ad}$, where $\ed$ is given by \Eq~(\ref{eq:ed}) and $\e_{\rm ad}=q^{\alpha-1}\eu$ accounts for adiabatic heating by shock compression $q$ with adiabatic index $\alpha$. The plasma flowing through the train of many shocks sustains $\e\gg 1$ and $\alpha=1+\kk^{-1}$. \Eq~(\ref{eq:ed}) for $\ed$ then simplifies, and we find its dissipation part:
\beq
\label{eq:g}
   g  \equiv \frac{\ediss}{\ed} = 1-\frac{[(3\kk-1)q^2+\kk+1]\,q^{1/\kk} }{(\kk+1)q^3+(3\kk-1)q}
   \quad (\eu\gg1). 
\eeq
At large $q\gg 10$, $g$ approaches unity. In the opposite weak-shock limit, expansion in $q-1$ gives
\beq
  g\approx \frac{(3\kk-1)(\kk^2-1)}{12\kk^3} (q-1)^3  \qquad (q-1\ll 1).
\eeq

Synchrotron losses peak downstream of each shock, and we estimate the thermal balance~(\ref{eq:RRL}) as
\beq
\label{eq:RRL1}
    \frac{\sT\Bbg^2\cd^3\ed^2}{2\pi\kk\, \me c\,\omega}  \sim \ediss.
\eeq
This gives
\beq
\label{eq:e_balance}
  \ed\sim g(q)\,\frac{2\pi \kk \me c\,\omega} {\sT\Bbg^2\cd^3}.
\eeq
This balance determines $\e$ that shocks with compression factors $q$ can sustain against radiative losses. Note that $\ediss$ is a small fraction of $\ed$ for $q<5$, and that $\e$ oscillates in the wave within a modest range $\eu\simlt\e \simlt\ed$ (Figures~\ref{fig:heat_L42} and \ref{fig:heat_L40}).

In \Eq~(\ref{eq:e_balance}), one can use  $\Bbg^2=\Bm^2/x^6$ where $x=r/\Rm$, $\Bm^2=\mu^2/\Rm^6=\mu^{-1}(8L/c)^{3/2}$, and $L$ is the initial power of the wave (before the dissipation). Then, we obtain
\beq
\label{eq:ed_diss}
    \ed \sim  \frac{1.7 \kk g x^6}{\cd^3}\, \frac{\me c^{5/2}\, \nu \mu}{\sT L^{3/2}}.
\eeq
Main dissipation occurs near $\Rm$, at $x\approx 0.94$-0.95 for $L=10^{40}\,$erg/s and at $x\approx 1.3$-$1.5$ for $L=10^{42}\,$erg/s. The two waves also have different shock strengths, $q\approx 2$ and $5$, which give $kg\approx 0.2$ and 1.2, respectively. The value of $\cd\sim 1.5$-$1.8$ is close to the peak $\c$ in the oscillation. Using these values, one can see that the estimate~(\ref{eq:ed_diss}) well explains $\ed$ observed in the simulation: we find $\ed\sim 200$ for $L=10^{42}\,$erg/s (Figure~\ref{fig:shock_L42}), and $\ed\sim 3\times 10^3$ for $L=10^{40}\,$erg/s (Figure~\ref{fig:shock_L40}).

\subsubsection{Duration of the main dissipation phase}

As one can see in Figure~\ref{fig:L}, the main dissipation phase of the powerful wave, $L=10^{42}\,$erg/s, extends over a significant range of radii $\delta r\sim\Rm$. By contrast, the weaker wave with $L=10^{40}\,$erg/s dissipates quite suddenly near $\Rm$, in a narrow range $\delta r\ll\Rm$. 

The length $\delta r$ is related to the number of particles passing through the wave, $\delta\N\sim 4\pi\N(\delta r/r)$. This number (and hence $\delta r$) can be estimated using energy conservation. The wave energy contained in one oscillation is $L/\nu$, and each particle receives energy $\ediss \me c^2$ from the shock. Hence, the number of particles it takes to damp the wave is
 \beq
    \delta\N_{\rm damp}\sim \frac{L}{\ediss \me c^2\,\nu} = \frac{L}{g \ed \me c^2\,\nu},
 \eeq
 and the corresponding damping length is
 \beq
 \label{eq:dr_damp}
    \frac{\delta r_{\rm damp}}{r}
    \sim \frac{\cd^3\, \sT L^{5/2}}{20 \kk g^2 x^6 \me^2c^{9/2}\mu\,\N \nu^2},
 \eeq
where we substituted $\ed$ from \Eq~(\ref{eq:e_balance}). In particular, for the wave with $L=10^{40}\,$erg/s this estimate gives $\delta r_{\rm damp}/r\sim 2\times 10^{-3}$, consistent with the simulation results (Figure~\ref{fig:shock_L40}). \Eq~(\ref{eq:dr_damp}) holds if $\delta r_{\rm damp}/\Rm\ll 1$. Note that $\delta r_{\rm damp}/\Rm$ grows with $L$ and saturates at $\sim 1$ for waves with large $L$ (as in Model~I) or low $\nu$.

\subsubsection{Validity of MHD description}
\label{MHD_validity}

The MHD description of waves fails when the particles become unmagnetized, i.e. their Larmor timescale becomes comparable to the fluid dynamical timescale measured in the fluid rest frame. The demagnetization can happen in kHz waves when they accelerate the plasma to huge Lorentz factors (Paper~I). In GHz waves, the fluid Lorentz factors are modest, but demagnetization may occur for a different reason: the ultra-relativistic temperature of the plasma increases its Larmor timescale. 

In both presented simulations, the MHD requirement $\omL\approx\omB/\e\gg \omega$ holds throughout the evolution of the wave. The ratio $\omL/\omega$ is shown in Figures~\ref{fig:shock_L42} and \ref{fig:shock_L40}. For waves with lower power $L<10^{40}\,$erg/s and the same $\nu=0.3\,$GHz the condition $\omL\gg\omega$ would become violated.

One can estimate $\omL\approx e\Bbg/\e \me c$ during the main dissipation phase using $\e\sim\ed$ given by \Eq~(\ref{eq:ed_diss}):
\beq
\label{eq:omL_om}
    \frac{\omL}{\omega} \sim  
    \frac{8^{3/4}\cd^3}{6\pi g x^9} \, \frac{e\sT\,L^{9/4}}{\me^2c^{17/4} \mu^{3/2} \nu^2 },
\eeq
where we used $\Bbg=\mu(x\Rm)^{-3}=x^{-3}\mu^{-1/2}(8L/c)^{3/4}$. In particular, for the wave with $\nu=0.3\,$GHz and $L=10^{40}\,$erg/s (Model~II) with isotropic plasma ($\kk=3$), one can substitute the numerical factors $g\sim 0.1$ and $x^9\sim 0.6$ evaluated above. The result is approximately consistent with the minimum $\omL/\omega\sim 7$ observed in the simulation (Figure~\ref{fig:shock_L40}).

\subsection{Extremely powerful waves}

GHz waves with power $L\gg 10^{42}$\,erg/s develop radiative shocks at $\Rm$ in the sense that the plasma behind each shock is radiatively cooled on a timescale shorter than the wave period. This leads  to nearly periodic dynamics as the plasma moves through the wave train: shock heating in each oscillation is followed by immediate strong cooling. The wave train remains approximately periodic until it reaches the ``heating radius'' $\Rh$ where synchrotron cooling weakens enough, so that the shocked plasma retains a significant fraction of the received heat before it crosses one wave period and becomes heated again in the next shock. At radii $r>\Rh$ the plasma begins to accumulate heat in the wave train, and then reaches a thermal balance (heating=cooling) at enthalpy $\e$ far exceeding the enthalpy gained in a single shock; this thermal balance was described in \Sect~\ref{balance}.

\begin{figure}[t]
\includegraphics[width=0.48\textwidth]{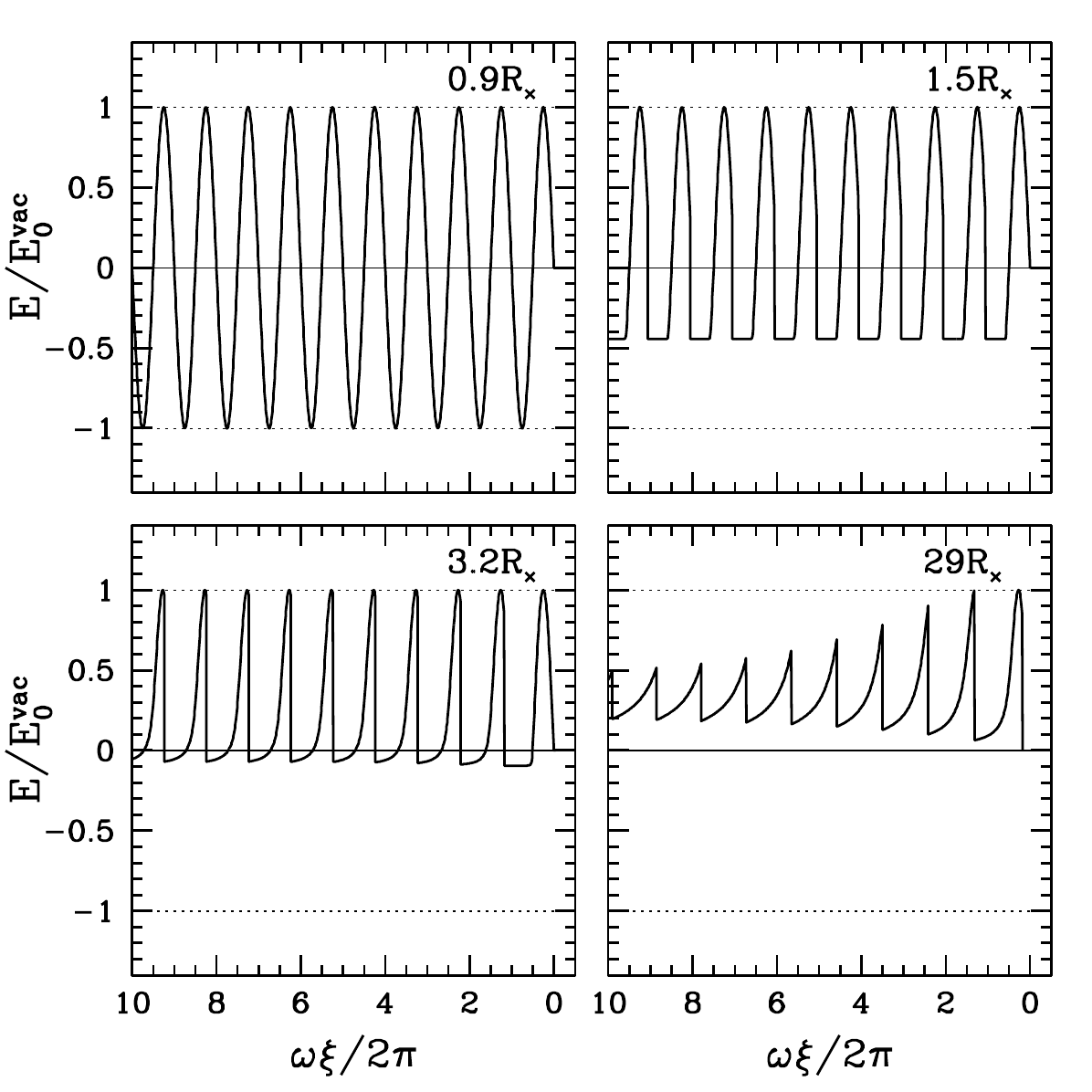} 
\caption{Same model as in Figure~\ref{fig:E_L42} except that here the wave has a higher power $L=10^{43}\,$erg/s. One can see the periodic plateaus in the snapshot at $r=1.5\Rm$. The transition to slow cooling occurs around $\Rh\approx 2\Rm$; then the periodic plateaus disappear, and the GHz wave experience strong damping, especially in its extended part further down the oscillation train (not included in the figure).}
\label{fig:E_L43}
 \end{figure}

\begin{figure}[t]
\includegraphics[width=0.48\textwidth]{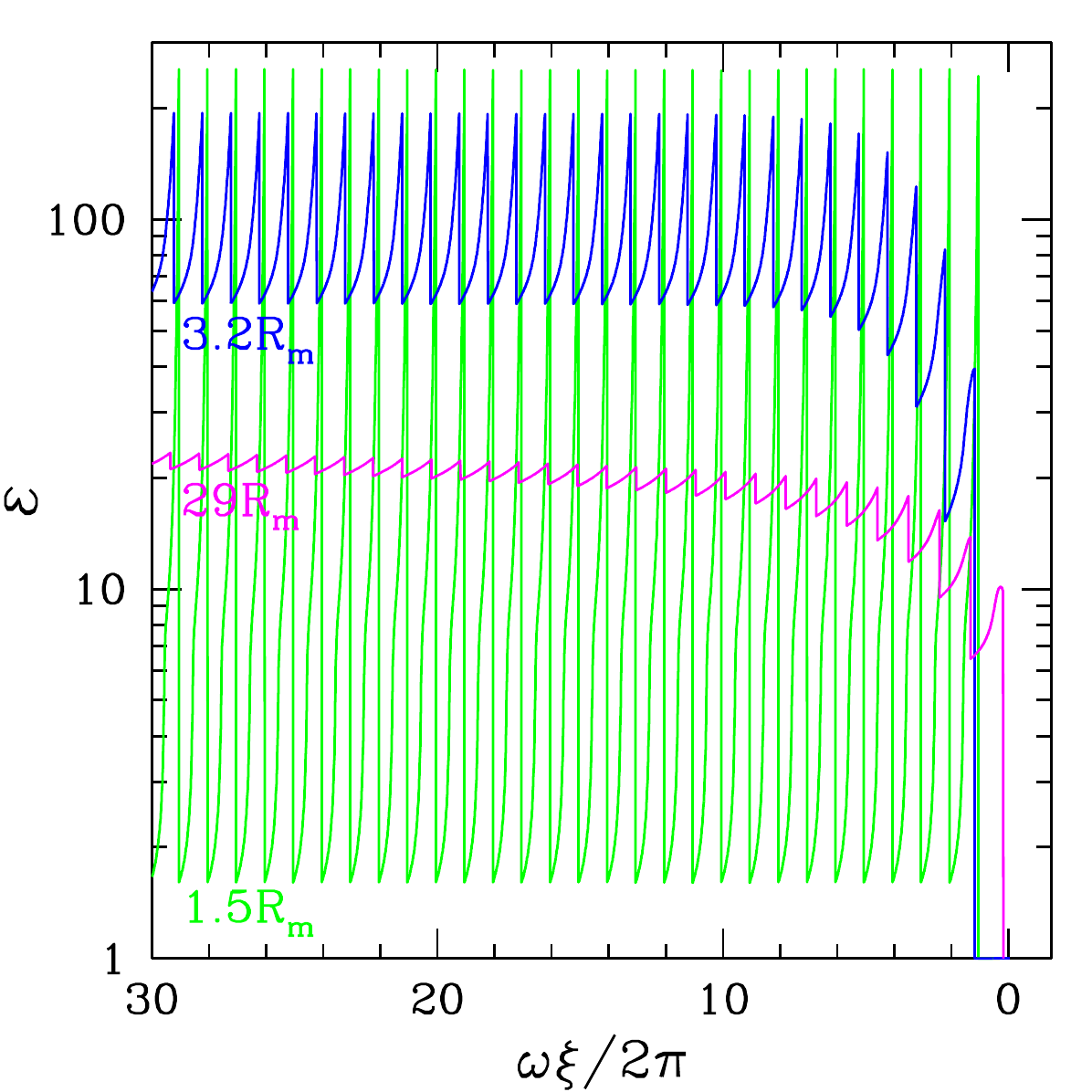} 
\caption{Evolution of the plasma internal energy profile $\e(\xi)$ in the wave shown in Figure~\ref{fig:E_L43}. The larger number of oscillations (30) are included in the figure to show the saturation of $\e$ near the heating=cooling balance when the wave propagates to $r\gg\Rh\approx 2\Rm$.}
\label{fig:heat_L43}
 \end{figure}

Figure~\ref{fig:E_L43} shows an example where the wave has a high initial power $L=10^{43}$\,erg/s (and the same frequency as in the other presented models, $\nu=0.3$\,GHz). The simulation followed the evolution of the wave train with 1000 oscillations, however the figure displays only its small leading part of 10 oscillations, for clarity.  One can see that at $r\approx\Rm$ the wave develops a periodic train of radiative ultra-relativistic shocks. Upstream of each shock the wave profile develops a plateau of $E\approx -\Bbg/2$ (which corresponds to $E^2\approx B^2$), and the plateau width quickly grows to $\Wp\sim \omega^{-1}$. The plateau forms because the MHD evolution ``shaves off'' the excess of $E^2$ to respect $E^2\leq B^2$, as explained in detail in Paper~I. The plasma accelerates as it crosses the plateau, then decelerates at the shock, and then radiates the heat received from the shock. The cycle repeats in each oscillation, and the wave train remains approximately periodic until it reaches radius $\Rh$. In this sample model, we find $\Rh\approx 2\Rm$.

An analytical estimate for the radius $\Rh$ is found by matching the results derived at $r<\Rh$ and $r>\Rh$. At radii $r<\Rh$, the plasma in each oscillation has a negligible memory of shock heating in previous oscillations; the particles periodically gain a Lorentz factor $\gamma_e\equiv \gamma\e$ and radiate the received energy away. The particles gain energy where $E^2\approx B^2$, i.e. at the plateau of  $E(\xi)\approx \Ep=-\Bbg/2$. The plateau evolution $\Ep\propto r^{-3}$ controls the energy loss of the electromagnetic wave, which in turn determines the energy gain of the plasma as it crosses the plateau (Paper~I), so particles crossing the plateau of width $\Wp\sim c/\omega$ gain the Lorentz factor
\beq
\label{eq:game1}
   \gamma_e \sim \frac{c }{\omega r}\,\sigbg \qquad (r\simlt \Rh).
\eeq
This energy is thermalized in the shock (and quickly radiated away) in each oscillation.

At radii $r\simgt \Rh$, the plasma energy per particle is set by the heating=cooling balance (\Eq~\ref{eq:e_balance}),\footnote{We omit a numerical factor ${\cal O}(1)$ set by the plasma dimensionality parameter ($\kk=2$ or $3$) and the average compression factor $\c\sim 1.4$-$1.7$ around the wave crest where main cooling occurs.} 
\beq
\label{eq:game2}
    \gamma_e\sim \frac{2\pi \me c\,\omega} {\sT\Bbg^2} \qquad (r\simgt \Rh).
\eeq
Here, we took into account that the shocks are strong in the extremely powerful waves, $q\gg 1$, which implies $g(q)\approx 1$. Now one can evaluate the transition radius $\Rh$ by matching \Eqs~(\ref{eq:game1}) and (\ref{eq:game2}), 
\beq
\label{eq:Rh}
   \Rh\approx \left(\frac{\sT\mu^4}{8\pi^2\me^2 c^2\N\omega^2}\right)^{1/10}
   \approx \frac{2.2\times 10^8\, \mu_{33}^{2/5}}{\N_{37}^{1/10} \nu_9^{1/5}}\,{\rm cm}.
\eeq
Only waves with sufficiently high power $L$ (and/or lower frequency $\nu$) form shocks in the fast-cooling regime $\Rh>\Rm$, and their profiles $E(\xi)$ develop the plateaus as they propagate between $\Rm$ and $\Rh$. The condition $\Rh>\Rm$ is satisfied when
\beq
\label{eq:Lh}
   L>\Lh\approx 1.5\times 10^{42}\,\mu_{33}^{2/5} \nu_9^{4/5}\,{\rm erg\,s}^{-1}.
\eeq
Such waves experience moderate energy losses ($\simlt 50\%$)  between $\Rm$ and $\Rh$. At radii $r>\Rh$, synchrotron cooling becomes slower than the wave oscillation, and then the GHz oscillations become nearly erased, similarly to the weaker waves with $L<\Lh$ (Figure~\ref{fig:L}). In summary, the GHz burst is strongly damped near $\Rm$ if $L<\Lh$ and near $\Rh$ if $L>\Lh$.


\section{Shock formation outside the equatorial plane}
\label{outside_equator}

Consider a spherical wave front expanding through the dipole magnetosphere. It has a radial wavevector $\bk$. Note that $\bBbg\perp\bk$ only in the equatorial plane $\theta=\pi/2$. The angle $\alpha$ between $\bBbg$ and $\bk$ is given by
\beq 
\label{eq:alpha}
   \tan\alpha=\frac{\Bbg^\theta}{\Bbg^r}=\frac{1}{2}\,\tan\theta.
\eeq
Shock formation is expected at radius $\rsh(\theta)$, which can be found from the condition $E^2=B^2$ with $E$ and $B$ evaluated for the wave propagation in vacuum (Paper~I):
\beq
\label{eq:rsh}
  \rsh(\theta)=\Rm\left(\frac{4-3\sin^2\!\theta}{\sin\theta}\right)^{1/2}=\Rm \frac{\sin^{1/2}\!\theta}{\sin\alpha}.
\eeq

The density parameter of the magnetosphere at $\theta\neq\pi/2$ is defined similarly to that in the equatorial plane,
\beq
   \N_\theta\equiv r^3\nbg(r,\theta).
\eeq
We have added subscript ``$\theta$'' to highlight a possible variation of density with $\theta$. It may be estimated assuming approximately uniform $e^\pm$ loading in the inner magnetosphere, where $\Bbg\simgt 10^{13}$\,G \citep{Beloborodov13a,Beloborodov20}. The created pairs outflow along $\bBbg$ with mildly relativistic speeds and annihilate when they approach the equatorial plane. This picture implies 
\beq
\label{eq:N_theta}
   \nbg\propto \Bbg, \qquad  \frac{\N_\theta}{\N} = \frac{\sin\theta}{\sin\alpha}.
\eeq

\subsection{Plasma velocity profile in the wave}
\label{v_profile}

The simplicity of equatorial waves described in the previous sections is due to the  simple relation between the wave electric field $E$ and the plasma speed $\beta$, which allows one to formulate the energy equation~(\ref{eq:NWE1}) for a single unknown function. In the equatorial wave, the plasma executes the radial $\bE\times\bB$ drift with $\beta=E/B=E/(\Bbg+E)$. By contrast, outside the equatorial plane the oblique $\bBbg$ implies that the wave will move the plasma in both $r$ and $\theta$. Furthermore, in addition to the $\bE\times\bB$ drift, the plasma can slide along the oblique $\bB$. Despite this complication one can still find analytically the relation between $\bb$ and $E$ across the wave profile.
 
The MHD condition $E^2<B^2$ implies existence of a ``drift frame'' $\tilde{\KF}$ (moving with velocity $\bbD=\bE\times\bB/B^2$) in which $\tilde{\bE}=0$. In this frame, the plasma has a pure sliding motion along $\tilde{\bB}$ with some speed $\tb$ and Lorentz factor $\tg=(1-\tb^2)^{-1/2}$. Transformation of the plasma four-velocity $u^\alpha=(\gamma,\bu)$ from frame $\tilde{\KF}$ to the lab frame gives
\beq
\label{eq:slide}
   \gamma=\tg\gD, \quad \bu=\tbu+\tg\buD, 
\eeq
where
\beq
    \gD=\frac{B}{\sqrt{B^2-E^2}}, \qquad \buD=\frac{\bE\times \bB}{B\sqrt{B^2-E^2}},   
\eeq
and $B=\sqrt{B_r^2+B_\theta^2}$. \Eq~(\ref{eq:slide}) shows the decomposition of $\bu$ into components parallel and perpendicular to $\bB$: $\bu_\parallel=\tbu$ and $\bu_\perp=\tg\buD$. 

The relation between the plasma Lorentz factor $\gamma=\gD(1+\tu^2)^{1/2}$ and the wave electric field $E$ will be found if we solve for $\tu(E)$. This can be done using 
\beq
\label{eq:upar}
   \frac{d}{dt}\left(\tu B\right)=\frac{d}{dt}\left(\bu\cdot\bB\right)=\bu\cdot\frac{d\bB}{dt}
   \approx u_\theta\,\frac{dB_\theta}{dt},
\eeq
where the derivative $d/dt$ is taken along the fluid streamline, and we used $\bB\cdot d\bu/dt =0$ (implied by \Eq~(\ref{eq:momentum})). The last (approximate) equality in \Eq~(\ref{eq:upar}) makes use of $dB_r/dt\ll dB_\theta/dt$, which holds for short waves. Indeed, the derivative 
\beq 
   \frac{d\bB}{dt} \approx \frac{d\bBw}{dt},
\eeq
is dominated by the fast oscillation of the wave field $\bBw\equiv \bB-\bBbg$ on top of the slowly varying $\bBbg$, and $\bBw$ in a short wave satisfies\footnote{Vector potential $\Aw$ for a short axisymmetric wave satisfies $\left.\partial_\xi \Aw \gg\partial_t \Aw \right|_\xi$ which implies $\Bw^r\ll\Bw^\theta$ and $\Bw^\theta\approx E$ (Paper~I). The small difference $\Bw-E$ is not negligible only in terms with the large derivative $\partial_\xi$; in particular, $\partial_\xi(E-\Bw)=\left.\partial_t\Bw\right|_\xi+ E/r$, as follows from induction equation $\partial_t \bE=-c\,\nabla\times\bBw$.}
\beq
 \label{eq:short}
    \Bw^r\ll\Bw^\theta\approx E.
\eeq 
Relations $BdB/dt\approx B_\theta dB_\theta/dt$ and $\tu_\theta=\tu B_\theta/B$ give
\beq
    \tu\,\frac{dB}{dt}\approx \tu_\theta\,\frac{dB_\theta}{dt} 
    \approx \tu_\theta\,\frac{dE}{dt}.
\eeq
Using these relations, we find from \Eq~(\ref{eq:upar})
\beq
\label{eq:upar1}
   B\,\frac{d\tu}{dt}  = \frac{d}{dt}(\tu B) - \tu \frac{dB}{dt}
   \approx (u_\theta-\tu_\theta)\frac{dB_\theta}{dt}\approx 
   \tg \uD^\theta \frac{dE}{dt},
\eeq
and hence
\beq
  \frac{d\tu}{dt}  \approx - \frac{\tg\,E\Bbg^r}{B^2\sqrt{B^2-E^2}}\,\frac{dE}{dt}.
\eeq
This gives a differential equation for $\tu(E)$ or $\tg(E)$:
\beq
\label{eq:tu}
   \frac{d\tg}{\tu } = \frac{d\tu}{\tg }
   =  -\frac{E\Bbg^r\,dE}{(\Bbg^2+2\Bbg^\theta E+E^2)\sqrt{\Bbg^2+2\Bbg^\theta E}}.
\eeq
It can be integrated with $\bBbg\approx const$ across the short wave profile, 
\beq
\label{eq:tg}
   \ln \left(\tu+\tg\right)  = I \equiv - \int_0^{E/\Bbg} s(\alpha,z)\,dz,
\eeq
where $z=E/\Bbg$, $\tan\alpha=\Bbg^\theta/\Bbg^r$, and 
\beq
   s\equiv  \frac{E\Bbg^r \Bbg}{B^2\sqrt{B^2-E^2}}= \frac{z\cos\alpha}{(1+2z\sin\alpha+z^2)\sqrt{1+2z\sin\alpha}}.
\eeq
Note that $z\rightarrow -(2\sin\alpha)^{-1}$ corresponds to $E^2\rightarrow B^2$. In this limit, we find 
\beq
  \tg=\frac{1}{\sin\alpha} \quad {\rm when~} E^2\rightarrow B^2.
\eeq

The obtained $\tu(E)$ determines $\gamma(E)$ and $\bu(E)$ according to \Eq~(\ref{eq:slide}). This solution describes the plasma velocity profile in a short wave in the oblique $\bBbg$.

\subsection{Shock formation}

As long as the plasma oscillating in the wave stays cold (i.e. until shock formation), the wave evolution obeys \Eq~(\ref{eq:NWE1}), which states $(\partial_t E)_\xi$ in terms of $(\partial_\xi\gamma)_t$. One can use the relation between $E$ and $\gamma$ found in \Sect~\ref{v_profile} to express $\partial_t E$ in terms of $\partial_t\gamma$, and then  \Eq~(\ref{eq:NWE1}) becomes a differential equation for $\gamma(t,\xi)$,
\beq
\label{eq:NWE2}
     \frac{\partial_t\gamma }{f} + 2\pi \rhobg c^2\, \partial_\xi\gamma =  \frac{2cE^2}{r},
\eeq
where function $f(E,\bBbg)$ is given by (see Appendix~\ref{cold_wave}),
\beq
\label{eq:f}
  f\equiv \frac{(\Bbg^2+\Bbg^\theta E) \gamma^3 - \tu\Bbg^r B \gamma^2}{B^4\tg^2}.
\eeq
 \Eq~(\ref{eq:NWE2}) is the generalization of \Eq~(\ref{eq:evol_gamma}) to polar angles $\theta\neq \pi/2$. It contains $\theta$ as a parameter and will show how the wave develops caustics at each $\theta$. Its characteristics $\xi_+(t)$ are determined by the coefficients of $\partial_t\gamma$ and $\partial_\xi\gamma$, 
\beq
\label{eq:C+_oblique}
  \frac{d\xi_+}{dt} =  2\pi \rhobg c^2  f,
\eeq
and the evolution of $\gamma$ along characteristics is given by
\beq
\label{eq:evol_oblique}
  \left.\frac{d\gamma}{dt}\right|_{C^+} = \frac{2cE^2}{r} f.
\eeq 
In the equatorial plane, these equations reproduce the results of the previous sections. Then, $f=\Bbg\gamma^3/B^3=(\c^3\Bbg^2)^{-1}$ and $d\xi_+/dt$ is reduced to \Eq~(\ref{eq:C+_cold}) while $(d\gamma/dt)_{C^+}$ becomes equivalent to \Eq~(\ref{eq:evol}) (taking into account the relation $\gamma=(\c^2+1)/2\c$).

The evolution equation for $rE$ along $C^+$ has the form (see Appendix~\ref{cold_wave}),
\beq
   \left.\frac{1}{E}\frac{d(rE)}{dt}\right|_{C^+} = {\cal O}\left(\frac{\gamma^3}{\sigbg}\right).
\eeq
Thus, waves with $\gamma^3\ll\sigbg$ satisfy $rE\approx const$ along the characteristics. This feature was demonstrated in the previous sections at $\theta=\pi/2$, and it also holds outside the equatorial plane.

Shocks appear at caustics in the flow of characteristics. The caustic location $\rc(\theta)$ and the plasma Lorentz factor at the caustic $\gc(\theta)$ are calculated in Appendix~\ref{cold_wave}. The calculation can be completed analytically for waves with $\zeta\ll 1$ and $\zeta\gg 1$. In these two limits, we find
\begin{eqnarray}
  \frac{\rc}{\rsh}\approx  \left\{\begin{array}{cr}
  1 & \qquad \zeta\ll 1 \\
  \displaystyle{ \left(\frac{8}{\zeta}\,\frac{\sin^3\!\alpha}{\sin\theta}\right)^{1/6} } & \qquad \zeta\gg 1
                             \end{array}\right.
\label{eq:rc}
\end{eqnarray}
\begin{eqnarray}
  \gc\approx  \left\{\begin{array}{cr}
  \displaystyle{
    \left(\frac{\sqrt{3/2}}{\zeta \sin\theta\sin\alpha}\right)^{1/2} } & \qquad \zeta\ll 1 \\
  1 & \qquad \zeta\gg 1
                                         \end{array}\right. 
\label{eq:gc}
\end{eqnarray}
where $\alpha(\theta)$ is given by \Eq~(\ref{eq:alpha}), and the parameter $\zeta$ is defined in the equatorial plane (\Eq~\ref{eq:zeta}).


\section{Transition to the kinetic regime}
\label{MHD_kinetic}

\subsection{Shock heating and de-magnetization of particles}
\label{de-mag}

Shock heating of the plasma to $\e\gg 1$ reduces the Larmor frequency $\omL=\omB/\e=e\Bbg/\me c\e$. If $\omL$ becomes comparable to $\omega$, the MHD description of the wave fails. Then, a kinetic description will be required, where particles individually interact with the wave. In \Sect~\ref{MHD_validity} we estimated $\omL/\omega$ in the equatorial waves. Below we evaluate this ratio in waves outside the equatorial plane, and find the boundary between the two damping regimes (kinetic vs. MHD shocks) on the $L$-$\theta$ plane.

The plasma internal energy $\e$ is controlled by the balance between synchrotron cooling (averaged over one oscillation) and shock heating. This thermal balance is described in \Sect~\ref{balance} for waves in the equatorial plane, and a similar balance can be stated at $\theta\neq \pi/2$. In particular, we can use \Eq~(\ref{eq:e_balance}) to evaluate $\e$. The choice of $\kk=2$ vs. $\kk=3$ weakly affects the numerical coefficient in \Eq~(\ref{eq:e_balance}); for definiteness, below we assume isotropic plasma ($k=3$). We will also use $\cd\sim 1$ because in the transition regime of interest $\kappa_d$ is close to unity (see Figure~\ref{fig:kappa_L40} for Model~II, which comes close to the MHD/kinetic transition). Then, we find 
\beq
\label{eq:omL_om1}
    \frac{\omL}{\omega}\sim \frac{e\sT\Bbg^3}{6\pi \me^2 c^2\omega^2 g},
\eeq
where the numerical factor $g=\ediss/\ed$ depends on the shock compression factor $q$. For the perpendicular shocks at $\theta=\pi/2$, $g(q)$ is stated in \Eq~(\ref{eq:g}). Note also that $g(\theta)\approx 1$ holds for waves with $\zeta\ll 1$, which develop relativistic shocks during the main damping phase.

The plasma response to the wave may be described as MHD drift if the full Larmor period $2\pi/\omL$ is shorter than the timescale for a large change of the field, which we take as 1/4 of the wave period. Thus, we roughly estimate that the transition between the MHD and kinetic regimes  occurs at $\omL/\omega\sim 4$. This corresponds to
\beq
   \Bbg\sim \BMHD\approx 4\left(g \frac{\me^2c^2\omega^2}{e\sT}\right)^{1/3}
   \approx 2\times 10^7 g^{1/3}\nu_9^{2/3}\,{\rm G}.
\eeq
The same condition may be stated as 
\beq
        \omB\sim 2\left(g\,\frac{c}{r_e}\,\omega^2\right)^{1/3},
\eeq
where $r_e=e^2/\me c^2$. Damping of the wave through MHD shocks begins when caustics form (at radius $\rc(\theta)$ given by \Eq~(\ref{eq:rc})), and most of the damping occurs at $r\sim\rc$. The approximate condition for the wave to be damped in the MHD regime at a given polar angle $\theta$ is 
\beq
   \Bbg(\rc)>\BMHD \quad ({\rm shock~damping}).
\eeq 

\begin{figure}[t]
\includegraphics[width=0.46\textwidth]{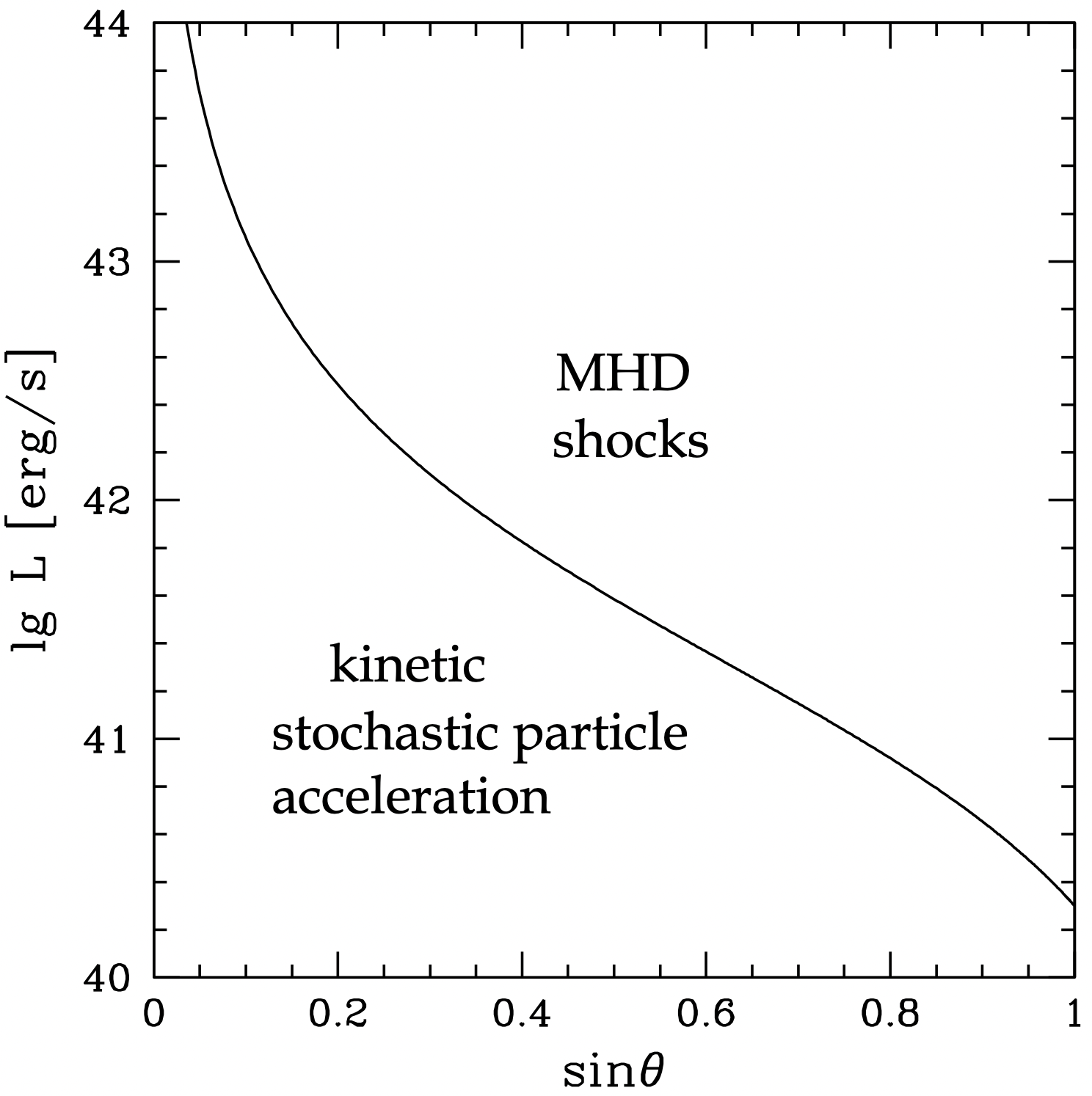} 
\caption{Two regions in the $L$-$\sin\theta$ parameter space: wave damping in the MHD regime (through shocks) and in the kinetic regime (through stochastic particle acceleration). Their approximate boundary $\LMHD(\theta)$ [or $\theta_{\rm MHD}(L)$]  is plotted here for waves with frequency $\nu=1$\,GHz propagating in the magnetosphere with dipole moment $\mu=10^{33}$~G~cm$^3$. The scaling of $\LMHD(\theta)$ with $\nu$ and $\mu$ is given in \Eq~(\ref{eq:LMHD_theta}).}
\label{fig:L_theta}
 \end{figure}

In particular, waves with $\zeta<1$ have $\rc\approx\rsh$ (\Eq~\ref{eq:rc}) and $g\sim 1$, and then we find 
\beq
   \frac{\Bbg(\rc)}{\BMHD} \sim \left(\frac{L}{c}\right)^{3/4}
    \frac{\sin^2\!\alpha}{\sqrt{\mu \sin\theta}} \left(\frac{e\sT}{\me^2c^2\omega^2}\right)^{1/3},
\eeq
where we used \Eq~(\ref{eq:rsh}) for $\rsh$. One can see that the MHD regime $\Bbg(\rc)>\BMHD$ holds at a given angle $\theta$ if the wave power $L$ exceeds a critical value $\LMHD$,  
\begin{eqnarray}
\nonumber
  \LMHD(\theta) &\sim& 
  c\mu^{2/3}
    \left(\frac{\me^2c^2\omega^2}{e\sT}\right)^{4/9}
     \frac{\sin^{2/3}\!\theta}{\sin^{8/3}\!\alpha}   \qquad (\zeta\simlt 1)   \\
     &\sim& 2\times 10^{40} \mu_{33}^{2/3}\nu_9^{8/9} 
      \frac{\sin^{2/3}\!\theta}{\sin^{8/3}\!\alpha}\,\frac{\rm erg}{\rm s}.
\label{eq:LMHD_theta}
\end{eqnarray}
At small polar angles $\theta\ll 1$, one can use $\alpha\approx \theta/2$ and see that $\LMHD\propto\theta^{-2}$. The MHD damping condition $L>\LMHD(\theta)$ may also be written as a condition on $\theta$ for a given $L$: $\theta>\theta_{\rm MHD}(L)$. For example, waves with frequency $\nu=1$\,GHz and power $L=10^{42}$\,erg/s will experience shock damping in the MHD regime at polar angles $\theta>\theta_{\rm MHD}\sim 0.3$ (Figure~\ref{fig:L_theta}).

\subsection{Charge starvation?}
\label{starvation}

Another condition for wave propagation in the MHD regime is a sufficiently high plasma density, $n> j/ec$, capable of sustaining electric current $j$ demanded by MHD. The electric current in an axisymmetric wave is toroidal, $\bj=(0,0,j_\phi)$, and
satisfies the relation,
\beq
   -E j_\phi = \rho c^2\,\frac{d\gamma}{dt} = \rho c^2(1-\beta_r)\frac{d\gamma}{d\xi}.
\eeq 
For waves with $\gamma^3\ll\sigbg$ one can use $d\gamma/d\xi\approx\partial_\xi\gamma$, so 
\beq
   \frac{j_\phi}{enc} \approx -\frac{\me c(1-\beta_r)}{eE}\,\partial_\xi\gamma.
\eeq 

In particular, consider waves in the equatorial plane with $\zeta\ll 1$. They accelerate the plasma to large $\gamma$. Large $\partial_\xi\gamma$ develops at caustic formation with $\beta\approx -1$ and $\gamma$ growing to $\gc\approx 24^{1/4}\zeta^{-1/2}$ in the narrow interval $\delta\xi\sim \zeta^{1/2}/\omega$, which gives
\beq
\label{eq:j_xi}
   |\partial_\xi \gamma| \sim 24^{1/4}\,\frac{\omega}{\zeta}.
\eeq
A similar gradient of $\gamma$ is sustained later ahead of the developed shock, when the upstream $\gamma$ approaches its maximum $\gamma\sim \zeta^{-1}$ while the width of the pre-shock acceleration region grows to $\delta\xi\sim \omega^{-1}$ (see Paper~I). Using \Eq~(\ref{eq:j_xi}), we obtain a rough estimate for the current density during the MHD evolution of the wave:
\beq
   \frac{j}{enc} \sim \frac{10}{\omB \omega \zeta},
\eeq
where we used $E\approx\Bbg/2$ near $\Rm$ and $\omB\equiv e\Bbg/\me c$. We conclude that charge starvation does not prevent the MHD evolution if
\beq
   \frac{c\,\sigm}{\omB \Rm}\simlt 0.1.
\eeq
Using $\sigm=\omB^2/\omega_p^2$, one can also rewrite this condition as $c\omB/\omega_p^2\Rm\simlt 0.1$. The same condition for shock formation was found in PIC simulations of magnetosonic waves \citep{Chen22b}.

If the wave had only one oscillation, with a single shock, it would heat the plasma up to $\e\sim\zeta^{-1}$ at the peak of the shock strength. Then, the MHD condition $\omL\simgt 4 \omega$ would be similar to $j/enc<1$ derived above. In a wave train with many shocks, the plasma is heated to $\e\gg \zeta^{-1}$. Therefore, the condition $\omL \simgt 4\omega$ examined in \Sect~\ref{de-mag} becomes more demanding than $j/enc<1$, i.e. MHD applicability is limited by de-magnetization of heated particles rather than by charge starvation.

\section{Damping in the kinetic regime}
\label{kinetic}

Radio waves propagating at a polar angle $\theta$ with power isotropic equivalent $L(\theta)<\LMHD(\theta)$ cannot be damped by shocks, because the magnetospheric particles become unmagnetized in the wave, i.e. reach $\omL\sim\omega$, before the wave experiences significant losses. The de-magnetization transition occurs close to the radius $\rsh$ with the background gyro-frequency $\omB(\rsh)= (e\mu/\me c\rsh^3)\sin\theta/\sin\alpha$. The transition happens when shock heating increases the plasma internal energy $\e$ (thermal Lorentz factor) to
\beq
   \etr \sim \frac{\omB(\rsh)}{\omega} 
   = \frac{\omB(\Rm)}{\omega}\,\frac{\sin^2\!\alpha}{\sin^{1/2}\!\theta}.
\eeq
The breaking of MHD description (i.e. breaking of the drift description of plasma response to the oscillating electromagnetic field) implies that the condition $E^2<B^2$ is no longer enforced. The part of the wave developing $E^2>B^2$ is no longer ``shaved off'' by the shock as described in Paper~I, because the shocks dissolve when $\omL\sim\omega$.

Subsequent evolution will further energize plasma particles as described below, leading to $\omL<\omega$ (note that damping of a small fraction of the wave energy is sufficient for plasma heating to $\omL<\omega$). In a wave with $L\ll\LMHD$, radiative losses are negligible at the transition $\omL\sim \omega$ ($\etr$ is below the radiative ceiling), however losses will become important with increasing particle energies, leading to efficient radiative damping of the wave. The main damping will develop in the regime of $\omL<\omega$, away from the cyclotron resonance $\omL=\omega$. 

When the wave exits the MHD regime at $r\approx\rsh$ and continues its propagation to $r>\rsh$ with the intact sine profile (which includes parts with $E^2>B^2$), it interacts with the plasma very differently from the MHD regime. As shown in \cite{Beloborodov22} (hereafter B22), magnetospheric particles exposed to the train of wave oscillations experience quick stochastic acceleration. This process is convenient to view in frame $\KF'$ boosted along $\bBbg$ so that the wavevector $\bk'$ becomes perpendicular to $\bBbg'=\bBbg$.\footnote{In this frame, plasma has no bulk motion perpendicular to $\bBbg$, after averaging over Larmor rotation. Note that in the kinetic regime, $\omL<\omega$, averaging over Larmor rotation also removes the wave oscillation, as the wave period is smaller than the Larmor period. (By contrast, in the MHD regime, $\omL\gg\omega$, the fluid velocity is defined on scales smaller than $\omega^{-1}$ and oscillates with the wave period.)}
This frame moves along $\bBbg$ with speed $\beta_F=\cos\alpha$ and Lorentz factor 
\beq 
   \gamma_F=\frac{1}{\sin\alpha}.
\eeq 
Hereafter all quantities with primes are measured in frame $\KF'$. Note that 
\beq
  \omega'=\omega\sin\alpha, \quad a'_0=a_0, \quad \bBbg'=\bBbg,
\eeq
where $a_0\equiv eE_0/\me c\,\omega$. The transition radius $\rsh$ corresponds to $2\Em'=\Bbg$ or $a_0=\omB/2\omega'$.

The particle interaction with the radio wave affects the four-velocity component $u'_\perp$ perpendicular to $\bBbg$. Indeed, in frame $\KF'$ particles experience no  force component along $\bBbg$, since $\bE'\perp\bBbg$ and $\bB'\parallel \bBbg$. As stochastic acceleration pumps $u'_\perp$, it tends to increase the particle pitch angle relative to $\bBbg$. At the same time, radiative losses reduce all components of $\bu'$, as the relativistic particle radiates along $\bu'$. Thus, the continued pumping of $u'_\perp$ combined with persistent radiative losses will drive the particle pitch angle toward $\pi/2$ in frame $\KF'$. Note also that $\gamma'\approx u'_\perp$ for the ultra-relativistic particle.
 
The description of particle acceleration by the wave in B22 applies in frame $\KF'$. Stochastic acceleration may be viewed as diffusion in $\gamma'$ with a diffusion coefficient 
\beq 
   D\sim \omL' \left(\frac{\gamma'a_0^3\omega'}{\omB}\right)^{2/3},
\eeq
where $\omL'=\omB/\gamma'$, and
\beq
   \omB=\frac{e\Bbg}{\me c}\approx 1.8\times 10^{13}\,\frac{\mu_{33}}{r_9^3}\,\frac{\sin\theta}{\sin\alpha}\; {\rm rad\,s}^{-1}.
\eeq
Mean expectation for the energy gain rate is given by (omitting a numerical factor $\sim 1$),
\beq
  \langle \dot{\gamma'}\rangle \sim \frac{D}{\gamma'}
     \sim \frac{a_0^2\,\omB^{1/3} {\omega'}^{2/3}}{{\gamma'}^{4/3}}. 
\eeq

In a quasi-steady state, stochastic acceleration becomes offset by radiative losses (which are Lorentz-invariant),
\beq
\label{eq:dgem}
   \langle \dot{\gamma}_{\rm em}\rangle = \frac{2r_e}{3c}\,a_0^2\,\omega'^2\langle\gamma'^2\rangle.
\eeq
Then, the mean expectation for the particle Lorentz factor $\langle\gamma'\rangle$ may be estimated by balancing $\langle \dot{\gamma'}\rangle $ with $\langle \dot{\gamma}_{\rm em}\rangle$,
\beq
\label{eq:mean}
    \langle\gamma'\rangle \sim \left(\frac{c}{r_e\omega'}\right)^{3/10}\left(\frac{\omB}{\omega'}\right)^{1/10}.
\eeq 
Its value for GHz waves is a few times $10^4$. One can also check the condition $\langle\gamma'\rangle > \gamma_\star\equiv(a_0^3\omega'/\omB)^{1/2}$ that is expected in stochastic acceleration (see B22). Note that $\langle\gamma'\rangle \propto r^{-3/10}$, and $\gamma_\star=a_0(\rsh)/\sqrt{2}$ is constant with radius. After some algebra, we find (at $\theta<\theta_{\rm MHD}<1$):
\beq
    \frac{\langle\gamma'\rangle}{\gamma_\star}  
    \sim\left(\frac{\theta_{\rm MHD}}{0.06}\right)^{3/5}
      \left(\frac{\theta_{\rm MHD}}{\theta}\right)^{3/4}\left(\frac{\rsh}{r}\right)^{3/10}.
\eeq

Direct calculation of unmagnetized particle motions in the wave demonstrates that they quickly develop chaos, forming a quasi-steady distribution around $\langle\gamma'\rangle$. B22 showed that the distribution extends from $\sim\gamma_\star$ up to the radiation-reaction limit,
\begin{eqnarray}
\nonumber
  \gRRL' &\approx& \left(\frac{c}{r_e\omega' a_0}\right)^{3/8}\left(\frac{\omB}{\omega'}\right)^{1/4} \\
  &\approx& \left(\frac{c}{r_e\omega' }\right)^{3/8}\left[\frac{4}{a_0(\rsh)}\right]^{1/8}
    \left(\frac{\rsh}{r}\right)^{3/8}.
\end{eqnarray}
For typical FRB parameters, $\gRRL'/\langle\gamma'\rangle\sim$ a few.

The characteristic timescale for particle acceleration to $\langle\gamma'\rangle$ is 
\beq
   t'_{\rm acc}\sim  \frac{\langle\gamma'\rangle}{\langle \dot{\gamma'}\rangle}
   \sim \frac{\langle\gamma'\rangle^{7/3}}{a_0^2\, \omB^{1/3} {\omega'}^{2/3}},
\eeq
which gives
\beq
   \omega' t'_{\rm acc}\sim \left(\frac{c}{r_e\omega'}\right)^{7/10} 
   \left(\frac{\omega'}{\omB}\right)^{1/10}  a_0^{-2}.
\eeq
The Lorentz-invariant quantity $N_{\rm acc} = \omega' t'_{\rm acc}/2\pi$ is the number of wave oscillations that the plasma should cross to approach the quasi-steady $\langle\gamma'\rangle$ (see Figure~4 in B22). The value of $a_0\equiv eE_0/\me c\omega$ may be expressed as  
\begin{eqnarray}
\nonumber
   a_0 &=& \frac{\omB(\rsh)}{2\omega'}\frac{\rsh}{r}=\frac{2^{5/4}\,e}{mc\mu^{1/2}\omega}\left(\frac{L}{c}\right)^{3/4}\,\frac{\sin\alpha}{\sin^{1/2}\!\theta} \\
   &\approx& 9 \times 10^4\,\frac{L_{42}^{3/4}}{\mu_{33}^{1/2}\nu_9}\,\frac{\sin\alpha}{\sin^{1/2}\!\theta} \frac{\rsh}{r}.
\label{eq:a0_damp}
\end{eqnarray}
Then, we find
\beq
   \omega' t'_{\rm acc} \sim 0.1\,
     \frac{\mu_{33}^{21/20} \nu_9^{7/5}}{L_{42}^{63/40}}
   \frac{\sin^{21/20}\!\theta}{\sin^{14/5}\!\alpha}\left(\frac{r}{\rsh}\right)^{17/10}.
\eeq
The wave train in a typical FRB with duration $\tau\sim 0.1$-$1$\,ms has $N=\omega\tau/2\pi\sim 10^5$-$10^6\gg N_{\rm acc}$. So, after the transition to the kinetic regime (which occurs near $\rsh$), particles exposed to the wave almost immediately develop a quasi-steady distribution with the mean expectation $\langle\gamma'\rangle$ (\Eq~\ref{eq:mean}), and continue to move through the wave train with $\langle\gamma'\rangle$.

The quasi-steady particle distribution established in the wave reflects the balance between stochastic acceleration and radiative losses. The radiative losses are irreversible and occur at the expense of the electromagnetic wave energy. The rate of energy loss per particle (\Eq~\ref{eq:dgem}) in the quasi-steady distribution is 
\beq
   \langle \dot{\gamma}_{\rm em}\rangle 
   \sim \frac{2}{3}\,a_0^2 \left(\frac{r_e}{c}\right)^{2/5}\omega^{6/5}\omB^{1/5}\sin^{6/5}\!\alpha.
\eeq
As the plasma crosses the wave train, it radiates the following energy per particle (measured in the lab frame):
\beq
   \frac{\Delta\E_e}{\me c^2}\sim \langle \dot{\gamma}_{\rm em}\rangle\, \tcross,
\eeq
where $\tcross$ is the time it takes the plasma to cross the wave train of duration $\tau$. The crossing time is determined by the radial component of the fluid velocity, 
\beq
  \beta_F^r=\beta_F\cos\alpha=\cos^2\!\alpha.
\eeq
This velocity describes the average radial motion of the magnetospheric plasma exposed to the wave (assuming a static $\bBbg$; the acceleration of background magnetic field lines gives a small correction, as explained below). The fluid speed relative to the wave front, is $1-\beta_F^r=\sin^2\!\alpha$, and so  
\beq
\label{eq:tcross}
\tcross\sim \frac{\tau}{\sin^2\!\alpha}.
\eeq

As the spherical wave propagates a radial distance $\delta r$, it interacts with $\delta\N\approx 4\pi\N_\theta\,\delta r/r$ of magnetospheric particles. The number of particles sufficient to damp the wave is $\delta\N_{\rm damp}=L\tau/\Delta\E_e$, and we find
\begin{eqnarray}
\nonumber
  \frac{\delta\N_{\rm damp}}{\N_\theta}
   &\approx& \frac{L\tau}{\N_\theta \Delta\E_e}
   \sim \frac{2\pi r^2  \sin^{4/5}\!\alpha}{\sT \N_\theta}\left(\frac{r_e^3\omega^4}{c^3\omB}\right)^{1/5} \\
   &\sim& 3\times 10^{-3}\,\frac{r_9^{13/5}\nu_9^{4/5}}{\mu_{33}^{1/5}\N_{37}}\,  \frac{\sin^2\!\alpha}{\sin^{6/5}\!\theta},
\end{eqnarray}
where we used $L/a_0^2=\me c\, \omega^2 r^2/2r_e$ and substituted $\N_\theta=\N\sin\theta/\sin\alpha$ (\Eq~\ref{eq:N_theta}). One can see that $\delta \N_{\rm damp}<\N_\theta$ for the relevant radii $r\simgt \rsh$, which implies quick damping of the wave near $\rsh$.

One caveat in the above calculation is that at $\theta\ll 1$ the plasma may fail to cross the entire wave train on the wave expansion timescale $r/c$. This happens where $\tcross>r/c$, which corresponds to $\alpha<\alpha_{\rm cross}\sim (c\tau/r)^{1/2}$. For the typical parameters, $\alpha_{\rm cross}\sim 0.1$. Thus, near the magnetic axis, the plasma becomes ``stuck'' in the wave train and surfs its leading part instead of crossing it. The damping process cannot be completed without plasma filling the entire wave train, and so one might conclude that the wave will escape in the cone of $\theta\approx 2\alpha <2\alpha_{\rm cross}$. 

However,  there is an additional process that efficiently fills the wave train with plasma. The accelerated particles surfing the leading part of the wave emit gamma-rays, which freely propagate across the wave train and load it with $e^\pm$ pairs via photon-photon collisions. This process triggers an $e^\pm$ avalanche \citep{Beloborodov21b}.

It is easy to verify that the spectrum of curvature photons emitted by the accelerated particles extends to the gamma-ray band. The characteristic frequency of curvature photons in frame $\KF'$ is given in B22,
\beq
  \omega'_c \approx 2a_0 {\gamma'}^2 \omega'.
\eeq
Particles with the average $\langle  \gamma' \rangle$ (\Eq~\ref{eq:mean}) radiate  photons of characteristic frequency 
\beq
   \langle  \omega'_c \rangle 
   \sim 2a_0 \left(\frac{c}{r_e}\right)^{3/5}\left(\omega'\omB\right)^{1/5}.
\eeq
At $r\approx\rsh$, one can substitute $a_0\approx \omB/2\omega'$ (\Eq~\ref{eq:a0_damp}) and find the characteristic photon energy,
\beq
  \epsilon'_c (\rsh) 
   \equiv \langle  \frac{\hbar\omega'_c}{\me c^2}\rangle \sim 10^3
\frac{L_{42}^{9/10}}{\mu_{33}^{3/5}\nu_9^{4/5}} 
           \frac{\sin^{8/5}\!\alpha}{\sin^{3/5}\!\theta}.
\eeq
This gives $\epsilon'_c (\rsh)\sim 30-100$ at polar angles where wave damping occurs in the kinetic regime. It is not far from the maximum $\epsilon_c'$ estimated in B22 for the most energetic particles in the distribution, near the radiation-reaction limit $\gRRL'$,
\beq
  \left.\frac{\hbar\omega'_c}{\me c^2}\right|_{\rm RRL}\approx \frac{1}{\alpha_f}\left(\frac{r_e\omega_B^2 a_0}{c\omega'}\right)^{1/4},
\eeq
where $\alpha_f=e^2/\hbar c\approx 1/137$. In particular, at $r\approx\rsh$, one can use $a_0(\rsh)\approx\omB/2\omega'$ to get 
\beq
   \left.\frac{\hbar\omega'_c}{\me c^2}\right|_{\rm RRL}
  \approx \frac{a_0^{3/4}}{\alpha_f}\left(\frac{4r_e\omega'}{c}\right)^{1/4}
  \approx 500 \frac{L_{42}^{9/16}}{\mu_{33}^{3/8}\nu_9^{1/2}}\frac{\sin\alpha}{\sin^{3/8}\!\theta}.
\eeq

The curvature radiation spectrum emitted by each particle extends to lower energies with  the power-law index $1/3$. Therefore, a significant fraction of the radiated power is in the MeV range, where  photons collide and convert to $e^\pm$ with a large cross section $\sim 0.1\sT$. Using estimates similar to \cite{Beloborodov21b}, one can verify that this process converts a significant fraction of the wave energy into secondary $e^\pm$ pairs, loading the wave with a large number of particles, far exceeding the initial particle number in the background magnetosphere. The   $e^\pm$ loading of the wave implies its inevitable damping.
 
Our choice of frame $\KF'$ neglected the fact that the wave exerts pressure on the background magnetosphere, driving its bulk acceleration and compression (see \cite{Beloborodov21b}). This additional effect slightly changes the fluid rest frame used to calculate stochastic particle acceleration (in this frame $\Ebg'=0$ and the plasma motion vanishes after averaging over Larmor rotation). This effect is moderate at radii of interest, $r\sim\rsh(\theta)$, where the wave becomes damped and deposits radial momentum into the magnetosphere. Near radius $\Rm=\rsh(\pi/2)$, deposition of wave momentum $\E/c$ into the magnetosphere results in its bulk acceleration to speed $\beta_p\approx \E/\Rm^3\Bbg^2\approx \tau c/8\Rm\sim 10^{-2}$ for typical parameters. At small $\theta$, bulk acceleration is stronger. However, at all polar angles, the damping radius $r\approx\rsh(\theta)$ corresponds to $E'_0\approx\Bbg/2$, i.e. damping occurs where the wave energy density in frame $\KF'$ is comparable to $\Bbg^2/8\pi$. Therefore, at $r\sim\rsh$ the wave is at best capable of mildly relativistic bulk acceleration and moderate compression of the background magnetic field. The corresponding change of frame $\KF'$ will not change our conclusion that GHz waves are efficiently damped by stochastically accelerating particles.

\cite{Qu22} argued that near the magnetic dipole axis the wave damping should be considered on open magnetic field lines with the background magnetospheric plasma flowing with a Lorentz factor $\gamma_{\rm bg}\simgt 10^3$. They suggested that the high $\gamma_{\rm bg}$ would render the damping mechanism inefficient and that radio bursts could escape in a broad solid angle around the magnetic axis. This possibility is, however, problematic:

(1) Solid angle $\delta\Omega$ where damping develops on open field lines is not broad. Damping of waves near the magnetic axis begins at $\rsh(\theta)\approx 2\Rm/\theta^{1/2}$. The open field lines occupy $\theta<\theta_{\rm open}(r)\approx (r/\RLC)^{1/2}$. One can see that damping develops on the open field lines for waves propagating at angles $\theta < (2\Rm/\RLC)^{4/5}$. The corresponding solid angle is
\beq
   \delta\Omega\approx \frac{1}{2}\left(\frac{2\Rm}{\RLC}\right)^{8/5} \sim 10^{-2}\,P^{-8/5}\mu_{33}^{4/5} L_{42}^{-2/5},
\eeq
where $P$ is the magnetar rotation period in seconds. This estimate assumes $R_{\rm LC}\gg \Rm$, which is satisfied for $P\simgt 0.1$\,s. The population of observed local magnetars have periods $P=2-12$\,s \citep{Kaspi17}, as expected from their fast spindown due to the strong fields. Hyper-active magnetars proposed as sources of repeating FRBs have ages $t\sim 10^9$\,s and periods $P\sim 1$~s \citep{Beloborodov17b}.

(2) Assuming a large $\gamma_{\rm bg}\sim 10^2-10^3$ would be reasonable for open field lines in ordinary (rotation-powered) pulsars, but not in magnetars. The $e^\pm$ plasma in the outer magnetosphere, on both open and closed field lines, experiences  drag exerted by resonant scattering of dense radiation flowing from the magnetar \citep{Beloborodov13a}. Therefore, the $e^\pm$ flow in the outer magnetosphere is expected to be mildly relativistic. The speed and density of $e^\pm$ on the open field lines of magnetars is estimated in \cite{Beloborodov20}.

(3) Particles tend to forget their pre-wave motion in the magnetosphere once they become exposed to the wave. As shown above, particles interacting with the wave in the kinetic regime ($\omL<\omega$) quickly establish a quasi-steady momentum distribution, with stochastic motions becoming perpendicular to $\bBbg$ in frame $\KF'$ defined by the condition $\bk'\perp\bBbg'=\bBbg$.

\cite{Qu22} also argued that the wave pressure on the magnetosphere could stretch out the magnetospheric field lines, making them more radial, so that the angle between $\bBbg$ and the wavevector $\bk$ is reduced, potentially helping the wave to escape. In fact, the wave pressure cannot significantly change the direction of $\bBbg$ in the wave. Note that near the damping surface $\rsh(\theta)$ the wave pressure perpendicular to $\bBbg$ is comparable to $\Bbg^2/8\pi$, so the background field resists strong changes. Furthermore, even a much stronger wave pressure would be unable to stretch radially the magnetic field lines inside the spherical wave packet, because its thickness is far smaller than radius, $c\tau\ll r$.

\newpage 

\section{Discussion}
\label{discussion}

\subsection{Summary of main results}

Our main conclusion is that FRBs are efficiently damped in the static dipole magnetosphere surrounding the magnetar at radii $10^8\,{\rm cm}<r <\RLC\sim 10^{10}\,$cm. 

We first discussed O-mode GHz waves. They can propagate across the magnetic field lines (and so have a chance to escape the closed magnetosphere) when the plasma density is low, so that it does not screen the wave electric field component $E_\parallel$ parallel to $\bBbg$. However, in this regime $E_\parallel$ accelerates particles to high energies and the wave experiences immediate radiative losses with an avalanche of $e^\pm$ creation. An O-mode FRB could escape only if it is beamed within a cone around the magnetic axis where losses are small. The escape cone is constrained by \Eqs~(\ref{eq:Omode_escape1}) and (\ref{eq:Omode_escape2}).

X-mode GHz waves with power $L$ become damped near the surface defined by
\beq
   \rsh(\theta)\approx 2.5\times 10^8 \mu_{33}^{1/2}\,L_{42}^{-1/4}
   \sqrt{\frac{4-3\sin^2\theta}{\sin\theta}} {\rm ~cm},
\eeq
where $\mu$ is the magnetic dipole moment of the magnetar. Damping develops because the wave approaches the condition $E^2\approx B^2$ which leads to particle energization. We have investigated this process in detail and found that it develops in two different regimes near the magnetic axis and near the equator. Our estimate for the boundary $\theta_{\rm MHD}(L)$ between the two regimes is shown in Figure~\ref{fig:L_theta}. 

(1) In the equatorial region $\sin\theta>\sin\theta_{\rm MHD}$ the entire wave evolution is well described by MHD. The MHD solution demonstrates that at $r\approx\rsh$ the wave train develops shocks in each oscillation. The resulting shock train heats the plasma to an ultra-relativistic temperature (specific internal energy $\e\sim 10^2-10^3$) at which heating becomes offset by synchrotron cooling. We have followed the wave evolution with a detailed simulation in the equatorial plane ($\theta=\pi/2$) and also described it analytically. The results show nearly complete damping of the GHz oscillations (Figure~\ref{fig:L}). The alternating component of the electromagnetic field gets suppressed by a factor of $\sim 10^{-3}$, and the wave train becomes transformed into a smooth and weak electromagnetic pulse of the same duration, with  the oscillating component wiped-out. We have also examined the MHD evolution at $\theta\neq \pi/2$ (where $\bBbg$ is oblique to the wave propagation direction) and verified that it leads to similar shocks.

Our method for solving this MHD problem employed characteristics $C^\pm$. It allows one to find the solution with realistic parameters of the magnetosphere (where magnetization $\sigbg$ can exceed $10^8$, see \Eq~(\ref{eq:sigbg})). We also exploited the fact that the wavelength $\lambda=c/\nu$ is far shorter than radius $r$ (the variation scale of $\Bbg$), and the wave duration $\tau\simlt 1\,$ms satisfies $\tau\ll r/c$ at radii of interest. This feature facilitates the solution, as it gives a simple integral along $C^-$ across the wave.

(2) In the polar regions $\sin\theta<\sin\theta_{\rm MHD}$, the wave damping also begins with MHD shocks developing at $\rsh$. However, here the shock heating quickly ends because the heated particles become unmagnetized in the wave, i.e. their Larmor frequency $\omL$ drops below the wave frequency $\omega$. This transition happens before significant damping, with the practically intact wave profile $E(t-r/c)$. As the wave continues to propagate to $r>\rsh$, it develops regions of $E^2>B^2$ in each oscillation, triggering stochastic particle acceleration described in B22: magnetospheric particles exposed to the GHz wave train develop a quick random walk in energy.\footnote{At the same time, the wave controls the plasma bulk flow: the sliding  along the oblique $\bBbg$ with the Lorentz factor $\gamma_F\approx(\sin\alpha)^{-1}$, as described in \Sect~\ref{kinetic}. The bulk flow may also be viewed as a result of the wave ponderomotive force along $\bBbg$. This force changes sign if the plasma bulk Lorentz factor exceeds $\gamma_F$, so the bulk flow is forced to have $\gamma_F$. Same result is found with the more detailed approach of B22, by solving the particle equation of motion in the wave and finding the particle trajectory.}
 As a result, the particles are forced into a quasi-steady energy distribution (with ``thermal'' Lorentz factors $10^4$-$10^5$), in which stochastic acceleration is balanced by radiative losses, quickly draining the wave energy. This damping effect of the wave-particle interaction may also be formulated as a large particle cross section for scattering the GHz wave to the gamma-ray band, $\sigma_{\rm sc}\simgt 10^8\sT$ (B22). The gamma-rays emitted by the accelerated particles produce copious secondary $e^\pm$ pairs, which fill the entire radio wave and assist its damping.

In both kinetic and MHD regimes, the magnetosphere at $r\simgt \Rm$ effectively acts as a pillow absorbing the wave, with most of the wave energy converted to hard radiation and a residual fraction feeding a low-energy magnetic explosion, ejecting the outer layers of the magnetosphere.

\subsection{Comparison with kHz waves}

The strong GHz radio waves (X-modes) evolve at $r\simgt \rsh$ differently from kHz waves studied in Paper~I.

(a) In the MHD propagation regime, radio waves accelerate the plasma to a bulk Lorentz factor $\gamma\propto \nu^{-1}$, which differs by $\sim 10^6$ between kHz and GHz waves. The moderate $\gamma$ in the GHz waves leads to moderately strong relativistic shocks, different from the monster shocks described in Paper~I. 

(b) The number of oscillations in the GHz wave train is $N=\tau/\nu=10^6\,(\tau/1{\rm \, ms})\,\nu_9^{-1}$. Shocks develop in each oscillation and their large number produces a huge cumulative damping effect on the wave, nearly completely erasing the GHz oscillations. By contrast, in kHz waves the monster shocks erase half of each oscillation. 

(c) Radiative cooling of the shock-heated plasma in GHz waves typically occurs on a timescale longer than the wave period. In particular, at $r\approx\Rm$ cooling is slow for waves with power $L$ below $\Lh$ given in \Eq~(\ref{eq:Lh}). In this case, a thermal balance is established in the wave packet only when the cumulative heating by multiple shocks enhances the plasma temperature, reducing its cooling timescale. By contrast, in kHz waves, each monster shock radiates the dissipated energy almost instantaneously.

(d) The critical wave power $\LMHD(\theta)$ for the transition to the kinetic regime scales as $\nu^{8/9}$ (\Eq~\ref{eq:LMHD_theta}). The transition is relevant for GHz waves and irrelevant for kHz waves. MHD fails in powerful kHz waves differently: when the plasma is accelerated to extremely high Lorentz factors its motion transitions to the two-fluid regime as explained in Paper~I.

\subsection{Mechanism of observed FRBs}

An observed FRB power $L$ requires a source with energy density $U\sim L/4\pi r^2 c \eta$ where $\eta\ll 1$ is the efficiency of GHz emission. The energy density around a magnetar is $U(r)\sim \mu^2/8\pi r^6$, and then the condition $U\gg L/4\pi r^2 c$ requires a source of size $r\ll (c\mu^2/2L)^{1/4}$. It is tempting to picture a compact GHz source confined inside the ultra-strong magnetosphere (e.g. \citealt{Lu20}). However, our results imply trouble for this scenario: the condition $r\ll (c\mu^2/2L)^{1/4}$ is practically the same as $r\ll\Rm$, and we find that the emitted waves experience strong damping when they try to escape through the outer magnetosphere at $r\simgt \Rm$. Damping occurs in both propagation regimes (MHD and kinetic).

Therefore, emission of observed FRBs must involve violent events that first relocate energy from radii $r\ll \Rm$ to outside the magnetosphere, where GHz waves can be released. This is accomplished by magnetospheric explosions, which  produce ultra-relativistic ejecta. The explosion transports a large magnetic energy $\E$ far outside the magnetosphere, e.g. $\E \sim 10^{44}$\,erg is expected in repeating FRBs from hyper-active, flaring magnetars \citep{Beloborodov17b}. It has been shown that the blast wave from the explosion can emit a GHz burst with energy $\EFRB\sim (10^{-4}-10^{-5})\,\E$ and sub-millisecond duration as radii $r\sim 1$\,AU \citep{Beloborodov17b,Beloborodov20}. The emission is generated by the well studied mechanism of ``shock maser precursor''  (e.g. \citealt{Sironi21}). A variation of the blast wave model involving a slow ion wind ahead of the explosion is discussed in \cite{Metzger19} and \cite{Beloborodov20}. In addition, \cite{Thompson23} recently proposed that the blast wave may emit radio waves via another mechanism if it expands into a turbulent medium (a  pre-explosion magnetar wind carrying a spectrum of perturbations). 

Ejecta from powerful magnetospheric explosions may themselves carry magnetosonic fluctuations with radio frequencies.\footnote{The ejecta serve as a new background $\Bbg\propto r^{-1}$ for small oscillations with amplitude $\Em\ll\Bbg$. Small oscillations can remain frozen in the  ejecta for a long time, as they expand from the magnetosphere with $E_0\propto\Bbg\propto r^{-1}$, keeping $E_0/\Bbg\ll 1$.}
At large radii, the fluctuations may decouple and leave the ejecta as free waves, forming a GHz burst. A model of this type was proposed by \cite{Lyubarsky20} and further investigated by \cite{Mahlmann22}. These works invoked the explosion interaction with the current sheet near the light cylinder as a source of ejecta fluctuations. 

Another possibility for FRB production is the precursor emission from the magnetospheric monster shocks described in Paper~I. The precursor will ride on top of the parent kHz wave that forms the monster shock, not in a static dipole magnetosphere, and therefore it could escape from small radii. This possibility is further discussed elsewhere.

Observational diagnostics for FRB models include the burst spectra, temporal structure, and polarization (see for example a recent discussion of polarization in \cite{Qu23}). The observed properties can be changed by the burst propagation through the magnetar wind \citep{Sobacchi22} and the surrounding nebula  \citep{Margalit18,Vedantham19,Gruzinov19}. The propagation effects will need to be disentangled from the intrinsic emission properties before conclusions can be made regarding the source.

When this paper was completed, the author became aware of the work by \cite{Golbraikh23}, who find the inability of FRBs to escape the magnetosphere using different considerations, by analyzing nonlinear wave-wave interactions.

\medskip
This work is supported by NSF AST-2009453, NASA 21-ATP21-0056, Simons Foundation \#446228, and NSF PHY-2206609. The author thanks the anonymous referee for careful reading of the paper and useful comments.

\newpage

\appendix


\section{Characteristics in relativistic MHD}
\label{char_MHD}

\subsection{MHD stress-energy tensor}

MHD fluid is described by the plasma mass density $\rho$, velocity $\bv=c\bb$, magnetic field $\bB$, and electric field $\bE$. Throughout this Appendix we will use the units of $c=1$. The stress-energy tensor $\Ttot^{\mu\nu}$ of the MHD fluid  includes contributions from the electromagnetic field ($\TEM^{\mu\nu}$) and plasma ($\Tp^{\mu\nu}$). Explicit expressions for $\TEM^{\mu\nu}$ in terms of $\bE$, $\bB$ and for $\Tp^{\mu\nu}$ in terms of $\rho$, $\bv$ are given below. Energy and momentum conservation in MHD is expressed by 
\beq 
\label{eq:MHD}
   \Ttot^{\mu\nu}_{\;\; ;\nu}
   =  \frac{1}{\sqrt{-g}}\, \partial_\nu \left(\sqrt{-g}\Ttot_\mu^\nu \right)-\frac{1}{2}\Ttot^{\alpha\beta}\partial_\mu g_{\alpha\beta}
   =-Q^\mu, \qquad  \Ttot^{\mu\nu}=\TEM^{\mu\nu}+\Tp^{\mu\nu},
\eeq
where semicolon denotes covariant derivative, $g_{\alpha\beta}$ is the spacetime metric, and $g\equiv {\rm det}g_{\alpha\beta}$. $Q^\mu$ represents radiative losses of the plasma. The losses typically have the form $Q^\mu=-Q u^\mu$, where $u^\mu =(\gamma,\gamma\bb)$ is the plasma four-velocity.

The electromagnetic stress-energy tensor is $\TEM^{\alpha\beta}=F^{\alpha\mu} F^\beta_\mu/4\pi - g^{\alpha\beta} F_{\mu\nu}F^{\mu\nu}/16\pi$ (e.g. \citealt{Landau75}), $F_{\mu\nu}=\partial_\mu A_\nu-\partial_\nu A_\mu$, $F_{\mu\nu}F^{\mu\nu}=2(B^2-E^2)$, and $A_\mu$ is the four-potential of the electromagnetic field. We express all field components in the normalized basis $(\be_r,\be_\theta,\be_\phi)$ in Minkowski space with coordinates $x^\mu=(t,r,\theta,\phi)$. This gives
\beq
\TEM^{\mu\nu} = \left(\begin{array}{cccc}
  \vspace*{1mm}
 \frac{E^2+B^2}{8\pi}  & -\frac{E_\phi B_\theta}{4\pi} & \frac{E_\phi B_r}{4\pi r} & 0 \\
  \vspace*{1mm}
   -\frac{E_\phi B_\theta}{4\pi} & \frac{E^2-B_r^2+B_\theta^2}{8\pi} & -\frac{B_\theta B_r}{4\pi r} & 0 \\
     \vspace*{1mm}
   \frac{E_\phi B_r}{4\pi r} & -\frac{B_rB_\theta}{4\pi r} & \frac{E^2+B_r^2-B_\theta^2}{8\pi r^2} & 0 \\
   0 & 0 & 0 & \frac{B^2-E^2}{8\pi r^2\sin^2\theta} 
                           \end{array}\right)
\label{eq:TEM}
\eeq

The stress-energy tensor of the plasma treated as an ideal (isotropic) fluid has the form,
\beq
\label{eq:Tp}
   \Tp^{\mu\nu}=\Hp u^\mu u^\nu + g^{\mu\nu}\Pp,
\eeq
where $\Pp$ is the plasma pressure, and $\Hp$ is its relativistic enthalpy density (including the fluid rest mass density $\trho$). Heating by Larmor-mediated shocks can result in a two-dimensional (2D) plasma, with $e^\pm$ thermal speeds $\bb_e\perp\tilde{\bB}$. Then, one can calculate $\Tp^{\mu\nu}$ as follows. First, find the (diagonal) stress-energy tensor in the fluid rest frame $\tilde{\KF}$ by viewing the plasma as a collection of cold $e^\pm$ streams with different $u_e^{\tilde{\mu}}$ and proper densities $d\trho/\gamma_e$, 
\beq
   \Tp^{\tilde{\mu}\tilde{\nu}}=\trho\, \langle \frac{u_e^{\tilde\mu} u_e^{\tilde\nu}}{\gamma_e}\rangle,
\eeq
where $\langle...\rangle$ means averaging over the distribution of $u_e^{\tilde\mu}$. The stress-energy tensor in the lab frame is $\Tp^{\mu\nu}=\Lambda^\mu_{\tilde\mu}\Lambda^{\nu}_{\tilde\nu}  \Tp^{\tilde{\mu}\tilde{\nu}}$ where $\Lambda^\mu_{\tilde\mu}$ is the Lorentz matrix for the boost from the fluid frame to the lab frame. This gives the general $\Tp^{\mu\nu}$ for plasmas with any anisotropy; in the isotropic case it is reduced to \Eq~(\ref{eq:Tp}). 

To avoid unnecessary distraction, our derivation of MHD characteristics will assume isotropic plasma. However, looking at the derivation, one will see that only the $t,r$ components of the plasma stress-energy tensor $\Tp^{\mu\nu}$ affect the final result, so only radial pressure $\Pp=\Tp^{\tilde r\tilde r}$ enters the wave propagation problem. The calculation of $\Tp^{\mu\nu}$ for anisotropic plasma in the equatorial plane gives the $t,r$ components of $\Tp^{\mu\nu}$ of the same form as in \Eq~(\ref{eq:Tp}), with $\Pp=\Tp^{\tilde r\tilde r}$ instead of isotropic pressure. Therefore, the final equations for characteristics hold for anisotropic plasma. The only important effect of anisotropy is that it changes the plasma equation of state --- the relation between energy density and radial pressure. This relation enters through $\gs$, which is given in \Sect~\ref{gs}.


\subsection{Equatorial waves}

We now focus on the wave dynamics in the equatorial plane $\theta=\pi/2$. By symmetry, $B_r=0$ and $v_\theta=0$ at $\theta=\pi/2$. We will use the following notation:
\beq
   E\equiv -E_\phi, \qquad B\equiv B_\theta.
\eeq
These definitions imply $E^2=\bE^2$ and $B^2=\bB^2$ at $\theta=\pi/2$; $E$ and $B$ may be positive or negative. The plasma four-velocity has the form, 
\beq
   u^\alpha=(\gamma,u,0,0), \qquad 
   u=\gamma\beta, \qquad
   \beta\equiv \beta_r = \frac{E}{B}, \qquad \gamma^2=\frac{1}{1-\beta^2}=\frac{B^2}{B^2-E^2}    \qquad \left(\theta=\frac{\pi}{2}\right).
\eeq

In the equatorial plane, there are two relevant components of the dynamical equation $\Ttot^{\mu\nu}_{\;\;;\nu}=-Q u^\mu$ with $\mu=t,r$. Instead of these two components we will use two projections:
\beq
\label{eq:proj}
   u_\mu\Ttot^{\mu\nu}_{\;\;;\nu}=Q,  \qquad 
  (u_t u_\mu+g_{t\mu})\Ttot^{\mu\nu}_{\;\;;\nu}=0.
\eeq
The vanishing of $B_r$ and $v_\theta$ in the equatorial plane implies $\Ttot^\theta_t=\Ttot^\theta_r=0$. However, their $\theta$-derivatives are not zero and will enter the conservation laws. For instance, divergence of the Poynting flux $\nabla\cdot(\bE\times\bB)/4\pi$ includes a term with $\partial_\theta B_r\neq 0$, as $B_r$ changes sign across the equatorial plane.

For the plasma stress-energy tensor (\Eq~\ref{eq:Tp}), one finds
\beq
\label{eq:proj_p}
  u_\mu T^{\mu\nu}_{{\rm p}\,;\nu}=-\Hp u^\nu_{;\nu}-u^\mu\partial_\mu(\Hp-\Pp),
\qquad
   (u_tu_\mu+g_{t\mu})T^{\mu\nu}_{{\rm p}\,;\nu}=-\Hp u^\nu\partial_\nu \gamma-\gamma u^\nu\partial_\nu \Pp+\partial_t\Pp.
\eeq
The term $u^\mu\partial_\mu(\Hp-\Pp)$ may be written as the sum of adiabatic part $\Hp u^\mu \partial_\mu \ln\trho$ and radiative part $-Q$. Note that $u_\mu T^{\mu\nu}_{{\rm p}\,;\nu}=Q$ regardless of the presence of $\TEM^{\mu\nu}$ in the system. This condition states the first law of thermodynamics (and for a cold flow it is reduced to conservation of the plasma rest mass). The divergence of four-velocity is
\beq
\label{eq:divu}
      u^\nu_{;\nu}=\frac{1}{\sqrt{-g}}\,\partial_\nu\left(\sqrt{-g}u^\nu\right)
      = \partial_t\gamma+\frac{1}{r^2}\partial_r(r^2u)
      +\frac{1}{r}\,\partial_\theta u_\theta 
      \qquad \left(\theta=\frac{\pi}{2}\right),
\eeq
where the component $u_\theta$ is taken in the normalized basis $(\be_r,\be_\theta,\be_\phi)$.

Next, consider the electromagnetic stress-energy tensor $\TEM^{\mu\nu}$ (\Eq~\ref{eq:TEM}). Direct calculation yields at $\theta=\pi/2$:
\beq
\label{eq:FFE_}
    u_\mu T^{\mu\nu}_{\rm f\; ;\nu}
    =-\frac{\sqrt{B^2-E^2}}{4\pi}\left(\partial_tB+\partial_r E+\frac{E}{r}\right), 
    \qquad  (u_tu_\mu+g_{t\mu}) T^{\mu\nu}_{\rm f \; ;\nu}
   = -\frac{E}{4\pi}\left(\partial_tE+\partial_rB+\frac{B}{r}+\frac{2A}{r^2}\right).
\eeq
The identity $\partial_t\nabla\times\boldsymbol{A}=\nabla\times\partial_t \boldsymbol{A}$ implies $\partial_t B+\partial_r E+E/r=0$ and $u_\mu T^{\mu\nu}_{\rm f\;\,;\nu}=0$. For our purposes it will be convenient to rewrite $\TEM^{\mu\nu}$ using the effective pressure $\PEM$ and the effective inertial mass density $\HEM$ defined by
\beq
   \PEM=\frac{B^2-E^2}{8\pi}=\frac{\tB^2}{8\pi}, \qquad \HEM = \frac{\tB^2}{4\pi}=2\PEM.
\eeq
This allows one to cast the $t$,$r$ components of $\TEM^{\mu\nu}$ in the form similar to ideal fluid:
\beq
\label{eq:TEM_fluid}
  \TEM^{tt}=\gamma^2\HEM - \PEM, \quad \TEM^{tr}=\gamma u\HEM, 
   \quad \TEM^{rr}=u^2\HEM+\PEM.
\eeq
Other relevant components of $\TEM^{\mu\nu}$ are
\beq
   \TEM^{\theta t}=-\frac{EB_r}{4\pi r}, \quad  \TEM^{\theta r}=-\frac{BB_r}{4\pi r}, 
   \qquad     T^{\,\phi}_{\rm f\, \phi}=\frac{\HEM}{2}=-T^{\,\theta}_{\rm f\, \theta}. 
\eeq
Then, we find in the equatorial plane
\begin{eqnarray}
\label{eq:proj_EM1}
 u_\mu T^{\mu\nu}_{ \rm f\, ;\nu}
   &=& -\HEM (\partial_t\gamma+\partial_r u) 
   - u^\mu\partial_\mu(\HEM-\PEM) -\frac{\HEM u}{r}, \\
\label{eq:proj_EM2}
 (u_tu_\mu+g_{t\mu}) T^{\mu\nu}_{\rm f\; ;\nu}
   &=& 
   - \HEM u^\mu\partial_\mu \gamma -\gamma u^\nu\partial_\nu \PEM+\partial_t \PEM  
   - \frac{ \HEM u\gamma}{r} + \frac{E\,\partial_\theta B_r}{4\pi r}.
\end{eqnarray}
Note that $\HEM u^\nu_{;\nu} + u^\mu\partial_\mu(\HEM-\PEM)=\sqrt{\HEM}\;(\sqrt{\HEM} u^\nu)_{;\nu}=(\tB/4\pi)(\tB u^\nu)_{;\nu}$. For short waves, the flow oscillation is nearly plane parallel, and the equations may be simplified: magnetic flux freezing gives $\tB\propto\trho$ and the continuity equation implies $(\tB u^\nu)_{;\nu}=0$; in the same approximation one can use $u^\mu\partial_\mu(\HEM-\PEM)=\HEM u^\mu\partial_{\mu}\ln\trho$. 

Substitution of \Eqs~(\ref{eq:proj_p}),  (\ref{eq:proj_EM1}), and (\ref{eq:proj_EM2}) into \Eqs~(\ref{eq:proj}) gives 
\beq
   \label{eq:proj_1}
  -\Htot(\partial_t\gamma+\partial_r u)  - u^\mu\partial_\mu(\Htot-\Ptot)
  -\frac{\HEM u}{r} - \frac{\Hp}{r}\left(2u+\partial_\theta u_\theta\right) = Q,
\eeq
\beq
  -\Htot u^\mu\partial_\mu \gamma 
   -\gamma u^\nu\partial_\nu \Ptot+\partial_t \Ptot  
   - \frac{\HEM u\gamma }{r} + \frac{E \partial_\theta B_r}{4\pi r}=0,
\label{eq:proj_2}
\eeq
where 
\beq
   \Ptot=\PEM+\Pp, \qquad \Htot=\HEM+\Hp.
\eeq
We will use $d\gamma=\beta du$ to express all derivatives of $u^\nu$ in terms of derivatives of $u=\gamma\beta$. \Eqs~(\ref{eq:proj_1}) and (\ref{eq:proj_2}) also contain derivatives of $\Ptot$ and $\Htot-\Ptot$. One can retain only derivatives of $\Ptot$ by defining 
\beq
\label{eq:bs}
    \bs^2\equiv\frac{u^\mu\partial_\mu\Ptot}{u^\mu\partial_\mu(\Htot-\Ptot)}
    =\frac{d\Ptot}{d(\Htot-\Ptot)},
\eeq
where differential $d$ is taken along the worldline of a fluid element. The quantity $\bs$ (and the characteristics $C^\pm$ below) will be defined in the adiabatic approximation, $Q\approx 0$. Then, \Eqs~(\ref{eq:proj_1}) and (\ref{eq:proj_2}) become
\begin{eqnarray}
    \Htot(\beta\partial_t u+\partial_r u) + \frac{\gamma}{\bs^2}(\partial_t\Ptot+\beta\partial_r\Ptot) 
 =  -\frac{\HEM u}{r} - \frac{\Hp}{r}\left(2u+\partial_\theta u_\theta\right),
\label{eq:pr1}
\end{eqnarray}
\beq
  \Htot (\partial_t u+\beta\partial_r u) 
     +\gamma(\beta\partial_t \Ptot+\partial_r \Ptot) 
   = -\frac{ \HEM \gamma}{r} + \frac{B\,\partial_\theta B_r}{4\pi r\gamma}.
\label{eq:pr2}
\eeq
We multiply \Eq~(\ref{eq:pr1}) by $\bs$ and add/subtract it from \Eq(\ref{eq:pr2}). This yields
\begin{eqnarray}
\label{eq:main}
   (1\pm\bs\beta)\left( \partial_\pm \ln\sqrt{\frac{1+\beta}{1-\beta}} \pm \frac{\partial_\pm \Ptot}{\bs \Htot}  \right)
   = \frac{\HEM}{r\Htot}\left(\frac{\partial_\theta B_r}{B}-1\mp\beta\bs\right)
   \mp \frac{\bs\Hp}{\gamma r \Htot} \left(2u+\partial_\theta u_\theta\right),
 \end{eqnarray}
where we used the identity
\beq
  \frac{du}{\gamma}=d\ln\sqrt{\frac{1+\beta}{1-\beta}},
\eeq
and defined
\beq
\label{eq:bpm}
   \partial_\pm\equiv \partial_t+\beta_\pm\partial_r, 
   \qquad \beta_\pm \equiv \frac{\beta\pm\bs}{1\pm\beta\bs}.
\eeq
The radial speed $\beta_\pm$ in the lab frame corresponds to propagation with speed $\pm\bs$ relative to the fluid. The derivatives $\partial_\pm$ are taken along the characteristics $C^\pm$. The characteristics are defined as the curves $r_\pm(t)$ that satisfy $dr_\pm/dt=\beta_\pm$.

\Eq~(\ref{eq:main}) is the MHD generalization of equations given by \cite{Johnson71} and \cite{McKee73}, which were derived for one-dimensional relativistic hydrodynamics. It is easy to verify that their hydrodynamical equations are recovered in the limit of a weak electromagnetic field $E,B\rightarrow 0$. In this limit, $\HEM/\Hp=0$, $\Hp/\Htot=1$, and $\partial_\theta u_\theta=0$ if the flow is spherically symmetric. Then, \Eq~(\ref{eq:main}) becomes equation II.b.20 in \cite{McKee73}. We are interested in the opposite, field-dominated, regime $\Htot\approx \HEM\gg\Hp$.

\subsection{Magnetically dominated limit ($\HEM\gg\Hp$)}
\label{domin}

In the magnetically dominated regime, one can simplify the MHD equations. \Eq~(\ref{eq:main}) becomes
\beq
\label{eq:main_}
  r(1\pm\beta\bs)\left( \partial_\pm \ln\sqrt{\frac{1+\beta}{1-\beta}} \pm \frac{\partial_\pm \Ptot}{\bs \Htot}\right)
  =  -1\mp\beta  -\frac{2\Bbg}{B} + {\cal O}\left(\frac{\Hp}{\HEM}\right),
\eeq
where on the r.h.s. we used $1-\bs={\cal O}(\Hp/\HEM)$. We also find 
\beq
    \frac{\partial_\pm \Ptot}{\bs \Htot} 
    = \frac{\partial_\pm \PEM}{\HEM} \left[ 1+{\cal O}\left( \frac{\Hp}{\HEM} \right) \right]
    = \partial_\pm \ln \tB \left[1+{\cal O}\left(\frac{\Hp}{\HEM}\right)\right],
\eeq
where we used $\HEM=2\PEM=\tB^2/4\pi$. Thus, \Eq~(\ref{eq:main}) simplifies to 
\beq
 \label{eq:strong}
   \partial_\pm J_\pm =
    \frac{-1\mp\beta + \partial_\theta B_r/B + {\cal O}(\Hp/\HEM)}{r(1\pm\beta\bs)},
 \qquad J_\pm=\ln\sqrt{\frac{1+\beta}{1-\beta}}\pm \ln \tB  \left[1+{\cal O}\left(\frac{\Hp}{\HEM}\right)\right].
\eeq
In the denominator we did not use the expansion $1\pm\beta\bs=1\pm\beta+{\cal O}(\Hp/\HEM)$, because $1+\beta$ can approach zero during the wave evolution. The numerator is never close to zero, since $\partial_\theta B_r$ has a finite negative value.

\subsection{Short waves}
\label{short_waves}

We are interested in short wave packets with wavelength $\lambda$ many orders or magnitudes shorter than $r$. The wave electromagnetic potential $\Aw=A-\Abg$ is related to the wave magnetic field $\bBw=\bB-\bBbg$ by $r\Bw^\theta = -\partial_r (r\Aw)$ and $r\Bw^r= \partial_\theta \Aw$. In short waves, $\partial_r \Aw\gg r^{-1}\partial_\theta \Aw$, and so $\Bw^r\ll \Bw^\theta$. This implies $B_r\approx \Bbg^r$ and
\beq
   \partial_\theta B_r \approx -2\Bbg \qquad \left(\theta=\frac{\pi}{2}\right).
\eeq 
Then, only derivatives $\partial_\pm$ are left in \Eq~(\ref{eq:strong}), i.e. the problem is reduced to ordinary differential equations. This enables simple integration for $J_\pm$ along $C^\pm$. 

The $C^-$ characteristics propagate radially inward, and cross the short wave packet on a timescale $t_-^{\rm cross}\ll r$. Therefore, the change of $J_-$ across the wave is small, $|\Delta J_-/J_-|\ll 1$, i.e. $J_-$ is approximately uniform across the wave and weakly changed  from its value in the unperturbed background just ahead of the wave, $J_-^{\rm bg}\approx -\ln\Bbg$. This gives the following relations:
\beq
\label{eq:J-}
  J_-=-\ln\Bbg \qquad \Rightarrow \qquad
   J_+ = \ln\c^2 +\ln\Bbg,    \qquad 
 \c\equiv \frac{\tB}{\Bbg} = \sqrt{\frac{1+\beta}{1-\beta}}, 
  \qquad B=\frac{\Bbg}{1-\beta}.
\eeq
The relation $B=\Bbg/(1-\beta)$ states the compression of the magnetic field in the lab frame by the factor $(1-\beta)^{-1}$. Plasma density is compressed by the same factor (\Eq~\ref{eq:short1}), consistent with magnetic flux freezing: $B/\rho=\Bbg/\rhobg$. Magnetic flux freezing also implies $\sigma/\sigbg=\c$, where $\sigma\equiv \HEM/\trho$.

The equation for $J_+$ evolution along $C^+$ (\Eq~\ref{eq:strong}) and the definition of $C^+$ ($\partial_+r=\beta_+$) give two coupled equations for $\beta(t)$ and $r(t)$ along $C^+$:
\beq
\label{eq:J+}
   \partial_+\ln  \sqrt{\frac{1+\beta}{1-\beta}}  = \frac{2\beta+{\cal O}(\Hp/\HEM)}{r(1+\beta\bs)},  \qquad
  \partial_+ r = \beta_+=1-\frac{(1-\beta)}{1+\beta\bs} (1-\bs),
\eeq
where we used $\partial_+\ln\Bbg=-3\beta_+/r$. These equations still contain the fast magnetosonic speed $\bs$, which is close to unity. Setting $\bs=1$ would correspond to FFE. It is the small term $1-\bs\approx (2\gs^2)^{-1}={\cal O}(\Hp/\HEM)$ that controls the MHD correction to FFE, and it is retained in the leading order in \Eqs~(\ref{eq:J+}). In particular, it controls the deviation of $\beta_+$ from unity, bending the $C^+$ characteristics from straight lines in spacetime. This is the main effect responsible for the deformation of the wave profile. When $1-\beta_+\ll 1$ (satisfied in GHz waves), one can simplify $1+\beta\bs\approx 1+\beta$ in the denominators in \Eqs~(\ref{eq:J+}). Retaining $\bs$ in the denominators is required in kHz waves (see Paper~I) because in that case $\beta_+$ significantly decreases below unity and even changes sign.

Substituting $\beta=(\c^2-1)/(\c^2+1)$, one can state \Eqs~(\ref{eq:J+}) in terms of $\c$. Using $\gs^2-1\approx\gs^2$, we obtain
\beq
 \label{eq:main1}
   \partial_+ \ln\c = \frac{1-\c^{-2}}{ r\,[1 + (2\gs\c)^{-2}]},  
   \qquad
   \partial_+ r = 1 - \frac{2}{(2\gs\c)^2+1}.
\eeq
For a cold plasma $\gs^2=\sigma=\c\sigbg$ (see below), and then \Eqs~(\ref{eq:main1}) reproduce \Eqs~(43), (45) in Paper~I.


\subsection{Fast magnetosonic speed}
\label{gs}

From the definition of $\bs$ (\Eq~\ref{eq:bs}), one finds
\beq
\label{eq:gs0}
    \frac{1}{\gs^2}=1-\bs^2=\frac{d(\Htot-2\Ptot)}{d(\Htot-\Ptot)}
   = \frac{d(\Hp-2\Pp)}{d(\PEM+\Hp-\Pp)},
\eeq
where we used $\HEM=2\PEM$. It is convenient to express $\gs$ in terms of $\sigma=\HEM/\trho$, where $\trho=\rho/\gamma$ is the proper rest-mass density of the plasma. For a cold plasma $\Hp=\trho$ and $\Pp=0$, and \Eq~(\ref{eq:gs0}) gives 
\beq
\label{eq:gs_cold}
   \gs=\sqrt{1+\sigma}, \qquad \bs=\sqrt{\frac{\sigma}{1+\sigma}}.
\eeq
For a hot plasma, 
\beq
   \Hp=\trho +\Up+\Pp,
\eeq
where $\Up$ is the thermal energy density (measured in the fluid frame). Note that $\Pp$ here is the plasma pressure in the radial direction; this fact becomes important if the plasma is anisotropic. 

A useful analytical expression for $\gs$ can be derived in the limit of $\HEM/\Hp\gg 1$, which corresponds to $\gs\gg 1$. Then, \Eq~(\ref{eq:gs0}) simplifies to
\beq
\label{eq:gs1}
  \frac{1}{\gs^2}=\frac{d(\Hp-2\Pp)}{d\PEM}\left[1+{\cal O}\left(\frac{\Hp}{H}\right)\right].
\eeq
It can be rewritten using the magnetic flux freezing condition for short waves, $\tB/\trho=\Bbg/\rhobg$, which implies
\beq
\label{eq:gs2}
   \frac{d\ln\PEM}{d\ln\trho}=2\frac{d\ln\tB}{d\ln\trho}=2
   \qquad \Rightarrow \qquad 
     \frac{1}{\gs^2}=\frac{\trho}{\HEM} \frac{d}{d\trho}(\trho+\Up-\Pp)
   \qquad (\gs\gg 1).
\eeq
It can be further simplified using the equation of state that relates $\Up$ and the radial pressure $\Pp$. For a hot plasma with mono-energetic particles this relation is
\beq
\label{eq:eos}
   \Pp = \frac{\Up}{\kk} \left(1+\frac{1}{\e}\right), \qquad \e\equiv \frac{\trho+\Up}{\trho},
\eeq
where $\kk=2$ if the thermal velocity distribution is two-dimensional (confined to the plane perpendicular to $\bB$), and $\kk=3$ if the plasma is isotropic. For a Maxwellian plasma, the same expression holds at all $\e$ with better than $5$\% accuracy. \Eq~(\ref{eq:gs2}) now becomes
\beq
\label{eq:gs3}
  \frac{1}{\gs^2} = \frac{\trho}{\kk\HEM} 
     \left[(\kk-1)\e+\frac{1}{\e}+\left(\kk-1-\frac{1}{\e^2}\right)\frac{d\e}{d\ln\trho}\right].
\eeq
When radiative losses are small during each wave oscillation, one can use the adiabatic law $d(\Up/\trho)=-\Pp d(1/\trho)$ along the plasma streamline. Plasma compression in a short wave is one-dimensional (in the radial direction), so only the radial pressure $\Pp$ enters the adiabatic law. It gives
\beq
\label{eq:gs_}
   \frac{d\e}{d\ln\trho}=\Pp=\frac{1}{\kk}\left(\e-\frac{1}{\e}\right)
   \qquad \Rightarrow \qquad   \frac{1}{\gs^2} = \frac{\trho}{\kk^2\HEM} \left[(\kk^2-1)\e+\frac{1}{\e^3}\right] 
  \qquad (\gs\gg 1).
\eeq


\section{Jump conditions for perpendicular shocks}
\label{jump}

Jump conditions are formulated in the shock rest frame. In this section, it will be denoted by $\KF'$, and quantities measured in frame $\KF'$ are denoted with a prime. Indices ``u'' and ``d'' will refer to the plasma in the immediate upstream and immediate downstream of the shock. The jump conditions state the continuity of particle flux $F'=\tn u'$, energy flux $\Ttot^{t'r'}$, and momentum flux $\Ttot^{r'r'}$. Magnetic flux freezing $\tB\propto \tn$ implies that proper density $\trho=\me\tn$ and magnetization parameter $\sigma\equiv\tB^2/4\pi\trho\propto \trho$ jump at the shock by the same  factor (denoted as $q$). The continuity of $F'$ yields
\beq
\label{eq:F'}
     \me F'=\uds\trhod=\uus\trhou, \qquad \frac{\sigd}{\sigu}= q \equiv \frac{\trhod}{\trhou}.
\eeq
The ultra-relativistic motion of the shock relative to the plasma implies $|\uus|>|\uds|\gg 1$. Since we know the evolution of $\trho=\c\rhobg$ along each $C^+$ in the simulation, the shock compression factor $q(t)$ will be known if we keep track which characteristics $\Cu^+$ and $\Cd^+$ terminate at the shock at time $t$. Note also that $q=\cd/\cu$. The colliding characteristics $\Cu^+$ and $\Cd^+$ also define the Lorentz factor of the upstream relative to the downstream, $\Grel=\gu\gd(1-\betau\betad)$. It can be expressed in the shock frame as $\Grel=(\uus/\uds+\uds/\uus)/2$ (using $|\uus|, |\uds|\gg 1$), which gives $\Grel=(q+q^{-1})/2$.

The $t'r'$ and $r'r'$ components of the total stress-energy tensor (plasma + electromagnetic field) have the ideal-fluid form, same as in the lab frame (\Eqs~(\ref{eq:Tp}) and (\ref{eq:TEM_fluid})) but with the fluid four-velocity measured in frame $\KF'$, $u^{\alpha'}=(\gamma',u',0,0)$. The continuity of $F'$, $\Ttot^{t'r'}$, and $\Ttot^{r'r'}$ gives
\begin{eqnarray}
  \frac{\Ttot^{t'r'}}{\me F'} = \gds\hd=\gus\hu, \qquad
   \frac{\Ttot^{r'r'}}{\me F'} = \uds\hd+\frac{\pd}{\uds}+\frac{\sigd}{2\uds}
                                  = \uus\hu+\frac{\pu}{\uus}+\frac{\sigu}{2\uus},
\end{eqnarray}
where $p\equiv\Pp/\trho$ and $h\equiv \e+p+\sigma$. We wish to find the downstream specific energy $\ed$ in terms of the upstream parameters and the shock compression factor $q$. We use the continuity of $\Ttot^{t'r'}$ to express $\hd=\hu\gus/\gds$, and \Eq~(\ref{eq:F'}) to exclude $\sigd=q\sigu$ and $\uds=\uus/q$. Then, the continuity of $\Ttot^{r'r'}$ gives 
\beq
\label{eq:pd}
  \pd=\left(\frac{1}{q}-\frac{\gus}{q^2\gds}\right){\uus}^2\hu-\left(q-\frac{1}{q}\right)\frac{\sigu}{2}
  +\frac{\pu}{q}.
\eeq
Ratio $\gus/\gds$ can be expressed in terms of $q=\uus/\uds$ and $\uus$. The calculation simplifies for magnetically dominated shocks, since we can use $|\uus|, |\uds|\gg 1$ and expand $\gus/\gds=q\,(1+{\uus}^{-2})^{1/2} (1+{\uds}^{-2})^{-1/2}$ in the small parameter ${u'}^{-2}\ll 1$. We expand up to the second-order terms $\sim {u'}^{-4}$; this is needed because larger terms get canceled with the term $(q-q^{-1})\sigu/2$ in \Eq~(\ref{eq:pd}). As a result, we find
\beq
\label{eq:1}
   2q\pd=(q^2-1)\eu+(q^2+1)\pu-(3q^2+1)\psi, \qquad \psi \equiv \frac{\hu(q^2-1)}{4{\uus}^2}.
\eeq
In the relation $\hd=\hu\gus/\gds$ it is sufficient to expand $\gus/\gds$ up to the liner order in ${u'}^{-2}$. This gives, after cancelation of two large terms proportional to $\sigu$,
\beq
\label{eq:2}
   \ed+\pd=q(\eu+\pu-2\psi).
\eeq
Note that $p$ and $\e$ are not independent --- they are related by the equation of state $p=(\e-\e^{-1})/\kk$ (\Eq~\ref{eq:eos}). Therefore, \Eqs~(\ref{eq:1}) and (\ref{eq:2}) form a closed set for two unknowns $\ed$ and $\psi$. We use \Eq~(\ref{eq:2}) to express $\psi$ in terms of $\ed$, substitute it into \Eq~(\ref{eq:1}), and obtain a quadratic equation for $\ed$:
\beq
\label{eq:quadratic}
    \left[(3\kk-1)q^2+\kk+1\right] \ed^2-\left\{\left[(\kk+1)q^2+3\kk-1\right]\eu+\frac{1-q^2}{\eu}\right\} q\, \ed+q^2-1=0.
\eeq
One should choose the larger root of the quadratic equation, as this branch satisfies $\ed=\eu$ at $q=1$.

Next, we find the shock speed in the lab frame $\bsh$ using the relation $\bsh=(\betau-\betaus)/(1-\betau\betaus)$. It gives 
\beq
\label{eq:vsh_1}
   1-\bsh=\frac{(1-\betau)(1+\betaus)}{1-\betau\betaus}\approx \frac{1-\betau}{1+\betau}\,(1+\betaus)
   \approx \frac{1}{2{\uus}^2\cu^2},
\eeq
where we used $1+\betaus \approx 1/2{\uus}^2 \ll 1+\betau$. Using the definition of $\psi$ (\Eq~\ref{eq:1}) and \Eq~(\ref{eq:2}), we find
\beq
\label{eq:vsh_}
   1-\bsh\approx \frac{2\psi}{\cu^2\hu(q^2-1)}
   =  \frac{1}{\cu^2\hu(q^2-1)}\left(\eu+\pu-\frac{\ed+\pd}{q}\right).
\eeq
Here, one can substitute $\hu\approx\sigu=\cu\sigbg\gg \eu+\pu$ and $p=(\e-\e^{-1})/\kk$ to obtain the final expression (\Eq~\ref{eq:vsh}) for $1-\bsh$ in terms of $\cu$, $\eu$, $q$, and the found $\ed$.

Note also that the shock four-velocity relative to the upstream is equal to $-\uus$ and related to $\bsh$ by \Eq~(\ref{eq:vsh_1}). The shock four-velocity relative to the downstream equals $-\uds=-\uus/q$. In the case of a cold upstream $\eu=1$ and $q\gg 1$, the above relations give ${\uds}^2=\sigu(3\kk-1)/[4(\kk-1)]$.


\section{Wave propagation and shock formation outside the equatorial plane}
\label{cold_wave}

\subsection{Equations for $\gamma(t,\xi)$ and $E(t,\xi)$}

At a given polar angle $\theta$, the coupled oscillation of $\gamma(t,\xi)$ and $E(t,\xi)$  in a cold GHz wave (before shock heating) is described by \Eq~(\ref{eq:NWE1}). One can use it to obtain a wave equation containing only derivatives of $\gamma$ and no derivatives of $E$. This can be accomplished using the relation between $E$ and $\gamma$ found in \Sect~\ref{v_profile}. As a first step, express $\partial_t E$ in terms of $\partial_t\gD$ using 
\beq
\label{eq:gD2}
    \gD^2-1=\frac{E^2}{B^2-E^2}\approx \frac{E^2}{\Bbg^2+2\Bbg^\theta E}.
\eeq
We take $\left.\partial_t\right|_\xi$ of both sides, use $\partial_t\bBbg=-3c\bBbg/r$, and  find (in this section, we do not use the units of $c=1$ and so retain $c$ in all equations)
\beq
 \label{eq:dtE}
    E\,\partial_t E = \frac{B^4}{\Bbg^2+\Bbg^\theta E}\,\frac{\partial_t\gD}{\gD^3}
    -\frac{3cE^2}{r}.
\eeq
Here, we substitute 
\beq
\label{eq:dtgD}
  \partial_t\gD = \frac{\partial_t\gamma-\gD\partial_t\tg}{\tg},
\eeq
which follows from $\gamma=\tg\gD$. It remains to evaluate $\partial_t\tg$. Note that $\tg$ is a function of angle $\alpha$ and $E/\Bbg$ (\Eq~\ref{eq:tg}), and $\alpha(\theta)$ is constant in $\partial_t\tg$, so we find
\beq
\label{eq:dtgD1}
  \frac{\partial_t\tg}{\tu} = \frac{\partial_t\tu}{\tg} 
  = -s\, \partial_t\left(\frac{E}{\Bbg}\right)
  \quad \Rightarrow \quad \gD\, \partial_t\tg = - \gD^2\tu \frac{E\Bbg^r}{B^3} \left(\partial_t E+\frac{3cE}{r}\right).
\eeq

Substitution of \Eqs~(\ref{eq:dtgD}) and (\ref{eq:dtgD1}) into \Eq~(\ref{eq:dtE}) gives 
\beq
 \label{eq:dtE1}
    E\,\partial_t E = \frac{\partial_t\gamma }{f} - \frac{3cE^2}{r}, 
    \quad {\rm where} \quad f\equiv \frac{(\Bbg^2+\Bbg^\theta E) \gamma^3 - \tu\Bbg^r B \gamma^2}{B^4\tg^2}.
\eeq
Substituting this result into \Eq~(\ref{eq:NWE1}), we obtain the equation for $\gamma(t,\xi)$ stated in the main text (\Eq~\ref{eq:NWE2}).

Derivation of the equation for $w=rE$ involves rewriting $d\gamma/d\xi$ on the r.h.s. of \Eq~(\ref{eq:NWE1}) in terms of $\partial_tw$ and $\partial_\xi w$. We here outline the steps of the derivation, omitting the algebra details, and give the final result. For waves with $\gamma^3\ll\sigbg$, it is reduced to a simple statement: $w\approx const$ along $C^+$. The formal derivation can start from $\gamma=\tg\gD$, use the expressions for $d\gD/d\xi$ and $d\tg/d\xi$ in terms of $dE/d\xi$ and $dB/d\xi$, and substitute $d/d\xi= \partial_\xi+(1-\beta_r)^{-1}[\partial_t + (v_\theta/r)\partial_\theta]$ for the derivative along the fluid streamline. $B\partial_\xi B=B_\theta\partial_\xi B_\theta+B_r\partial_\xi B_r$ can be expressed in terms of the derivatives of $E$ (or $w$) using the induction equation $\left.\partial_t\bB\right|_{\boldsymbol{r}} = \left.-c\nabla\times\bE\right|_t$ rewritten in coordinates $(t,\xi)$. The final result is 
\beq
   \left.\frac{1}{E}\frac{d(rE)}{dt}\right|_{C^+} 
   = \frac{\Bbg^2}{2\sigbg E^2} \left( cE^2 f - r\gD^3W\right)
   = {\cal O}\left(\frac{\gamma^3}{\sigbg}\right),
\eeq
where 
\begin{eqnarray}
\nonumber 
    W = \frac{\tg E^2}{B^3}\left[\frac{cEB_\theta}{rB} -\partial_t B+\frac{cB_r\partial_\theta(E\sin\theta)}{rB\sin\theta}\right] + \frac{E}{B^2(1-\beta_r)}\left[ \left(\tg-\frac{\tu\Bbg^r }{\gD B}\right)\left(\partial_t E+\frac{v_\theta}{r}\partial_\theta E\right) 
    - \frac{\tg E}{B}\left(\partial_t B+\frac{v_\theta}{r}\partial_\theta B\right)\right].
\end{eqnarray}

\subsection{Shock formation}

The ratio of \Eqs~(\ref{eq:C+_oblique}) and (\ref{eq:evol_oblique}) that govern the $C^+$ flow gives
\beq
\label{eq:xi+}
  \frac{d\xi_+}{d\gamma}=\frac{\pi mc\, \nbg r}{E^2}=\frac{\pi m c \,\N_\theta}{\mu^2 \K^2}
 \qquad \Rightarrow \qquad  \xi_+
     =\xiin + \frac{\pi m c\,\N_\theta}{\mu^2 \K^2}\,(\gamma-1),
\eeq
where we substituted $E\approx \mu \K/r$ along $C^+$ (which holds for waves with $\gamma^3\ll\sigbg$) and used the initial condition $\gamma=1$ at $\xi_+=\xiin$. The stated relation between $\xi_+$ and $\gamma$ holds along each $C^+$. Note that the plasma Lorentz factor $\gamma=\tg\gD$ is a known function of $E$ and $\bBbg$, and $E=\mu K(\xiin)/r$. Thus, one can express $\xi_+$ in terms of $\xiin$, $r$, and $\theta$. 

Vacuum wave propagation would correspond to $\xi_+=\xiin$ and $r=c(t-\xiin)$ along $C^+$. The MHD correction $\xi_+-\xiin$ may be evaluated using iteration, by substituting the vacuum solution for $r$,
\beq
\label{eq:rvac}
  r=r_{\rm vac}=c(t-\xiin). 
\eeq
Then, the r.h.s. of \Eq~(\ref{eq:xi+}) becomes a known function of $\xiin$ and $t$ (and $\theta$, which is constant along $C^+$).

Deformation of the $C^+$ flow with time, which eventually leads to shocks, is described by $(\partial\xi_+/\partial\xiin)_t$. Viewing $\gamma$ as a composite function $\gamma[E(t,\xiin),\bBbg(t-\xiin)]$, we can write
\beq
  \left.\frac{\partial\gamma}{\partial \xiin}\right|_t 
  =  \left.\frac{\partial\gamma}{\partial E}\right|_{\bBbg}  \left.\frac{\partial E}{\partial \xiin}\right|_t + \left.\frac{\partial\gamma}{\partial \bBbg}\right|_{E} 
 \cdot  \left.\frac{\partial\bBbg}{\partial \xiin}\right|_{t} 
  = \left.\frac{\partial\gamma}{\partial E}\right|_{\bBbg} \left(\frac{\mu \dot{\K}}{r}+\frac{\mu \K}{r^2}\right)
  + \left.\frac{\partial\gamma}{\partial \bBbg}\right|_{E} \cdot \frac{3c\bBbg}{r}.
\label{eq:dg_dxiin}
\eeq
Here $\dot{\K}\equiv dK/d\xiin$, and $(\partial\gamma/\partial E)_{\bBbg}$ can be found from \Eqs~(\ref{eq:tu}) and (\ref{eq:gD2}),
\beq
   \left.\frac{\partial\gamma}{\partial E}\right|_{\bBbg}=Ef,
\eeq 
where $f$ is defined in \Eq~(\ref{eq:dtE1}). For short waves, the term containing $\dot{\K}$ is dominant in \Eq~(\ref{eq:dg_dxiin}), and the other terms are negligible. Thus, we find
\beq
\label{eq:dxi+_dxiin}
   \left.\frac{\partial\xi_+}{\partial \xiin}\right|_t 
   = 1 + \frac{ \pi m c \,\N_\theta \dot{\K}}{\mu^2 \K^3}\left[ E^2f-2(\gamma-1)\right].
\eeq

A caustic appears on $C^+$ that first reaches the condition $(\partial\xi_+/\partial\xiin)_t=0$. It can be found by calculating time $\tv(\xiin)$ at which $(\partial\xi_+/\partial\xiin)_t$ vanishes, and then identifying the characteristic $\xiin^c$ with the minimum $\tv$. The result will also determine the caustic time $\tc=\tv(\xiin^c)$ and the plasma Lorentz factor at the caustic $\gc$. The calculation can be done numerically. Below we derive the result analytically in two limits, $\gc\gg 1$ and $\gc-1\ll 1$.

\subsubsection{Caustics with $\gc\gg 1$}

The limit of $\gamma\gg 1$ corresponds to $\gD\gg 1$ and $E^2\rightarrow B^2$. Note that the ratio $\gamma/\gD=\tg$ remains finite: $\tg^{-1}\approx \sin\alpha$.
\Eq~(\ref{eq:short}) implies
\beq
\label{eq:B2}
    B^2-E^2\approx \Bbg^2+2\Bbg^\theta E. 
\eeq
Hence, $E^2\rightarrow B^2$ corresponds to $B\approx -E \approx \Bbg^2/2\Bbg^\theta$. In this limit, the function $f$ given in \Eq~(\ref{eq:dtE1}) simplifies:
\beq
\label{eq:f1}
  E^2 f\approx \frac{(\Bbg^2+\Bbg^\theta E)\gamma^3}{E^2\,\tg^2} 
   \approx \frac{2\gamma^3\sin^2\!\alpha}{\tg^2}  \approx 2\gamma^3\sin^4\!\alpha \quad (\gamma\gg 1),
\eeq
and \Eqs~(\ref{eq:C+_oblique}) simplify to
\beq
\label{eq:C+_oblique1}
  \left.\frac{d\gamma}{dt}\right|_{C^+} \approx \frac{4c\,\gamma^3\sin^4\!\alpha}{r},  
   \qquad 
  \frac{d\xi_+}{dt} \approx  \frac{4\gamma^3\sin^6\!\alpha}{\sigbg}.
\eeq 
Waves with $\gamma^3\ll\sigbg$ have $d\xi_+/dt\ll 1$, which implies $dt\approx dr/c$, so \Eq~(\ref{eq:C+_oblique1}) can be integrated for $\gamma(r)$,
\beq
\label{eq:gC+}
   \frac{1}{\gamma^2} \approx 8\sin^4\!\alpha\,\ln\frac{\rK}{r} \approx 8x\sin^4\!\alpha,
   \qquad x\equiv\frac{\rK-r}{\rK},
\eeq
where we used $\ln[(1+x)^{-1}]= x+{\cal O}(x^2)$ for $x\ll 1$, which corresponds to $\gamma\gg 1$. Substitution of the obtained $\gamma(r)$ and $r=c(t-\xi_+)$ into \Eq~(\ref{eq:xi+}) gives a cubic equation for $r(t,\xiin)$. Its solution verifies that $r_{\rm vac}-r\ll \rK-r$ when $\gamma^3\ll\sigbg$, and so one can use $r=r_{\rm vac}$ (\Eq~\ref{eq:rvac}) in $x$, i.e. 
\beq
   x=\frac{\rK-c(t-\xiin)}{\rK}.
\eeq
The integration constant $\rK$ in \Eq~(\ref{eq:gC+}) defines the radius where $\gamma$ would diverge, however the characteristic will become terminated at the shock before reaching $r=\rK$. The radius $\rK(\xiin)$ can be found for each $C^+$ with $\K<0$ from $B^2-E^2\approx \Bbg^2+2\Bbg^\theta E = 0$ using $rE = \mu \K$: 
\beq
\label{eq:rK}
   r\Bbg^2 + 2\Bbg^\theta \mu \K = 0 \quad \Rightarrow \quad 
   \rK^2 = - \frac{\sin\theta}{2\K \sin^2\!\alpha}.
\eeq

Substituting \Eq~(\ref{eq:f1}) into \Eq~(\ref{eq:dxi+_dxiin}) and noting that $E^2f\gg 2(\gamma-1)$ when $\gamma\gg 1$, we find
\beq
\label{eq:dxi+_dxiin_}
   \left.\frac{\partial\xi_+}{\partial \xiin}\right|_t 
   = 1 + \frac{2\pi \me c \,\N_\theta \dot{K} \gamma^3 \sin^4\!\alpha}{\mu^2 \K^3}.
\eeq
The $C^+$ characteristic reaches $(\partial\xi_+/\partial\xiin)_t=0$ when 
\beq
\label{eq:gv}
   \gamma^3 = -\frac{\mu^2 \K^3}{2\pi \me c \,\N_\theta \dot{\K} \sin^4\!\alpha}.
\eeq
Using the obtained solution for $\gamma(x)$ (\Eq~\ref{eq:gC+}) we find that $(\partial\xi_+/\partial\xiin)_t=0$ is reached at time 
\beq
   \tv(\xiin)=\xiin +\frac{\rK}{c} - \frac{\rK}{2^{7/3} c \K^2} \left(\frac{\pi \me c \,\N_\theta \, \dot{\K}}{\mu^2 \sin^2\!\alpha}\right)^{2/3}.
\eeq 
The minimum of $\tv$ can be found from $d\tv/d\xiin=0$. Using $2d\rK/\rK=-d\K/\K$ (\Eq~\ref{eq:rK}), we obtain
\beq
  \frac{d\tv}{d\xiin} = 1 - \frac{(t-\xiin)\dot{\K}}{2\K} 
  - \frac{\rK}{2^{4/3} c} \left(\frac{\pi \me c \,\N_\theta}{\mu^2 \sin^2\!\alpha}\right)^{2/3} 
  \frac{\K\ddot{\K}-3\dot{\K}^2}{3\K^3\dot{\K}^{1/3}}.
\eeq
Here, $(t-\xiin)\dot{\K}/2\K\approx \rK \dot{K}/2c\K\gg 1$, and the condition $d\tv/d\xiin=0$ becomes
\beq
  2^{1/3}\dot{\K}^{4/3} = \left(\frac{\pi \me c \,\N_\theta}{\mu^2 \sin^2\!\alpha}\right)^{2/3} 
  \frac{3\dot{\K}^2-\K\ddot{\K}}{3\K^2}.
\eeq
This equation determines the Lagrangian coordinate $\xiin^c$ of the caustic in the $C^+$ flow with a given $\K(\xiin)$. In particular, for a wave with an initial sine profile, $\K=\Km\sin(\omega\xiin)$, it gives 
\beq
   \cos^{4/3}(\omega\xiin^c) = \left(\frac{\pi \me c \,\N_\theta\, \omega}{\sqrt{2} \mu^2\Km^2 \sin^2\!\alpha}\right)^{2/3} 
  \left[\frac{1}{3}+\cos^2(\omega\xiinc)\right]
  \approx \frac{1}{3}\left(\frac{\pi \me c \,\N_\theta\, \omega}{\sqrt{2} \mu^2\Km^2 \sin^2\!\alpha}\right)^{2/3}.
\eeq
The last (approximate) equality took into account the condition $\gc\gg 1$  ($E\approx -B$), which implies that $\xiinc$ is close to the minimum of the sine profile of $\K(\xiin)$, and so $\cos(\omega\xiinc)\ll 1$. The obtained $\xiinc$ determines $\dot{\K}(\xiinc)=\omega\Km\cos(\omega\xiinc)$, and then from \Eq~(\ref{eq:gv}) we find the plasma Lorentz factor at the caustic, 
\beq
\label{eq:gc1}
    \gc \approx \left(\frac{3}{2}\right)^{1/4}
    \left(\frac{\mu^2\Km^2}{\pi \me c \,\N_\theta\, \omega\sin^2\!\alpha}\right)^{1/2}
    \approx 
    \frac{(3/2)^{1/4}}{\sqrt{\zeta \sin\theta\sin\alpha}}.
\eeq
In the last equality we subsituted $\N_\theta$ given by \Eq~(\ref{eq:N_theta}) and used the parameter $\zeta=\mu^2\Km^2/\pi mc\N$ defined in the equatorial plane (\Eq~\ref{eq:zeta}). At $\theta=\pi/2$, \Eq~(\ref{eq:gc1}) reproduces $\gc=(2\cc)^{-1}$ given in \Eq~(\ref{eq:caustic}) and derived in Paper~I. The obtained extension to $\theta\neq \pi/2$ shows that $\gc$ increases outside the equatorial plane. Recall that this result was derived assuming $\gc \gg 1$. This regime holds for waves with $\zeta\ll 1$, as one can see from \Eq~(\ref{eq:gc1}).

\subsubsection{Caustics with $\gc-1\ll 1$}

The $C^+$ flow with $\gamma-1\ll 1$ can be described using expansion in variable $z\equiv E/\Bbg$, $|z|\ll 1$. From \Eqs~(\ref{eq:tg}) and (\ref{eq:B2}) we find
\beq
   \tu={\cal O}(z^2),  \qquad  \tg=1 + {\cal O}(z^4), 
  \qquad B=\Bbg\left[1+z\sin\alpha+{\cal O}(z^2)\right].
\eeq
This determines $\bD=E/B$, $\gD$, and
\beq
   \gamma=\tg\gD = 1 + \frac{z^2}{2}\left(1-2z\sin\alpha\right)+{\cal O}(z^4).
\eeq
Then, \Eq~(\ref{eq:f}) gives
\beq
   E^2f= z^2(1-3z\sin\alpha)+{\cal O}(z^4).
\eeq
Substituting these expansions into \Eq~(\ref{eq:dxi+_dxiin}), we obtain
\beq
\label{eq:dxi+_dxiin2}
   \left.\frac{\partial\xi_+}{\partial \xiin}\right|_t 
   = 1 - \frac{ \pi \me c \,\N_\theta \dot{\K}}{\mu^2 \K^3}\,z^3\sin\alpha
   = 1 - \frac{ \pi \me c \,\N_\theta \dot{\K}}{\mu^2}\,\frac{\sin^4\!\alpha}{\sin^3\!\theta}\, r^6.
\eeq
Here, we substitute $z\equiv E/\Bbg = \mu \K/r\Bbg = r^2\K\sin\alpha/\sin\theta$ and find that $\partial\xi_+/\partial\xiin$ vanishes when the characteristic reaches the radius
\beq
   r_{\rm v}(\xiin)=\left(\frac{\mu^2\sin^3\theta}{\pi \me c\,\N_\theta \dot{\K} \sin^4\alpha}\right)^{1/6}.
\eeq
The caustic appears where $r_{\rm v}$ is minimum, i.e. where $\dot{\K}(\xiin)$ reaches its maximum $\dot{\K}=\omega\Km$ (which occurs at $\xiin=0$). Thus, we find that the caustic appears when the wave reaches the radius
\beq
   \rc = \left(\frac{\mu^2\sin^3\theta}{\pi \me c\,\N_\theta\, \omega\Km\sin^4\alpha}\right)^{1/6}.
\eeq
Using the definition of $\zeta$ (\Eq~\ref{eq:zeta}), $\rsh$ (\Eq~\ref{eq:rsh}), $\N_\theta$ (\Eq~\ref{eq:N_theta}), and $\Km^{-1}=2\Rm^2$, one can rewrite $\rc$ as stated in \Eq~(\ref{eq:rc}).

 \newpage

\bibliography{ms}

\end{document}